\documentclass[hyper]{JHEP3}

\usepackage{graphicx}
\usepackage{epstopdf}

\input{epsf}
\usepackage{epsfig}
\usepackage{amssymb}
\usepackage{amsfonts}
\usepackage{amsbsy}
\usepackage{amsmath}

\def\IC{\mathbb{C}}

\def\IZ{{\mathbb{Z}}}
\def\IR{{\mathbb{R}}}
\def\IP{\mathbb{P}}

\def\ICP{\mathbb{CP}}

\def\CM {{\cal M}}

\def\CN {{\cal N}}

\def\CL {{\cal L}}

\def\CO {{\cal O}}

\def\one{{\hbox{ 1\kern-.8mm l}}}

\title{D-brane Deconstructions in IIB Orientifolds}

\author{Andr\'es Collinucci$^1$, Frederik~Denef$^{3,2}$ and Mboyo Esole$^3$\\

$^1$ Institute for Theoretical Physics, Vienna University of Technology, \\
Wiedner Hauptstr. 8-10, 1040 Vienna, Austria\\

$^2$ Instituut voor Theoretische Fysica, KU Leuven, \\
Celestijnenlaan 200D, B-3001 Leuven, Belgium \\
\\
$^3$ Jefferson Physical Laboratory, Harvard University, \\
Cambridge, MA 02138, USA\\
}

\abstract{With model building applications in mind, we collect and develop basic techniques to analyze the landscape of D7-branes in type IIB compact Calabi-Yau orientifolds, in three different pictures: F-theory, the D7 worldvolume theory and D9-anti-D9 tachyon condensation. A significant complication is that consistent D7-branes in the presence of O7$^-$ planes are generically singular, with singularities locally modeled by the Whitney Umbrella. This invalidates the standard formulae for charges, moduli space and flux lattice dimensions. We infer the correct formulae by comparison to F-theory and derive them independently and more generally from the tachyon picture, and relate these numbers to the closed string massless spectrum of the orientifold compactification in an interesting way. We furthermore give concrete recipes to explicitly and systematically construct nontrivial D-brane worldvolume flux vacua in arbitrary Calabi-Yau orientifolds, illustrate how to read off D-brane flux content, enhanced gauge groups and charged matter spectra from tachyon matrices, and demonstrate how brane recombination in general leads to flux creation, as required by charge conservation and by equivalence of geometric and gauge theory moduli spaces.

}

\begin{document}

\section{Introduction}

Type IIB O3/O7 orientifold flux compactifications and their F-theory avatars currently appear to be the most promising corner of the string theory landscape for controlled, realistic model building. The virtues of this class of models include:
\begin{itemize}
 \item Complex structure moduli, dilaton and D7-brane moduli can be stabilized classically at high mass scales by RR, NSNS and D7 worldvolume fluxes, and plausible stabilization mechanisms for the K\"ahler moduli based on quantum corrections have been proposed \cite{Kachru:2003aw,Balasubramanian:2005zx} and studied in detail, as reviewed in \cite{Douglas:2006es,Grana:2005jc,Denef:2008wq}.
 \item There is a very high degree of discrete tunability of physical parameters, which helps in producing controlled models. In particular the cosmological constant can in principle be discretely tuned to become extremely small, easily of the order of the measured cosmological constant or less \cite{Bousso:2000xa,Denef:2004ze}.
 \item The classical geometry of the compactification manifold remains Calabi-Yau after turning on fluxes, up to warping \cite{Giddings:2001yu}. This means in particular that many of the powerful techniques from algebraic geometry can still be used to describe these vacua.
 \item Strongly warped throats of Klebanov-Strassler type \cite{Klebanov:2000hb} occur naturally, generating large scale hierarchies \cite{Giddings:2001yu}.
 \item Slow roll inflation can be accommodated, at least in fine tuned local models \cite{Kachru:2003sx,Baumann:2007np,Baumann:2007ah,Krause:2007jk}.
 \item The F-theory description provides $g_s$ corrections to the geometry which smooth out the O7 singularities \cite{Sen:1996vd}.
 \item There is a rich set of explicit D-brane constructions possible in these models, useful for particle physics model building; for an extensive resp.\ short review see \cite{Blumenhagen:2006ci,Marchesano:2007de}. More recently, \cite{Donagi:2008ca,Beasley:2008dc} have initiated a program of model building in F-theory. This framework naturally incorporates GUTs and promises to be a powerful and elegant unifying geometrical framework for string phenomenology.
 \end{itemize}
Despite all this, there are still significant gaps in our understanding of the landscape of type IIB orientifold  vacua, in particular in our understanding of the landscape of D7-brane configurations and its interplay with moduli stabilization and other global issues.

Simple, flat, intersecting D7/D3-brane constructions are well understood by now in the context of toroidal orientifolds, as reviewed in \cite{Blumenhagen:2006ci}, but these models are just a tiny subset of all possibilities, miss some desirable features such as warping and do not incorporate complete moduli stabilization.

The combinatorics of globally well defined D-brane models is systematically understood in the framework of boundary states of CFTs at Gepner points of general type II Calabi-Yau orientifold compactifications \cite{Blumenhagen:2003su,Blumenhagen:2004cg,Brunner:2003zm,Brunner:2004zd,Dijkstra:2004cc,
Anastasopoulos:2006da,GatoRivera:2007yi}, but this analysis is by definition done at a special, nongeometric point in closed string moduli space and most of these studies did not yet address moduli stabilization and the parameter discretuum, as the tools to do this in this nongeometric regime have only been developed fairly recently \cite{Becker:2006ks}.

Rather general geometrical D-brane setups have been studied systematically in ``bottom up'', local, noncompact Calabi-Yau varieties decoupled from gravity (for example in \cite{Donagi:2008ca,Beasley:2008dc,Aldazabal:2000sa,Berenstein:2001nk,
Verlinde:2005jr,Buican:2006sn,Malyshev:2007zz,Heckman:2007zp,Wijnholt:2007vn}). Similarly, the models for slow roll inflation mentioned above are also essentially local models, with compactification effects only relatively crudely taken into account. The same is true for many recent models of dynamical supersymmetry breaking in string theory (such as \cite{Franco:2006es,GarciaEtxebarria:2006rw,Ooguri:2006pj,Ooguri:2006bg,Argurio:2006ny,Aganagic:2006ex,Giveon:2007fk,
Argurio:2007qk,Aharony:2007db,Aganagic:2007kd,Aganagic:2007py,Aganagic:2007zm}).

Eventually, all of these constructions need to be embedded in finite volume, fully moduli stabilized compactifications with broken supersymmetry, a tiny positive cosmological constant, the correct spectrum of couplings and masses, effective scalar potentials admitting slow roll inflation, and so on. Completing local models to genuine string vacua reduces the number of possible constructions from a continuous infinity to a finite, discrete set, and may destroy the desired features the local or continuously tuned model was designed to have. For example, an almost flat potential on the moduli space of a local model, suitable for slow roll inflation, will generically develop steep slopes in the direction of additional moduli induced by the embedding. An intersecting brane model will often be forced into recombination by turning on moduli stabilizing background fluxes, breaking the desired gauge symmetries. Global tadpole cancelation constraints generically require the presence of additional D-branes, potentially intersecting the local construction and producing additional, unwanted matter fields charged under the standard model gauge group.

Beyond these immediate model building concerns, one would like to address questions such as how many --- if any --- string vacua compatible with observational constraints we can expect to exist, and, more ambitiously, what the proper notion of naturalness or even probability is in the context of the string theory landscape. In other words, we would like to know whether the requirement of the existence of a consistent UV (quantum gravity) completion is sufficiently constraining to be predictive at low energies. To analyze these questions, statistical methods have been proposed and developed \cite{Douglas:2003um,Ashok:2003gk,Denef:2004ze,Denef:2004cf,Acharya:2005ez,Douglas:2005df,Dienes:2006ut,Blumenhagen:2004xx,Gmeiner:2005vz,
Gmeiner:2005nh,Gmeiner:2006vb,Douglas:2006xy}. Statistical analysis crucially relies on global constraints and topological parameters, such as tadpole cancelation, brane charges and flux lattice dimensions. For example, the number of critical points $DW=0$ of an ensemble of flux superpotentials, for IIB RR/NSNS, F-theory or D7 worldvolume fluxes compatible with D3 tadpole cancelation without adding anti-D3 branes, in a region $\CM$ of the appropriate moduli space, is estimated\footnote{This estimate is reliable in the regime $4 \pi e Q \gg b$, which is often not satisfied in unrestricted F-theory flux ensembles. For further discussion see \cite{Denef:2008wq} section 6.3.} by a formula of the general form (see \cite{Denef:2008wq} section 6 for a derivation):
\begin{equation} \label{vaccount}
 N_{\rm vac} \approx \frac{(2 \pi Q)^{b/2}}{(b/2)!} \, \int_{\CM} e(D) \, ,
\end{equation}
where $-Q=-Q^{(F)}$ is the total curvature induced D3-brane charge of the compactification (in F-theory unit conventions, see footnote \ref{IIBFconv}), $b$ is the dimension of the flux lattice and $e(D)$ is the euler density of the connection $D$ appearing in the critical point condition $DW=0$. Distributions of discrete D-brane data such as enhanced gauge groups and charged matter content have so far only been studied in very limited and simple ensembles. In particular, no systematic studies of the statistics of D-brane configurations in general Calabi-Yau orientifold flux vacua have been done. On the other hand, contrary to what is sometimes tacitly assumed, in typical IIB orientifold or F-theory compactifications, virtually \emph{all} of the degeneracy of flux vacua actually comes from worldvolume fluxes in the D-brane sector (the ``open string landscape'' \cite{Gomis:2005wc}). For instance in the example we will study throughout the paper, the IIB orientifold obtained as the weak coupling limit of F-theory on the elliptic fibration over $\ICP^3$, we will see that $Q=972$ and that the number of bulk (RR and NSNS) fluxes is $b_{\rm bulk} = 300+300=600$, while the number of D7-brane worldvolume fluxes is $b_{\rm brane} = 23320$. As a result, the above estimate gives $N_{\rm vac} \sim 10^{500}$ for the bulk sector, but $N_{\rm vac} \sim 10^{2000}$ for the D-brane sector!

For all of these reasons, it would seem desirable to develop a more systematic top-down, approach to the D-brane sector of generic, compact type IIB O3/O7 orientifold vacua in the geometric regime, in a way suitable for concrete model building including moduli stabilization and for statistical analysis. In fact most of the necessary ingredients for such an approach are already present in the literature, although often in a somewhat abstract and formally involved form. For example, there exists an extensive categorical framework for D-branes in Calabi-Yau manifolds \cite{Brunner:1999jq,Douglas:2000gi,Aspinwall:2004jr,Herbst:2008jq}, and in principle the construction of vacua with certain desired properties is, at sufficiently large Calabi-Yau volume, a well defined problem in standard algebraic geometry. There have indeed been several model building studies already where fully compact geometries were considered for nontoroidal Calabi-Yau O3/O7 orientifolds, sometimes even including complete moduli stabilization. This includes (with varying degrees of thoroughness and with varying degrees of assumptions and modeling) \cite{Blumenhagen:2002wn,Blumenhagen:2003vr,Lust:2004fi,Marchesano:2004xz,Denef:2004dm,Denef:2005mm,Lust:2005bd,
Lust:2005dy,Lust:2006zg,Diaconescu:2005pc,Diaconescu:2006nk,Diaconescu:2007ah,Conlon:2005ki,Conlon:2006gv,
AbdusSalam:2007pm,Blumenhagen:2007sm}. In particular in \cite{Blumenhagen:2007sm} the interaction between the particle physics D-brane sector and moduli stabilization mechanisms was studied, and it was found that the two could \emph{not} be considered independently, providing further motivation for a holistic (as opposed to modular) approach to D-brane model building.

However, as we will see, the \emph{generic} D7-brane sector in general Calabi-Yau orientifold compactifications presents a number of complications and puzzles. By ``generic'' we mean a D7-brane with minimal possible gauge group --- in the case at hand this means no continuous gauge group at all. This differs from the usual D7-image-D7 pairs or invariant multiple brane stacks, which do have nontrivial gauge groups. A single D7-image-D7 pair, which has gauge group $U(1)$, will typically be mutually intersecting. This leads to the possibility of recombination into a single component, generic D7, whereby the $U(1)$ gets Englert-Brout-Higgsed. One of the complications alluded to above is now that, as we will see, in the presence of an O7$^-$, such a generic D7 will be \emph{singular}: Locally near the O7$^-$, the D7 is forced to retain the structure of a D7-image-D7 pair, but globally it becomes a single connected object, and this causes the occurrence of pinch point singularities. Locally, these singularities are modeled by the so-called \emph{Whitney Umbrella} in $\IC^3$ (defined in section \ref{sec:singstruct} and illustrated in fig.\ \ref{whitney}).

We will confirm that generic D7-branes are locally Whitney Umbrellas in three independent ways: by taking the weak coupling limit of F-theory (also noted in \cite{Braun:2008ua}), by requiring Dirac quantization in perturbative IIB string theory, and from the shape of tachyon condensates of consistent D9-anti-D9 systems. As a result of the presence of these non-orbifold singularities, the usual index formulae computing the quantities appearing in (\ref{vaccount}) --- the curvature induced D3-charge $-Q$, the number of D7 moduli $n$ and the worldvolume flux lattice dimension $b$ --- fail. Going to nongeneric, nonsingular D-brane configurations does not help to compute these numbers, as these  configurations are in fact typically in different physical components of the moduli space of the theory, with different (smaller) values for D3-charge, flux lattice dimension and number of moduli.


Another general problem is to match D-brane gauge theory degrees of freedom to geometric moduli. A puzzle which arises here is that the number of gauge theory degrees of freedom of $N$ coincident branes scales as $N^2$, while the number of geometric moduli of a degree $N$ supersymmetric 4-cycle in general scales as $N^3$.


Dealing with the above issues has been a carefully sidestepped problem in much of the existing model building literature --- for example in the ``better racetrack'' models of \cite{Denef:2004dm} the D7-brane sector was left largely unspecified and in \cite{Denef:2005mm} potential problems with D7-branes were circumvented by considering a rather special compactification in which all D7-branes were rigid and coincident with the O7-planes. To the best of our knowledge, in the existing model building literature, only nongeneric D7-branes such as brane-image-brane pairs and invariant, nonsingular stacks have been considered.

It is clear though that in order to discuss D7 moduli stabilization by fluxes, or do systematic studies of the landscape of IIB orientifold compactifications, statistical or otherwise, one must face and fully resolve these issues.

One approach would be to directly work in F-theory instead of in its weakly coupled IIB limit, computing for example the D3-tadpole $Q$ using techniques such as those employed in \cite{Klemm:1996ts,Andreas:1999ng,Denef:2008wq}. However, if one wants to just build type IIB orientifold models, it is often desirable not to have to make the detour of finding a suitable F-theory completion. One could also avoid IIB orientifolds altogether, but in strongly coupled F-theory compactifications there is no general canonical distinction between bulk and localized degrees of freedom such as in the weakly coupled IIB limit, making the identification of gauge theory content somewhat difficult in compact models. This to date remains to be fully understood for F-theory compactifications down to four dimensions. Furthermore, one cannot use string perturbation theory to compute for example quantum corrections to the K\"ahler potential, some of which play a crucial role in the moduli stabilization scenario of \cite{Balasubramanian:2005zx}.

We believe it will therefore be useful to set up a concrete computational framework to systematically analyze the D7-brane sector in type IIB O3/O7 orientifolds in the weak coupling, large volume limit, collecting and developing various complementary approaches, in a maximally accessible way. This is the goal of this paper. Specifically, we use the F-theory, D7 and D9-anti-D9 tachyon condensation pictures to arrive at the following:
\begin{itemize}
  \item We derive concrete formulae for topological physical quantities such as the curvature induced D3-charge, the number of moduli and the dimension of the flux lattice, for any component of the moduli space. These are in complete agreement with the F-theory results when the latter are available, and with ``K-theory'' results obtained from D9-anti-D9 annihilation.
      Using fixed point index theorems, we furthermore relate in an interesting way for example $Q$ and the numbers of various massless particles in four dimensions.
 \item Starting from the D9-anti-D9 tachyon condensation picture, we explain how to compute in general the worldvolume and flux content of the resulting D7-branes and how this can change under brane recombination, as required by charge conservation.

     To do this we generalize and adapt to orientifolds the analysis and constructions of \cite{Gaiotto:2005rp,Gaiotto:2006aj}, where flux configurations on space-localized D4-branes were studied in (non-orientifolded) toric Calabi-Yau manifolds and traced back to D6-anti-D6 bound states --- a special case of the general D-brane bound state constructions formalized in \cite{Douglas:2000gi} (and reviewed in \cite{Aspinwall:2004jr}).

\item We illustrate with simple examples how to concretely compute charges, moduli, gauge groups and matter content in the D9-anti-D9 picture, using only polynomial manipulations.

    It was pointed out in \cite{Gaiotto:2005rp} that the presence of fluxes induced by brane recombination resolves the apparent discrepancy between geometric and gauge theoretic moduli mentioned earlier, and indeed the equivalence of gauge theory and geometry is manifest in the tachyon condensation picture.

    We also encounter some interesting subtleties, such as the necessity to have an even number of D9-image-D9 pairs in the presence of O7$^-$ planes, which turns out to be crucial to match the F-theory results.

  \item We give a method for explicit construction of D7 flux vacua using holomorphic curves, which is much more tractable than solving the superpotential critical point condition (which in our basic example involves extremizing a tadpole-respecting integral linear combination of 23320 3-chain periods over 3728 variables...). This is analogous to the constructions used in \cite{Gaiotto:2005rp,Gaiotto:2006aj,Gaiotto:2006wm,Denef:2007vg} to build and enumerate flux configurations on D4-branes in non-orientifolded Calabi-Yau manifolds.
\end{itemize}

A summary of our results and some ready-to-use formulae can be found in section \ref{sec:summary}, which the reader may consult before going to the detailed discussion.

This paper is meant to be expository, and throughout we focus on a particular example to illustrate the main ideas, making some generalizations as we go, as well as separately in section \ref{sec:generalizations}. We will not try to be complete; for much more advanced and comprehensive K-theoretic and categorical descriptions of orientifolds we refer to the upcoming works \cite{mooretexas,hori}. Our approach will be more elementary. All the geometrical tools we will rely on can be found in section 5 of the lecture notes \cite{Denef:2008wq}, and we will frequently refer to it.

\section{F-theory picture}

\subsection{Introduction} \label{sec:miniintro}

We start by briefly sketching the F-theory picture \cite{Vafa:1996xn} and its relation to IIB string theory and M-theory. An extensive introduction to this can be found in the lecture notes \cite{Denef:2008wq}.

F-theory can be defined as type IIB string theory continued away from its weak string coupling limit, just like M-theory can be defined as type IIA continued away from weak coupling. In practice it usually refers to type IIB compactifications on a manifold $B$ with dilaton-axion $\tau = C_0 + e^{-\phi}$ varying over $B$, described in the supergravity approximation and by giving $\tau$ the interpretation of the modular parameter of a 2-torus $T^2$ fibered over $B$. This interpretation gives an elegant and useful geometrization of such backgrounds: Supersymmetric configurations correspond to Calabi-Yau manifolds $Z$ elliptically fibered (with section) over $B$, the $SL(2,\IZ)$ gauge symmetry of IIB string theory corresponds to the geometrical $SL(2,\IZ)$ reparametrization symmetry of the $T^2$, and $(p,q)$ 7-branes on $B$ correspond to degeneration loci on $B$ where a $(p,q)$ 1-cycle of the $T^2$ degenerates. The surprising effectiveness of this higher dimensional geometric picture is best explained by noting that M-theory on the same elliptically fibered manifold $Z$, in the limit of vanishing fiber size, is dual to type IIB on the base $B$ of the elliptic fibration. Indeed M-theory on a small $T^2$ equals weakly coupled IIA on a small circle, which is T-dual to type IIB on a large circle. Applying this fiberwise to $Z$ yields type IIB compactified on $B \times S^1$, with the $S^1$ decompactifying in the limit in which the original fiber is sent to zero size. This duality allows one to use the geometrical objects of M-theory to define and analyze F-theory compactifications.

Of particular interest to us are F-theory compactifications on Calabi-Yau fourfolds, i.e.\ type IIB on their three complex dimensional base manifolds. For genuine $SU(4)$ holonomy fourfolds, this gives rise to an effective $\CN=1$ supergravity theory in four dimensions. The scalars in the massless chiral multiplets in this theory arise from the K\"ahler moduli of $B$, the complex structure moduli of $Z$ (which includes the IIB dilaton-axion and 7-brane moduli), D3-brane moduli and various axions.

In the following, we will give some physically important topological numbers associated to a particular example of an F-theory fourfold compactification. This includes the curvature induced D3 tadpole, the number of D7-brane moduli in the weak IIB coupling limit, and the number of fluxes that can be turned on. In the next section we will use these results as guidance to infer rules on how to compute these numbers directly in the perturbative IIB orientifold picture, and in particular how to deal with the D7 worldvolume singularities complicating the analysis there.

\subsection{Fourfold data and D3 tadpole} \label{sec:fourfold}

The starting point of our main example is F-theory on the Calabi-Yau
fourfold elliptically fibered over $\ICP^3$, which is described by
the equation
\begin{equation} \label{CY4eq}
 Z: y^2 = x^3 + f(u) \, x \, z^4 + g(u) \, z^6
\end{equation}
with projective $\IC^*$ equivalences
\begin{eqnarray}
 (u_1,u_2,u_3,u_4,x,y,z) &\simeq& (\lambda u_1,\lambda u_2,\lambda
u_3,\lambda u_4,\lambda^8 x,\lambda^{12} y,z) \\
 &\simeq& (u_1,u_2,u_3,u_4,\mu^2 x,\mu^3 y,\mu z),
\end{eqnarray}
where $(u_1,u_2,u_3,u_4) \neq (0,0,0,0)$ and $(x,y,z) \neq (0,0,0)$.
Here $f(u)$, $g(u)$ are homogeneous polynomials of degrees
16 resp.\ 24. Note that the projective equivalences define a
$W\ICP^3_{2,3,1}$ fiber bundle over $\ICP^3$, the fiber being
parametrized by $(x,y,z)$ and the base by $(u_1,u_2,u_3,u_4)$. At
fixed $u$, (\ref{CY4eq}) describes an elliptic curve in
$W\ICP^3_{2,3,1}$, hence this equation indeed defines an elliptic
fibration over $\ICP^3$. Moreover the elliptic fibration has a
section, obtained by putting $z=0$, which is up to projective
equivalences the surface $(x,y,z)=(1,1,0)$ with $\vec u \in \ICP^3$
arbitrary.

A derivation of the geometric data of this example using relatively basic methods can be found in sections 5.7 and 5.9 of \cite{Denef:2008wq}, or can be extracted from the general and more advanced discussion in \cite{Klemm:1996ts}.
We quote here the relevant results.

The nontrivial Hodge numbers of $Z$ are
\begin{equation}
 h^{1,1}=2,\quad h^{2,1}=0, \quad h^{2,2}=15564,\quad
 h^{3,1}=3878,
\end{equation}
so the number of complex structure moduli is $h^{3,1}=3878$ (which
can also be computed directly by counting the number of coefficients
of $f$ and $g$ modulo $GL(4,\IC)$ coordinate transformations: ${16+3
\choose 3} + {24+3 \choose 3} - 16 = 3878$), and
\begin{equation}
 \chi = 23328, \qquad b_4 = 23322.
\end{equation}
A particular basis for $H^{1,1}(Z,\IZ)$ is $\{K_1,K_2\}$, where $K_1$ is Poincar\'e dual to the divisor $[u_1 = 0]$, and $K_2$ to the divisor $[z=0] + 4 \, [u_1=0]$. This is in fact a basis for the K\"ahler cone, i.e.\ we can parametrize the K\"ahler form on $Z$ as $J_Z=\xi^1
K_1 + \xi^2 K_2$ with $\xi^1, \xi^2 > 0$. In the F-theory limit, the size of the
base is $(\xi^1)^3/6$ and the size of the elliptic fiber is $\xi^2 \to 0$.

The curvature induced D3-tadpole\footnote{\label{IIBFconv} There is a difference in standard conventions used in F-theory and in type IIB orientifolds. In F-theory, one mobile D3 has one unit of D3-charge. In the corresponding IIB description, this corresponds to a D3 together with its orientifold image on the Calabi-Yau double cover of the base of the elliptic fibration, and this usually gets assigned charge 2. Thus $Q^{(\rm IIB)}_c \equiv 2 \, Q^{(\rm F)}_c$. We furthermore define the sign of D3-charge such that D7 and +D3 branes are mutually supersymmetric.} is $-Q^{(F)}_c$, with \cite{Sethi:1996es}
\begin{equation} \label{Fcurvindcharge}
 Q_c^{(F)} = \frac{\chi(Z)}{24} = 972 \, .
\end{equation}
This together with the charge of mobile D3-branes and 4-form fluxes $G$ must add up to zero:
\begin{equation} \label{Ftadpole}
 - Q_c^{(F)} + N_{D3} + \frac{1}{2} \int_Z G \wedge G = 0 \, .
\end{equation}
Reproducing this number $Q_c^{(F)}$ from the perturbative IIB orientifold picture will be our first task in the next section.

\subsection{Weak coupling orientifold limit and D7 moduli} \label{sec:orlimit}

The IIB dilaton-axion $\tau$ is identified with the modular parameter of the elliptic fiber, and determined by
\begin{equation}
 j(\tau) = \frac{ 4 \cdot (24 f)^3}{\Delta} \, \qquad \Delta := 4 f^3+ 27 g^2 = 0 \, ,
\end{equation}
where $j(\tau)$ is Klein's modular invariant function
$j(\tau)=e^{-2\pi i \tau} + 744 + \CO(e^{2 \pi i \tau})$. The
7-branes are localized where the fibration degenerates, i.e.\ at
\begin{equation}
 \Delta(u) = 0 \, .
\end{equation}
To make contact with the weak coupling perturbative IIB orientifold picture, we
follow Sen's procedure \cite{Sen:1996vd,Sen:1997gv}. We parametrize, without loss of
generality,
\begin{eqnarray}
 f &=& -3h^2 + \epsilon \eta,\nonumber\\
 g &=& -2h^3 + \epsilon h \eta - \epsilon^2 \chi/12,
\end{eqnarray}
where $h$, $\eta$ and $\chi$ are a homogeneous polynomials of
degrees 8, 16 and 24 in the $u_i$, and $\epsilon$ is a constant. When $\epsilon
\to 0$ keeping everything else fixed, one finds
\begin{equation}
 \Delta \approx - 9 \, \epsilon^2 h^2 (\eta^2 - h \chi),
 \qquad j(\tau) \approx \frac{(24)^4}{2} \, \frac{h^4}{\epsilon^2(\eta^2 - h
 \chi)}.
\end{equation}
Thus, in this limit, $g_s = \frac{1}{{\rm Im} \, \tau} \sim -\frac{1}{\log |\epsilon|} \to 0$ everywhere
except near $h=0$, and the $\epsilon \to 0$ limit can therefore be
interpreted as a IIB weak coupling limit. A monodromy analysis
\cite{Sen:1997gv} reveals that in this limit the two components of
$\Delta=0$ should be identified with an O7-plane and a D7-brane in a Calabi-Yau orientifold as
follows:
\begin{equation} \label{O7D7eq}
 O7: h(u)=0 \, , \qquad D7: \eta(u)^2 = h(u) \, \chi(u) \, ,
\end{equation}
where the Calabi-Yau 3-fold is given by the equation
\begin{equation} \label{CYeq}
 X: \xi^2 = h(u)
\end{equation}
with $\IC^*$ equivalence $(u_1,u_2,u_3,u_4,\xi) \simeq (\lambda
u_1,\lambda u_2,\lambda u_3,\lambda u_4,\lambda^4 \xi)$,
orientifolded by the involution
\begin{equation}
 \sigma: \xi \to - \xi.
\end{equation}
The CY threefold $X$ is a double cover of $\ICP^3$ branched over
$h(u)=0$; moding out by $\sigma$ gives back $\ICP^3$. It has
$h^{2,1}=149$ complex structure deformations, given by the
coefficients of $h$ modulo $GL(4,\IC)$ coordinate transformations,
and $h^{1,1}=1$ K\"ahler deformation. In addition to this, there are
D7-brane moduli, counted by the number of inequivalent deformations
of (\ref{O7D7eq}), i.e.\ ${16+3 \choose 3} + {24+4 \choose 3} - {8+3
\choose 3} - 1 = 3728$. The first subtraction comes from the
fact that we can shift $\eta \to \eta + h \psi$ with $\psi$ an
arbitrary degree 8 polynomial and shift $\chi$ accordingly, without
changing the form of the equation (\ref{O7D7eq}), and the last
subtraction corresponds to overall rescaling of the coefficients. As
a check note that indeed the number of D7 moduli plus the number of
3-fold complex structure moduli plus one for the dilaton-axion
equals 3878, the number of fourfold complex structure moduli.

In conclusion, we find that the number of D7 moduli is
\begin{equation}
 {\rm dim} \, \CM_{\rm D7} = 3728 \, .
\end{equation}
Reproducing this number from IIB orientifold data will be our second task in the next section.

\subsection{Fluxes}

In M-theory one can turn on general 4-form fluxes $G \in H^4(Z,\IZ)$
on $Z$.\footnote{In the present example $c_2(Z)$ is even, so there is no half integral shift of $G$.} At first sight one would therefore conclude that the number of F-theory fluxes is $b_4$. However, this is not quite correct. Because the M-F duality described in section \ref{sec:miniintro} turns one M-theory elliptic fiber direction
into a IIB 4d spatial direction, not all $G$-fluxes dualize to 4d
Poincar\'e invariant fluxes in IIB.
Roughly speaking we need one and only one leg of the flux to be on the elliptic fiber. A more precise way of thinking about this
is in terms of the domain walls that source the fluxes (see also \cite{Denef:2008wq}). In
M-theory these are M5 branes wrapping 4-cycles $S$ in $Z$ --- this
will produce a flux $G$ Poincar\'e dual to $S$. Now if $S$ wraps the
complete elliptic fiber and a complex curve $C$ in the base, then in IIA
this becomes a D4 wrapping $S$ and the $S^1$ which gets T-dualized to go to IIB, resulting in a D3
wrapping $C$. This is a \emph{string} in the four noncompact
dimensions, and will clearly not produce a Poincar\'e
invariant flux. Similarly, if $S$ is completely transversal to the
elliptic fiber, then we end up with a IIB KK monopole wrapped on $S$,
which is again a stringlike object in four dimensions. If on the other hand $S$ wraps only a 1-cycle of the elliptic fiber, we end up with a 5-brane domain wall in IIB, which sources appropriate Poincar\'e invariant fluxes.

Now, when $S$ is the intersection of two regular divisors in $Z$, it will always
be of one of the two unacceptable types. Since divisor classes are in $H^{1,1}$, fluxes Poincar\'e dual to these are elements of $H^{1,1} \wedge H^{1,1}$. Therefore, to get good F-theory flux vacua, we should take our fluxes $G \in H^4$ to be perpendicular to $H^{1,1} \wedge H^{1,1}$,\footnote{If $Z$ has reduced holonomy or has singularities such as those giving rise to enhanced gauge symmetries, the situation is more subtle. We will not analyze these cases here.} i.e.\ be contained in
\begin{equation}
 H^{4\prime}(Z,\IZ) = \{ G \in H^4(Z,\IZ) \, | \, \int_Z G \wedge K_A \wedge K_B = 0 \quad \forall A,B
 \} \, ,
\end{equation}
where $\{ K_A \}_A$ is a basis of $H^{1,1}(Z)$.

Naively this reduces the lattice of allowable fluxes by three in our example, but there is in fact a quadratic relation $K_2^2 = 4 \, K_1 K_2$ for the basis introduced above (as derived in \cite{Denef:2008wq} section 5.7), so we get a reduction by two instead. In conclusion, the dimension of the lattice of allowable F-theory fluxes is
\begin{equation} \label{b4perp}
 b = b_4' = b_4-2 = 23320.
\end{equation}
In the weak coupling perturbative IIB picture, part of these fluxes correspond to bulk NS-NS and R-R fluxes, and the remainder to D7 $U(1)$ worldvolume fluxes. Reproducing the number $b$ above from IIB orientifold data will be our third task in the next section.

\section{Weak coupling geometric IIB picture}

\subsection{The orientifold and the D7 double intersection property} \label{sec:theorientifold}

As we saw in the previous section, the weak coupling orientifold description of our example starts from a Calabi-Yau $X$ given by the equation $\xi^2 = h(u)$ in $W\ICP^4_{1,1,1,1,4}$, where $h$ a
homogeneous polynomial of degree 8. Note that the most general
degree 8 hypersurface in this space can indeed brought in this form,
since terms linear in $\xi$ can be absorbed by a coordinate
transformation shifting $\xi$. The Hodge numbers of $X$ are
$h^{1,1}=1$, $h^{2,1}=149$. The space of D7 charges $H^2(X,\IZ)$ is
generated by the hyperplane class $H = [a_i u_i = 0]$. It
satisfies $H^3=2 \, \omega$ where $\omega$ is the unit volume
element on $X$. The second Chern class\footnote{See section 5.5 of \cite{Denef:2008wq} for a review on how to compute Chern classes of complete intersections in toric varieties} of $X$ is $c_2(X) = 22
\, H^2$.

The orientifold involution is
\begin{equation}
 \sigma: \xi \to - \xi.
\end{equation}
This is combined with $(-1)^{F_L} \Omega$ acting on string states,
where $\Omega$ is worldsheet orientation reversal and $F_L$ is the
spacetime fermion number in the leftmoving sector.

There is an O7-plane at $\xi=0$, whose Poincar\'e dual cohomology
class is $[O7]=4H$. We choose the orientifold projection such that
it produces $SO$ groups for D7-branes coincident with the
O7, so the O7 is an O7$^-$.\footnote{We define an O$p^-$ in general to be
an O$p$-plane such that stacks of D$p$ branes coincident with it have orthogonal gauge
groups and O$p^+$ for the case in which the gauge group is symplectic. The gauge
groups for lower dimensional branes follow from this; for example we get
symplectic groups for D3-branes coincident with an O7$^-$ and again orthogonal groups for D($-1$) branes on
an O7$^-$ \cite{Gimon:1996rq}. \label{fn:gaugegroups}}
To cancel the D7 charge tadpole induced by the O7, we need to add D7
branes for a total of D7 charge $[D7]=32 H$. Such a D7 can be
described as the zero locus of a homogeneous polynomial equation of
degree 32. To respect the orientifold $\IZ_2$ symmetry $\sigma$,
only even powers of $\xi$ can occur, which in turn can be eliminated
by (\ref{CYeq}). Hence the most general $\IZ_2$ symmetric
holomorphic surface in the class $32 H$ is described by an equation
of the form $P_{32}(u)=0$, where $P_{32}(u)$ is a homogeneous
polynomial of degree $32$ on $\ICP^3$, and naively we might be tempted to conclude the space of possible supersymmetric D7 embeddings to be given by
\begin{equation}
 D7: P_{32}(u) \qquad (\mbox{naive}) \, .
\end{equation}
Recall however that in the weak coupling limit of F-theory, we found
the D7 equation to be of the form (\ref{O7D7eq}):
\begin{equation} \label{D7eq}
 D7: \eta(u)^2 = \xi^2 \chi(u) \qquad (\mbox{F-theory}) \, ,
\end{equation}
where $\eta$ and $\chi$ are homogenous polynomials of degree 16 and
24. This is certainly not the most general degree 32 equation, as
can be seen for example by counting parameters: as we saw in section
\ref{sec:orlimit}, the number of distinct deformations of
(\ref{D7eq}) is 3728, while the most general $\sigma$-symmetric
surface $P_{32}(u)=0$ has 6544 deformations. Moreover, as will
become clear below, a generic $\IZ_2$-symmetric surface would give a
contribution to the D3 tadpole and would allow a number of
worldvolume fluxes which would both be in gross disagreement with
the F-theory results.

What characterizes the D7 worldvolumes described by
(\ref{D7eq}) is that for generic $\eta$ they all have \emph{double
point} intersections with the O7-plane: when we put $\xi=0$,
(\ref{D7eq}) reduces to $\eta^2=0$, so all zeros are double zeros.
Another way of saying this is that locally (away from $\chi=0$), the
D7 looks like a D7-image-D7 pair, as is manifest by writing the
equation as $\xi=\pm \eta/\sqrt{\chi}$. Globally however we
generically have a single connected brane, and (\ref{D7eq}) is in
fact the only possible globally well-defined equation which has this
local property.

\EPSFIGURE{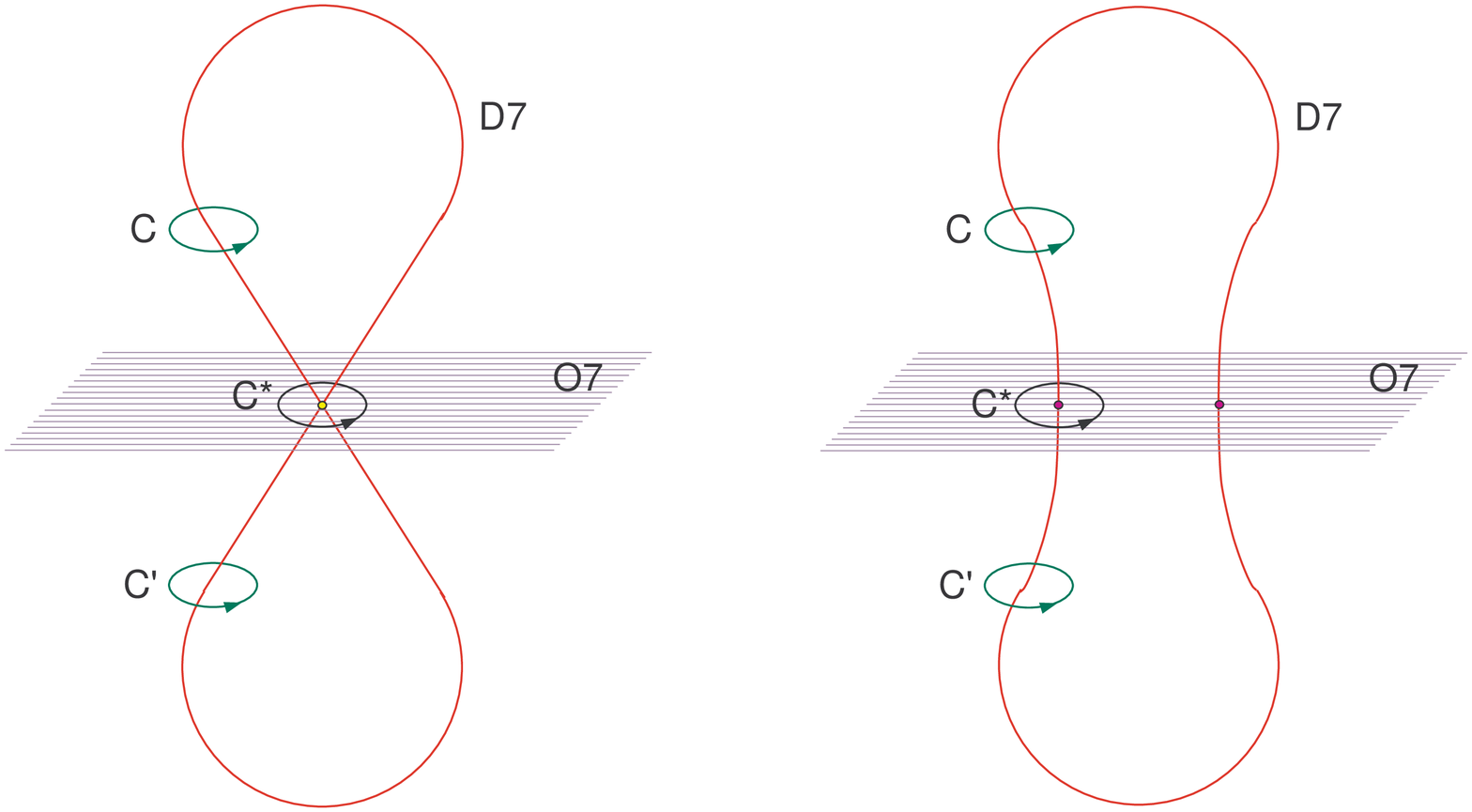,height=7cm,angle=0,trim=0 0 0 0}%
{Cartoon of an allowed D7 embedding on the left, and of a forbidden
embedding on the right. $C$ is a loop transporting a D($-1$) and
$C'$ its orientifold image. $C^*$ is a loop transporting a single
D($-1$) stuck on the orientifold plane.
  \label{D7D7}}

This double intersection property was independently noted in \cite{Braun:2008ua}.

\subsection{Perturbative IIB explanation of the double intersection} \label{sec:pertexpl}

How can we see the necessity of such double intersections with
$\xi=0$ directly within the perturbative type IIB picture? Put differently, what
is wrong with \emph{single} D7-O7 intersections? The answer, it turns out, is that such
single intersections violate Dirac quantization! To see this, first consider a D($-1$) probe at a point $p$ near
the D7 away from the O7. Its path integral phase is $e^{2 \pi i
C_0(p)}$. When parallel transporting it on a loop $C$, it will pick
up an additional phase $e^{2 \pi i n_7}$, where $n_7$ is the number
of D7-branes enclosed in the loop. Since $n_7$ is integral, this
equals 1, so the phase $e^{2 \pi i C_0(p)}$ is unambiguous, as it
should. We can move the D($-1$) to the O7, where it will coincide
with its orientifold image, with the same result. However, for the
O7$^-$ orientifold projection we have chosen, this is not the
minimal D($-1$) there exists on the O7: as we recalled in footnote
\ref{fn:gaugegroups}, D$(-1)$ branes coincident with the O7 give
orthogonal gauge groups; a brane-image brane pair corresponds to $O(2)$,
but $O(1)$ is possible too: this represents a D($-1$) stuck on the
O7, with charge \emph{half} of that of the bulk D($-1$) we have been
considering so far. When taken around a loop, such a ``half''
D($-1$) will thus pick up a phase $e^{\pi i n_7}$. Hence for the
phase of the D($-1$) to be unambiguous, $n_7$ must be \emph{even}, that is,
the D7 must intersect the O7 only in \emph{double} points.\footnote{As usual,
probe arguments of this sort are a little slick: one can always object
that it might just be the probe which is inconsistent, not the background. A more direct
argument can presumably be given along the lines of the analysis of \cite{Gimon:1996rq}, where
the relation between Dirac quantization and certain kinds of brane doublings was pointed out as well.
Essentially, the analysis of \cite{Gimon:1996rq} implies that smooth coincident brane stacks
transversal to the orientifold plane have ${\it USp}(2n)$ gauge groups, and in our case, the generic double intersection brane can be thought of as
a ${\it USp}(2)$ stack Higgsed by charged deformation fields forced to vanish on
the orientifold plane by the orientifold projection.}

The choice of orientifold projection was crucial for this argument. Had we
chosen the O7$^+$ projection, which gives respectively gauge groups
${\it USp}$, ${\it O}$ and ${\it USp}$ for D7, D3 and D($-1$) branes coincident with
the O7$^+$, there would be no half D($-1$) branes living on the O7
(since the minimal ${\it USp}$ group is ${\it USp}(2)={\it SU}(2)$), and therefore no
need for double intersections. This choice of orientifold projection would have
given positive D7-charge to the O7, so this would not have corresponded to the F-theory
case we started off with, and in fact it would have been incompatible with supersymmetry and D7-tadpole cancelation.

(Very) special cases of consistent D7 brane embeddings are global
brane-image-brane configurations, obtained by taking $\chi = \psi^2$
with $\psi$ a degree $12$ polynomial in equation (\ref{D7eq}). The
D7 and its image are then given by D7$_{\pm}$: $\eta = \pm \xi
\psi$, and both components are generically smooth. However this is
only a dimension 1423 subvariety of the full 3728 dimensional moduli
space, and moreover as we will see such configurations do not
reproduce the (zero flux) F-theory D3-tadpole. Thus if we want to
explore more than a tiny fraction of the landscape of this
compactification, we are forced to consider the most general
consistent D7 embedding, and deal with its singularities.

\subsection{Singularity structure of the D7} \label{sec:singstruct}

From the previous discussion we take that the only allowed
D7-worldvolumes $S$ in $X$ of D7-charge $2 m H$ are described by
equations of the form
\begin{equation} \label{D7eqgenm}
 S: \quad P_{2m} := \eta^2 - h \chi=0 \, ,
\end{equation}
inside $X: P_X:=\xi^2 - h = 0$. Here $(h,\eta,\chi)$ are polynomials of degree $(8,m,2m-8)$ on
$\ICP^3$. To saturate the D7-tadpole, one can add a number of such
D7-branes with charges $2 m_i H$, given by equations $P_{2 m_i}=0$,
satisfying $\sum_i 2 m_i = 32$. Such a combined system is described
by replacing the equation above with $\prod_i P_{m_i} = 0$,
and can be rewritten in the form $\tilde{\eta}^2 - \xi^2
\tilde{\chi} = 0$, with $\tilde{\eta}$ and $\tilde{\chi}$ suitable
degree 16 and 24 polynomials, consistent with the F-theory result.
For many purposes we can however forget about the D7 tadpole constraint
and just consider (\ref{D7eqgenm}) with arbitrary $m$. Unless stated
otherwise, we will moreover assume $m>4$. The advantage of allowing
general $m$ is that it allows sharper matching of topological data
in the various pictures we will present.

The singularities of the algebraic surface $S$ are those points for
which the rank of the gradient matrix $(d P_{2m},d P_X)$ deviates
from its generic value 2. It is not hard to see that for generic
polynomials  $(h,\eta,\chi)$, this happens precisely at the
intersection of the D7 with the O7, i.e.\ on the curve
\begin{equation}
 C: \quad \eta = 0 \quad \cap \quad h=0 \quad \cap \quad \xi^2 - h = 0.
\end{equation}
These are of course nothing but the double point intersections we
discussed earlier: away from $\chi=0$ the D7 looks locally like a D7
and its $\IZ_2$ image: $\xi = \pm \eta/\sqrt{\chi}$. Globally this
is generically not the case because the sheets get interchanged when
circling around $\chi=0$.

The projection $S_B$ of $S$ to the $\ICP^3$ base, given by the first
equation in (\ref{D7eqgenm}), does not have singularities at generic
points of the projection of $C$ to the base. However it does have
double point singularities, locally isomorphic to the
singularity $x^2+y^2+z^2 = 0$ in $\IC^3$, in the $8m(2m-8)$ isolated
points
\begin{equation}
 pp: \, h=\eta=\chi=0.
\end{equation}

\EPSFIGURE{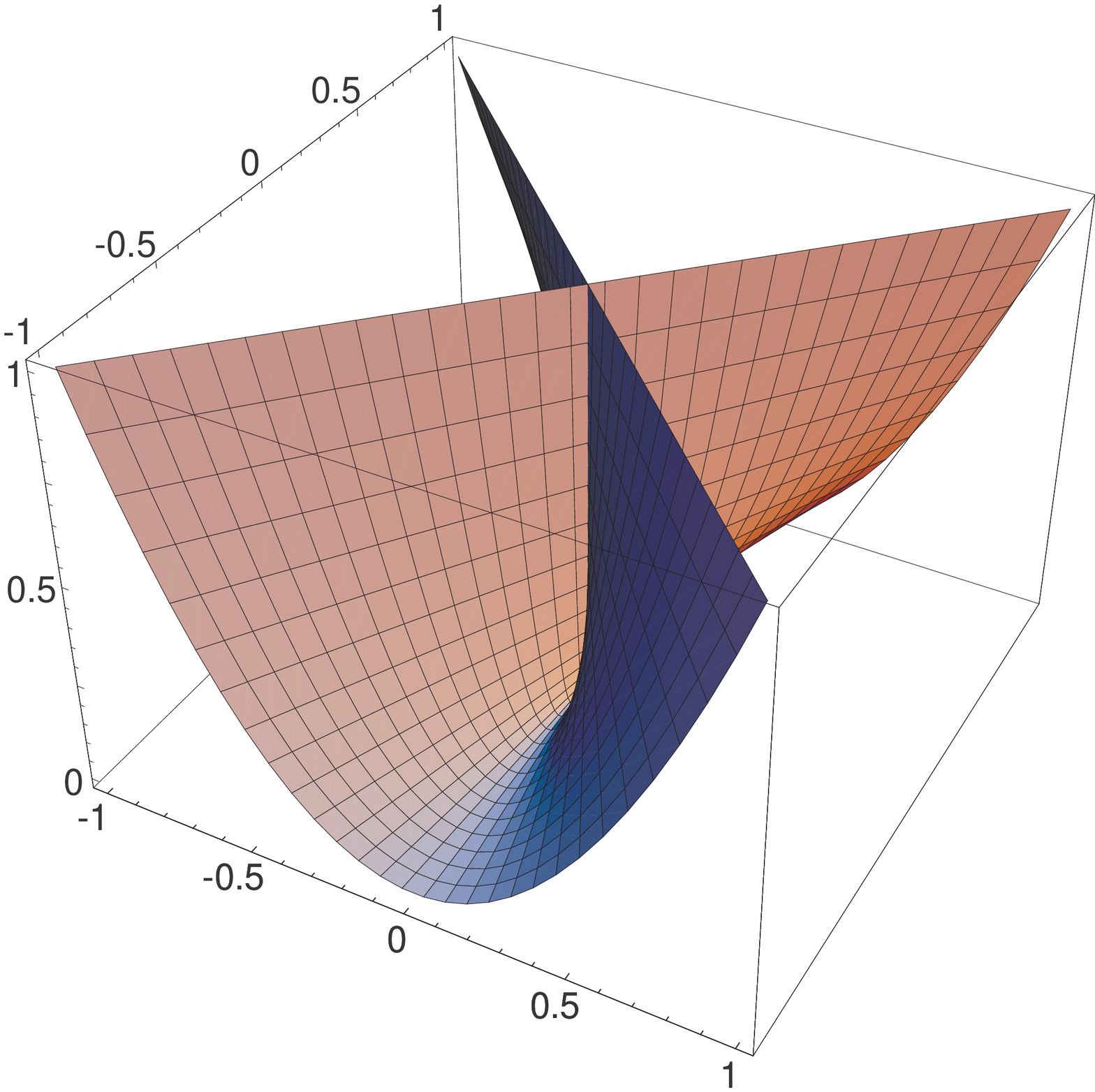,height=5cm,angle=0,trim=0 0 0 0}%
{The Whitney umbrella in $\IR^3$.
  \label{whitney}}

These singular points lift to pinch point singularities on $S$,
locally isomorphic to the so-called (complexified) ``Whitney
umbrella'':
\begin{equation}
 x^2 = z y^2
\end{equation}
in $\IC^3$ (we identified $x=\eta, y=\xi, z=\chi$, so the O7 is at
$y=0$). This can alternatively be seen as the embedding of $\IC^2$
into $\IC^3$ given by $(s,t) \mapsto (st,s,t^2)$. The Whitney
umbrella has a curve of double points at $x=y=0$ pinching off at
$x=y=z=0$. Its projection to $\IR^3$ is illustrated in fig.\
\ref{whitney}. A detailed analysis of such surfaces can be found in \cite{GH} p.\ 618-621, and
\cite{wikiwhitney}) currently gives an aesthetically pleasing and succinct description.

Due to the presence of these singularities, computing topological
quantities such as charges, deformation moduli and flux lattice
dimensions from topological data such as Hodge numbers becomes significantly more subtle than in the smooth
case. We now turn to these issues.

\subsection{RR charges}

\subsubsection{Generalities}

We start by reviewing the known formulae for RR charges of
\emph{smooth} D-branes and O-planes. For future reference we will be
more general than necessary to treat our example. However, we will be far
from \emph{completely} general; in particular we will not use the proper K-theoretic
framework, which is fine for our purposes. For a comprehensive treatment of
the geometry and topology of orientifolds, we refer to \cite{mooretexas}.

We normalize worldvolume, NSNS and RR potentials such that their
fluxes are integrally quantized (up to possible overall shifts), or
equivalently such that large gauge transformations act by integral
shifts on period integrals of flat potentials. Then a D-brane with
worldvolume $W$ carrying (possibly nonabelian) flux $F$ couples to
the total RR potential $C = C_0 + C_2 + C_4 + C_8 + C_{10}$ as \cite{Morales:1998ux,Stefanski:1998yx, Scrucca:1999uz,Minasian:1997mm}
\begin{equation}
 S^{\rm Dbrane}_{W,C} = 2 \pi \int_W C \wedge e^{-B} \, {\rm Tr} \, e^F
 \sqrt{\frac{\widehat{A}(TW)}{\widehat{A}(NW)}},
\end{equation}
where $\widehat A$ is the A-roof characteristic class defined e.g.\ in section 5.5 of \cite{Denef:2008wq}
and $TW$ and $NW$ are the tangent resp.\ normal bundles of $W$. We work in units with $\ell_s := 2 \pi \sqrt{\alpha'} = 1$. The flux $F$ obeys integral
quantization conditions shifted by what can be thought of as a half
integral diagonal $U(1)$ flux $\Delta F$ which reduces modulo $\IZ$
to $w_2(N_W)$ \cite{Minasian:1997mm,Freed:1999vc}. More generally for a sheaf
$E$, ${\rm Tr} e^F$ is replaced by ${\rm ch}(E)$.

For spacetime filling D-branes translationally invariant in the
$\IR^{1,3}$ directions this coupling defines a cohomology class
$\Gamma \in H^*(X)$, which we will call the charge of the D-brane,
by requiring that for all \emph{closed} $B, C$ we have
\begin{equation} \label{Dbranechargedef}
 S^{\rm Dbrane}_{W,C} = 2 \pi \int_{\IR^{1,3} \times X} C \wedge e^{-B} \, \Gamma.
\end{equation}
In the case at hand, this identifies (D9,D7,D5,D3) charges with
elements in respectively $(H^0(X),H^2(X),H^4(X),H^6(X))$.

More concretely, using the fact that $X$ is Calabi-Yau, one gets for
a D9-brane stack carrying a holomorphic vector bundle $E$
\begin{equation} \label{D9charge}
 \Gamma_{D9} = {\rm ch}(E) \left( 1+\frac{c_2(X)}{24} \right),
\end{equation}
and for a stack of $N$ D7-branes wrapped on a smooth surface $S$
carrying a holomorphic vector bundle $E$
\begin{equation} \label{GamD7}
 \Gamma_{D7} = N [S] + \biggl( \int_S D_A \wedge {\rm ch}_1(E) \biggr) \, \tilde{D}^A
 + \biggl( \frac{N \chi(S)}{24} + \int_S {\rm ch}_2(E) \biggr) \, \omega
\end{equation}
where $[S] \in H^2(X)$ is Poincar\'e dual to the homology class of
$S$, $D_A$ is a basis of $H^2(X)$ and $\tilde{D}^A$ the dual basis of
$H^4(X)$, $\chi(S)$ is the Euler characteristic of $S$, and $\omega$
the unit volume element of $X$, which serves as a basis of $H^6(X)$.
Thus the first term represents D7 charge, the second D5 charge and
the last one D3 charge.

The Euler characteristic of a smooth surface $S$ in a Calabi-Yau $X$
can be computed from the adjunction formula as
\begin{equation} \label{eulerformula}
 \chi(S) = \int_X S^3 + c_2(X) \, S.
\end{equation}
In our example this becomes for $S=n H$, $\chi(nH) = 2 n^3 + 44 n$.

For a stack of $N$ D5-branes wrapped on a curve $C$ carrying a
holomorphic vector bunlde $E$ we have
\begin{equation}
 \Gamma_{D5} = N [C] + \left(\int_C c_1(E)\right) \, \omega.
\end{equation}
Finally, for a single D3-brane (as opposed to an anti-D3-brane) we
take the convention
\begin{equation}
 \Gamma_{D3} = - \omega.
\end{equation}
Note the minus sign, which we chose such that D3-brane probes
preserving the supersymmetry of the orientifold backgrounds of
interest to us have positive D3-charge, as is common in the
model building literature. In other words, we will take the basis element of
$H^6(X)$ with respect to which we express D3-charge to be $-\omega$. (The other choice of sign is
actually more natural for a variety of reasons (for example requiring T-duality to map positive branes to positive branes), but we will stick to the above choice.)

An O$p$-plane on $V$ couples to the RR potentials as
\cite{Morales:1998ux,Stefanski:1998yx,Scrucca:1999uz,Brunner:2003zm,Brunner:2004zd,mooretexas}
\begin{equation}
 S^{\rm O^{\pm}plane}_{V,C} = \pm 2 \pi \int_V C \wedge 2^{p-4} \,
 \sqrt{\frac{L(\frac{1}{4}TV)}{L(\frac{1}{4}NV)}},
\end{equation}
where $L$ is the Hirzebruch $L$-genus (again defined e.g.\ in section 5.5 of \cite{Denef:2008wq}). Again we define a charge $\Gamma$ by pairing with closed forms:
\begin{equation} \label{Oplanechargedef}
 S^{\rm Oplane}_{V,C} = 2 \pi \int_{\IR^{1,3} \times X} C \wedge \Gamma.
\end{equation}
In particular for an O7$^-$ wrapping a single component smooth
divisor $U$, as in our example, one gets, using the adjunction
formula and the fact that $X$ is Calabi-Yau:
\begin{equation} \label{GamO7}
 \Gamma_{O7^-} = -8 [U] + \frac{\chi(U)}{6} \, \omega.
\end{equation}
where we recall $\chi(U)=U^3+c_2 U$.


With these definitions, the RR tadpole cancelation condition for backgrounds with flat B-field can be
formulated as
\begin{equation}
 e^{-B} \Gamma_D + \Gamma_O = 0,
\end{equation}
where $\Gamma_D$ denotes the total charge of all D-branes and
$\Gamma_O$ the total charge of all O-planes.

For non-flat B-fields or B-fields with torsion these formulae need to be modified \cite{mooretexas}.

\subsubsection{Orientifold actions} \label{sec:oractions}

An O3/O7 orientifold is produced by a holomorphic involution
$\sigma$ combined with $\Omega (-1)^{F_L}$. The action on the
massless closed string fields is \cite{Brunner:2003zm}
\begin{equation} \label{sigmaonRRNSNS}
 C_{0,4,8} \to \sigma^* C_{0,4,8}, \quad C_{2,6,10} \to - \sigma^*
 C_{2,6,10}, \quad B \to - \sigma^* B, \quad g \to \sigma^* g \, .
\end{equation}
The action on a worldvolume gauge field living on an orientifold-invariant D3- or D7-brane stack is \cite{Gimon:1996rq}
\begin{equation} \label{sigmaonA}
 A \to - M \sigma^*A^t M^{-1} \, ,
\end{equation}
where $M$ is a symmetric or antisymmetric constant unitary matrix, depending on the choice of orientifold projection and the stack under consideration.\footnote{For a D$(p-4k)$-brane stack coincident with an O$p^\pm$-plane we have $M^t=\pm (-1)^k M$, and for a D7 stack wrapping a smooth 4-cycle transversal to an O7$^\pm$, we have $M^t= \mp M$. For a stack wrapped on a generic $O7^-$-transversal cycle of the singular kind we discussed in section \ref{sec:singstruct}, we get $M^t=M$. These rules are most easily and universally derived in the tachyon condensation picture, as will be discussed in section \ref{sec:enhanced}. \label{GGrules}} If symmetric, the surviving four dimensional gauge group is orthogonal, and if antisymmetric it is symplectic. If $M=M^t$ one can choose a Chan-Paton basis such that $M={\bf 1}$ or alternatively if the rank of the stack is even $M={\scriptsize \left( \begin{array}{cc} 0&{\bf 1} \\ {\bf 1} & 0   \end{array} \right)}$. If $M=-M^t$ one can choose a basis such that $M={\scriptsize \left( \begin{array}{cc} 0&i {\bf 1} \\ -i {\bf 1} & 0   \end{array} \right)}$. Here ${\bf 1}$ is a unit matrix of the appropriate dimension. For a ${\rm D}p$-${{\rm D}p}'$ pair where ${{\rm D}p}' = \sigma({\rm D}p)$, the action can be taken to be
\begin{equation} \label{sigmaonA2}
 A \to - \sigma^* {A'}^t \, .
\end{equation}
This leads to unitary gauge groups.

Consistent with the above actions on the fields, the charge vectors $\Gamma$ transform as
\begin{equation} \label{Orchargetransf}
 \Gamma \to -\sigma^* \Gamma^*, \qquad \mbox{i.e.} \quad \Gamma^{(2k)} \in
 H^{2k}(X) \to (-1)^{k+1} \sigma^* \Gamma^{(2k)}.
\end{equation}
Here $\Gamma^*$ is obtained from $\Gamma$ by flipping the sign of
the 2- and 6-form components, i.e.\ $\Gamma^*:=\sum_k (-1)^k
\Gamma^{(2k)}$ where $\Gamma^{(2k)}$ is the $(2k)$-form component of
$\Gamma$. Thus the lattice of invariant charges is
\begin{equation}
 (D9 \oplus D7 \oplus D5 \oplus D3)_{\rm invar} \quad = \quad 0 \oplus H^2_+ \oplus
 H^4_- \oplus H^6
\end{equation}
where $H^{2k}_{\pm}$ is the $\pm 1$ eigenspace of $\sigma^*$ on
$H^{2k}$. In our example $H^2_+=H^2$ and $H^4_-=0$.

\subsubsection{Failure of naive charge formula for generic consistent D7-branes}

We now return to the example. Our goal is to guess a formula for the
D3 charge of the generic consistent D7-branes described by
(\ref{D7eqgenm}), by comparing to the known F-theory tadpole
cancelation condition (\ref{Ftadpole}) at zero flux $G$. Reasonably
assuming that zero flux in F-theory corresponds to zero worldvolume
and bulk fluxes in IIB, we see that a configuration without any
worldvolume or bulk flux and 972 mobile D3-branes plus their
orientifold images must be tadpole free. In other words, minus the
curvature induced D3 charge on the O7 and the degree 32 D7 should
add up to $2 \times 972 \, \omega = 1944 \, \omega$.

Since the O7 is smooth for generic $h$, the charge of the O7 is
given directly by (\ref{GamO7}) and (\ref{eulerformula}), using
$U=4H$, $H^3=2\omega$ and $c_2(X)=22H^2$:
\begin{equation} \label{O7charge}
 \Gamma_{O7} = -32 H + \frac{152}{3} \, \omega.
\end{equation}
Now, if we recklessly ignored the fact that the degree 32 surface
$S$ wrapped by the D7 is singular, we would find, using (\ref{eulerformula}), that $\chi(S) = 66944$
and thus from (\ref{GamD7}) with $N=1$ and $E$ trivial\footnote{Note
that we can indeed take $E$ trivial because the degree of $S$ is
even, so $w_2(N_S) = c_1(NS) \, {\rm mod} \, 2 = [S] \, {\rm mod} \,
2 = 0$.}
\begin{equation}
 \Gamma_{D7} = 32 H + \frac{8368}{3} \, \omega,
\end{equation}
and
\begin{equation}
 \Gamma_{D7} + \Gamma_{O7} = 2480 \, \omega.
\end{equation}
This overshoots the F-theory value $1944 \, \omega$. One could
contemplate the possibility that this type (supersymmetric) IIB
configuration actually secretly corresponds to an
F-theory vacuum with nonzero $G$-flux, but one quickly sees that
this is not possible since the equations of motion imply $G=*G$ \cite{Becker:1996gj,Gukov:1999ya,Giddings:2001yu}, so flux always adds positively to the D3-charge in (\ref{Ftadpole}), which would make the discrepancy even worse. Thus we conclude that we have been too reckless indeed in ignoring the singularity of $S$, in particular
by using (\ref{eulerformula}).

On the other hand, if we take our D7 to be of the special global
brane-image-brane type discussed at the end of section
\ref{sec:theorientifold}, i.e.\ $\chi=\psi^2$, we can reliably
compute the charge, since the two component surfaces are generically
smooth divisors in the class $16H$, and the total charge is just the
sum of the charges of these components. This gives
\begin{equation} \label{D7D7prime}
 \Gamma_{D7} + \Gamma_{O7} = 792 \, \omega,
\end{equation}
undershooting the F-theory value. This is not inconsistent, but
suggests that the sector in the moduli space of D7-branes physically
connected to this configuration maps in F-theory to a sector with
nonzero $G$-flux. We will later on show that this
interpretation is indeed correct. But this computation also shows
that we cannot compute the charge of the generic D7 by going to this
particular, well-controlled brane setup; we must deal with the
singular generic D7 directly.

\subsubsection{Modified charge formula} \label{sec:modcharge}

To formally preserve the usual charge formula (\ref{GamD7}) also for
the generic, singular D7 in the orientifold, we need to come up with
a physical definition of the Euler characteristic of $S$ which
reproduces the correct D3 charge. We will denote this modified Euler
characteristic by $\chi_{o}(S)$, where the subscript $o$ is for
orientifold, i.e.\ we define $\chi_o(S)$ for generic allowed\footnote{By ``generic allowed'' here and in the following we mean a D7 described by a generic polynomial of the restricted form (\ref{D7eqgenm}).} $S$ by
\begin{equation} \label{chiodef}
 \int_X \Gamma_{{\rm pure} \, D7} =: \frac{\chi_o(S)}{24},
\end{equation}
where $\Gamma_{{\rm pure} \, D7}$ is the charge of a pure (trivial
$E$) D7 wrapping $S$.

Consistency with the F-theory D3-tadpole formula (\ref{Fcurvindcharge}) requires that
\begin{equation} \label{chioFpred}
 2 \, \chi(Z) = \chi_o(D7) + 4 \, \chi(O7) \, ,
\end{equation}
which in our example gives $\chi_o(S_{32}) = 45440$ for generic allowed surfaces $S_{32}$ of degree 32.

For a D7 $S$ which globally splits in a smooth D7 $S_1$ and its
orientifold image $S_1'$, the correct identification is simply
$\chi_o(S_1 \cup S_1') = \chi(S_1) + \chi(S_1')=2 \, \chi(S_1)$,
where the Euler characteristics of the components are just the
topological ones. This additivity merely represents charge
additivity, or put differently, the fact that the induced D3 charge
is the integral of a local density (the euler density) over the
separate component brane worldvolumes. Note that $\chi_o(S_1 \cup
S_2)$ differs from the topological Euler characteristic of the
union: $\chi_{\rm top}(S_1 \cup S_2) = \chi_{o}(S_1 \cup S_2) -
\chi(S_1 \cap S_2)$.

\EPSFIGURE[h]{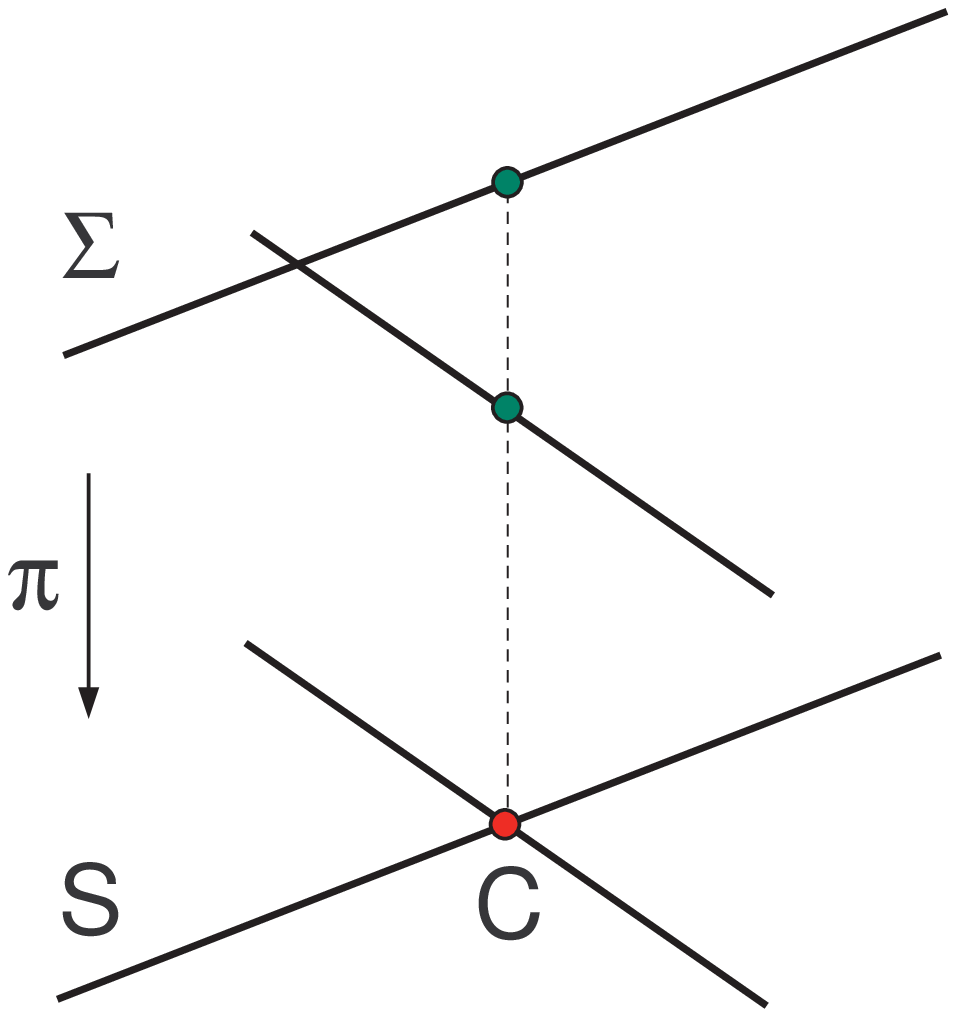,height=7cm,angle=0,trim=0 0 0 0}%
{Cartoon of parametrization of $S$ by $\Sigma$ near the double point
locus $C$, away from the pinch points.
  \label{Sigma}}

Now, as we saw, a generic D7 does not globally split in a
brane-image-brane pair, but nevertheless it splits locally; near the
curve of double point singularities $C$, away from the pinch points,
$S$ just looks like a D7 transversally intersecting its image on the
O7. This suggest we should parametrize the worldvolume of $S$ such
that the parametrization splits the two branches meeting on the
curve $C$ of double points, similar to the parametrization of the
Whitney umbrella $x^2=z y^2$ by $(x,y,z)=(st,s,t^2)$, which splits
the double point curve $x=y=0$ into $(s,t)=(0,\pm \sqrt{z})$. Modulo
subtleties due to the intrinsically singular pinch points, we can
then expect the proper $\chi_o(S)$ to be the Euler characteristic of
the parameter surface.

Mathematically, such a splitting can be realized by blowing up the
double point curve $C$. This blowup is of an auxiliary nature though
--- we are not really physically resolving space, we are just
defining a convenient worldvolume parametrization. Concretely,
consider the D7 described by (\ref{D7eqgenm}) for arbitrary $m$. We
want to blow up the curve $C: \eta=\xi=0$. This goes as follows.
Extend the set of coordinates $(u_1,u_2,u_3,u_4,\xi)$ with two new
coordinates $s$ and $t$, and mod out by a new $\IC^*$ rescaling
$(s,t) \to (\lambda s,\lambda t)$, so $(s,t)$ parametrize a $\IP^1$.
Furthermore impose the equation
\begin{equation} \label{blowupeq}
 t \, \xi = s \, \eta(u).
\end{equation}
In order for this to be compatible with the original projective
rescalings of $(u_i,\xi)$, we assign weights $(0,m-4)$ to $(s,t)$.
In summary we have the following toric $U(1)$ charges (or $\IC^*$
weights):
\begin{center}
\begin{tabular}{rccccccc}
 & $u_1$ & $u_2$ & $u_3$ & $u_4$ & $\xi$ & $s$ & $t$ \\
 $Q^1$:& 1 & 1 & 1 & 1 & 4 & 0 & $m-4$ \\
 $Q^2$:& 0 & 0 & 0 & 0 & 0 & 1 & 1
\end{tabular}
\end{center}
Note that away from the curve $C$, (\ref{blowupeq}) uniquely
determines a point $(s,t)$ in the $\IP^1$, so nothing changes. Each
point on $C$ on the other hand is replaced by a $\IP^1$.

Now let $\Sigma$ be the closure of our D7-divisor $S$ defined by
(\ref{D7eqgenm}) but with the curve $C: \eta=\xi=0$ removed, in this
blown up space. Explicitly, after gauge fixing\footnote{We can do
this since $s=0$ implies $\xi=0$, which is incompatible with the
equation for $S$ with $\xi=\eta=0$ removed.} $s \equiv 1$, this is
\begin{equation} \label{Sigmadef}
 \Sigma: \quad \xi t = \eta \quad \cap \quad t^2 = \chi \quad \cap
 \quad \xi^2 = h
\end{equation}
in $W\ICP^5_{1,1,1,1,4,m-4}$. Then the blow-down projection map
\begin{equation}
 \pi:\Sigma \to S: (\vec{u},\xi,t) \mapsto (\vec{u},\xi)
\end{equation}
is one to one except at $\xi=0$, $t\neq 0$, i.e.\ the curve of
double points $C$ away from the set of pinch points
$pp:t=\xi=h=\eta=\chi=0$, where it is two to one. At the pinch
points, it is again one to one. (Note that this parametrization
reduces to the $(s,t)$ parametrization given above when we apply
this prescription to the Whitney umbrella.) This means $\Sigma$
parametrizes $S$ through the projection $\pi$ in the way we were
after. Hence we expect
\begin{equation} \label{choformula}
 \chi_o(S) = \chi(\Sigma) + k \, n_{pp},
\end{equation}
where $k$ is some constant, to be determined, and $n_{pp}$ the number of pinch points. The second term represents a possible correction due to the isolated pinch point singularities, where the parametrization degenerates.

We now verify this and determine $k$. Since the coordinate surface $\Sigma$ is a smooth complete
intersection in weighted projective space, it is straightforward (using the adjunction formula, cf.\ section 5.5 of \cite{Denef:2008wq}) to compute its Chern classes and hence its Euler characteristic
$\chi(\Sigma)=\Sigma \cdot c_2(\Sigma)$. The total Chern class is, denoting the divisor class $[\xi=0]$ by $[\xi]$ and similarly for the others:
\begin{eqnarray} \label{chernSigma}
 c(\Sigma) &=& \frac{(1+[u])^4 \, (1+[\xi]) \, (1+[t])}{(1+[h]) \, (1+[\eta]) \, (1+[\chi])}  \\
 &=& \frac{(1+\bar H)^4 \, (1+4 \bar H) \, (1+(m-4) \bar H)}{(1+8 \bar H) \, (1+m \bar H) \, (1+(2m-8)\bar H)} \\
 &=& 1+ (4-2m) \, \bar H + (54-20 \, m+4\, m^2) \, {\bar{H}}^2 \, ,
\end{eqnarray}
where $\bar H$ is the hyperplane class $[u_1=0]$ in
$W\ICP^5_{1,1,1,1,4,m-4}$, which satisfies $\bar H^5=\frac{1}{4(m-4)}$ (cf.\ \cite{Denef:2008wq} section 5.2). Hence, using $\Sigma = 8m(2m-8) \, \bar{H}^3$,
\begin{eqnarray} \label{chiSigma}
\chi(\Sigma) = 16 \, m^3 - 80 \, m^2 + 216 \, m \, .
\end{eqnarray}
Furthermore $n_{pp}=8m(2m-8)$. For $m=16$ we thus get
$n_{pp}=3072$ and $\chi(\Sigma) = 45440 + 3072$. Hence
(\ref{choformula}) exactly reproduces the F-theory result
(\ref{chioFpred}) provided we take
\begin{equation}
 k=-1.
\end{equation}
For $S$ of general degree $2m$ we get, using (\ref{choformula}) with
$k=-1$ and (\ref{chiSigma}),
\begin{equation} \label{chioexample}
 \chi_o(S_{2m}) = 16 \, m^3 - 96 \, m^2 + 280 \, m.
\end{equation}
As a nontrivial consistency check, one can verify that if we have
two generic D7-branes of degrees $m_1$, $m_2$ with $m_1+m_2=16$, the
total D3-charge of the compactification is integral as it should; in
contrast taking e.g.\ $k=0$ in (\ref{choformula}) would violate
integrality for certain values of $m_i$.

In conclusion, we propose the following modified formula for the
curvature induced D3 charge on a D7 of the kind we are considering:
\begin{equation} \label{Q3cformula}
 (\Gamma_{{\rm pure} \, D7})_{D3} = \frac{\chi_o(S)}{24} \, \omega, \qquad \chi_o(S) = \chi(\Sigma)
 - n_{pp} \, ,
\end{equation}
where $\Sigma$ is the parametrization manifold of the D7 splitting
the curve of double points, and $n_{pp}$ the number of pinch
points, where the parametrization degenerates.

In section \ref{sec:D9aD9} we will give a direct derivation of this formula from D9-anti-D9 tachyon condensation.

\subsubsection{Mathematical characterization of $\chi_o(S)$}

The   Euler characteristic of a smooth space is  defined
without any  ambiguities.   It is always the standard topological Euler
characteristic, which satisfies specific additive and multiplicative
properties. The topological Euler characteristic can be computed as the
integral of the top Chern class.   For a singular variety,
there are many non-equivalent generalizations of the notion of Euler
characteristic  that can be viewed as the integral of a suitably generalized top Chern class.      Interestingly, $\chi_o(S)$ does not equal any of these common Euler
characteristics. However, it \emph{is} a sensible Euler
characteristic  that can be described  as a top Chern
class integral  \cite{Aluffi:2007sx}.

If  $Z$ in an elliptic fibration  over a base $B$ of  arbitrary dimension,  the  corresponding Euler characteristic  $\chi_o$  satisfies  the relation
$$
2\chi(Z)=\chi_o(S)+4\chi(O),
$$
where $S$ and $O$ are the two hypersurfaces defined  using Sen's weak coupling limit  along the lines that we followed in  F-theory to define the hypersurfaces wrapped  respectively by  the D7 brane and the O7 plane.  It is interesting to note that the previous  relation can be shown to hold  without assuming  the  Calabi-Yau condition and without any restrictions on the  dimensionality of   $Z$  \cite{Aluffi:2007sx}.

 The mathematical definition of  $\chi_o(S)$   \cite{Aluffi:2007sx}  is closely related to the one of
the stringy  Euler characteristic, or more generally to
motivic integration.   This is discussed at length in
\cite{Aluffi:2007sx}.  In a sense,   $\chi_o(S)$ is a non-trivial generalization of  the stringy Euler characteristic to   spaces admitting singularities in codimension one. The  usual stringy Euler characteristic is defined only for so-called  {\em normal spaces }  which  are always smooth in codimension 1 although they could admit singularities in higher  co-dimensions.

\subsection{Flux lattice}

In general a single D7-brane wrapped on a divisor $S$ can carry
internal $U(1)$ fluxes $F$. These take values in the shifted
integral lattice $\frac{c_1(S)}{2} + H^2(S,\IZ)$. If the D7 is a
$\IZ_2$-invariant brane in the O3/O7 orientifold we are considering,
then the only fluxes which survive the orientifold projection are
those which satisfy $\sigma^* F = - F$. So in some sense we expect
the allowed fluxes on $S$ to be given by $H_-^2(S,\IZ)$ (we can drop
the half integral shift since $c_1(S)$ is always even in the case at
hand). However, again, because $S$ is generically singular, it is
not immediately clear how to define $H_-^2(S,\IZ)$, and naive
attempts ignoring the singularity result in gross discrepancies with
F-theory.

Again, the problem is solved by considering the $\Sigma$-parametrization of the D7-worldvolume. The key is to
view the fluxes $F$ as living on the smooth surface $\Sigma$ instead of on the singular $S$. Note that the orientifold involution $\sigma$ on $S$ induces on $\Sigma$ the involution
\begin{equation}
 \sigma: \quad \xi \to -\xi, \quad t \to -t.
\end{equation}
We define $H^2_-(\Sigma)$ to be the part of $H^2(\Sigma)$ odd under
$\sigma^*$. We propose
\begin{equation}
 \mbox{Lattice of $U(1)$ wordlvolume fluxes} = H^2_-(\Sigma,\IZ) \, .
\end{equation}
As a nontrivial check, we will now show that the dimension of this
lattice plus the lattice of bulk RR and NSNS fluxes precisely equals the
dimension of the F-theory flux lattice in our example.

To compute $b^2_- \equiv \dim H^2_-(\Sigma)$, we use the Lefshetz
fixed point index formula \cite{GH}, which states
\begin{equation} \label{leffix}
 \sum_k {\rm Tr}_{H^k(\Sigma)}\, (-1)^k \sigma^* \, = \, \chi(\Sigma^\sigma)
\end{equation}
where $\Sigma^\sigma$ is the fixed point locus of the involution.
Noting that the fixed points of $\sigma$ are nothing but the pinch
points $pp$, we see that in the case at hand this becomes $1 + b^2_+
- b^2_- +1 = n_{pp}$. Here we also used that $b_1(\Sigma)=0$, which
follows from the Lefshetz hyperplane theorem \cite{GH}.
Combining this with $\chi(\Sigma)=2+b_2^++b_2^-$ hence results in
\begin{equation} \label{b2minformula}
 b^2_- = \frac{1}{2} \left( \chi(\Sigma) - n_{pp} \right) =
 \frac{\chi_o(S)}{2} = 8 \, m^3 - 48 \, m^2 + 140 \, m.
\end{equation}
For a tadpole-saturating D7 we have $m=16$, so in that case $b_2^-=22720$.

To compare to the dimension of the F-theory flux lattice, we should
add this to the dimension of the lattice of RR and NSNS fluxes. From
(\ref{sigmaonRRNSNS}), it follows that this equals (two times)
$b^3_-(X) = \dim H^3_-(X,\IZ)$. We always have
$b^{3,0}_-(X)=b^{3,0}(X)=1$. Moreover, because none of the complex
structure moduli of $X$ are projected out by the orientifolding, we
also have $h^{2,1}_-(X)=h^{2,1}(X)$. So $b^3_-(X)=b^3(X)=300$. This
can also be verified by the Lefshetz fixed point index formula. Thus
we find for the total flux lattice dimension
\begin{equation}
 b=b_2^-(\Sigma) + 2 b_3^-(X) = 23320.
\end{equation}
This is in precise agreement with the F-theory result (\ref{b4perp}).

\subsection{Moduli} \label{sec:D7modulicount}

A similar story applies to counting deformation moduli of the D7. In
non-orientifolded CY 3-folds, there is a one to one correspondence
between infinitesimal holomorphic deformations of the D7 and
holomorphic (2,0)-forms on the divisor $S$ wrapped by the D7.
The map is obtained by contracting the holomorphic section of the normal bundle
to $S$ corresponding to the deformation with the holomorphic
(3,0)-form $\Omega_3$ on the Calabi-Yau $X$. Equivalently, if $S$ is
described locally by equations $f(x)=0$ (so in particular when
$f(x)$ extends to a global homogeneous polynomial $P(x)$),
deformations are described by variations $\delta f$ of $f$, and we
can locally write the associated $(2,0)$-form $\rho[\delta f]$ as a
Poincar\'e residue, defined in section 5.6 of \cite{Denef:2008wq}:
\begin{equation}
 \rho[\delta f] = \frac{1}{2\pi i} \oint_{f=0} \frac{\delta
 f}{f} \, \Omega.
\end{equation}
Conversely, given a holomorphic (2,0)-form $\rho$ on $S$, $\rho
\wedge df$ extends to a holomorphic (3,0)-form in an infinitesimal
neigborhood of $S$, hence we can uniquely define $\delta f[\rho]$ by
writing $\rho \wedge df = \Omega \, \delta f + \CO(f)$ (so
$\rho=\rho[\delta f]$).

In an orientifolded CY 3-fold, infinitesimal deformations of the D7
respecting the $\IZ_2$ symmetry $\sigma$ are given by
$\sigma$-symmetric holomorphic sections of the normal bundle to the
D7. Contracting this vector field with $\Omega_3$ produces a
$(2,0)$-form on the D7 worldvolume which is
$\sigma$-\emph{anti}symmetric, since $\Omega_3$ itself is
$\sigma$-antisymmetric. Thus one would naively expect a one to one
correspondence between elements of $H^{2,0}_-(S)$ and deformations
of the D7. However this is not quite correct, since, as discussed in
section \ref{sec:theorientifold}, not all $\sigma$-symmetric
deformations of the D7 are actually allowed. Moreover, at the locus
$C$ of double points, the normal bundle to $S$ is not even
well-defined.

This leaves us with the question what the proper analog is which
\emph{does} correctly count the number of deformations. A natural
guess is
\begin{equation}
 \mbox{Number of D7 deformation moduli} = h^{0,2}_-(\Sigma) \, ,
\end{equation}
with $\Sigma$ defined
in (\ref{Sigmadef}). We will justify this in general below, but let
us first check that this indeed reproduces the correct counting in
our example.

For our example, we can compute the number of allowed D7
deformations directly --- for a degree $2m$ surface $S$ it is given
by the number of inequivalent deformations of $(\eta_m,\chi_{2m-8})$
in (\ref{D7eqgenm}). For $m > 4$, this is
\begin{equation}
 N_{\rm D7 \, def} = \mbox{${m+3 \choose 3} + {2m-8+3 \choose 3} - {m-8+3 \choose
 3} -1$} = \frac{4}{3} \, m^3 - 8 \, m^2 + \frac{59}{3} \, m,
\label{D7moddirect}
\end{equation}
where the first subtraction comes from the equivalence $(\eta,\chi)
\simeq (\eta + h \psi,\chi + 2 \eta \psi + h \psi^2)$ with $\psi$ of
degree $m-8$ and the second one from overall rescaling of the
coefficients. When $m<8$ the first equivalence is absent, but as
long as $m>4$ the subtracted binomial is zero in this case, so the
counting is still correctly given by (\ref{D7moddirect}).

To compute $h^{2,0}(\Sigma)$, we can use the formula for the holomorphic Euler characteristic given in section 5.8 of \cite{Denef:2008wq}:
\begin{equation} \label{h02form}
  1 + h^{2,0}(\Sigma) = \chi_0(\Sigma) = \int_\Sigma {\rm Td}(\Sigma) = \frac{1}{12}
\int_\Sigma c_1^2 + c_2 \, ,
\end{equation}
where we also used the Lefshetz hyperplane theorem to conclude $h^{1,0}(\Sigma)=0$.
For our example, plugging (\ref{chernSigma}) in (\ref{h02form}) then gives an explicit formula for $h^{2,0}$.

However, this is not yet what we need: We want $h^{2,0}_-(\Sigma)$.
This can be computed using the
holomorphic Lefschetz fixed point formula. In full generality, this can be stated
as follows. Let $V$ be a vector bundle on some manifold $M$ and let $g:V \to
V$ be a holomorphic symmetry descending to a holomorphic symmetry
$g:M \to M$. Then
\begin{equation} \label{holoLef}
 {\rm Tr}_{H^{0,*}(M,V)} \, (-)^p \, g = \int_{M^g} {\rm
 ch}_g(V) \frac{{\rm Td}(M^g)}{{\rm ch}_g(\wedge_{-1}
 \overline{N}_{M^g})}.
\end{equation}
Here $p$ is the form degree, ${\rm ch}_g(V) := {\rm Tr} \, g \, e^F$, $M^g$ is the fixed
point locus of $g$, $\wedge_{-1} N := 1 - N + N \wedge N - N \wedge
N \wedge N + \cdots$, and $\overline{N}_{M^g}$ is the complex
conjugate of the normal bundle to $M^g$. Specializing this to $g$
equal to a holomorphic involution $\sigma$ (think of an orientifold
involution possibly acting also on the Chan-Paton indices of $V$),
let us make this formula a little less obscure. First, in terms of
the Chern roots $\lambda_m^{(V)}$ of $V$, we can write
\begin{equation} \label{chernsigma}
 {\rm ch}_\sigma(V) = \sum_m \langle m | \sigma | m \rangle \,
 e^{\lambda_m^{(V)}},
\end{equation}
where the $|m\rangle$ are the Chan-Paton eigenvectors corresponding
to the $\lambda_m$. Moreover, since $\sigma$ acts as $-1$ on the
normal bundle to $M^\sigma$, we have in terms of the Chern roots
$\lambda_n^{(NM^\sigma)}$ of the normal bundle to $M^\sigma$:
\begin{equation} \label{wedgeminus1}
 {\rm ch}_\sigma(\wedge_{-1} \overline{N}_{M^\sigma}) = \prod_n
 \left(1+e^{-\lambda_n^{(NM^\sigma)}}\right).
\end{equation}
In the case at hand, $M=\Sigma$, $V$ is trivial, and the fixed point set
$\Sigma^\sigma$ of $\sigma:(\xi,t)\to(-\xi,-t)$ is a set of isolated
points, namely the pinch points $pp$, so ${\rm
Td}(\Sigma^\sigma)=1$. Furthermore from (\ref{wedgeminus1}) and the
fact that the normal bundle has rank 2 and the Chern roots are zero
(because $\dim \Sigma^\sigma=0$) it follows that ${\rm
ch}_\sigma(\wedge_{-1} \overline{N}_{\Sigma^\sigma})=2 \times 2 =
4$. Hence (\ref{holoLef}) becomes
\begin{equation}
 1 + h^{0,2}_+(\Sigma) - h^{0,2}_-(\Sigma) = \int_{\Sigma^\sigma} \frac{1}{4} = \frac{n_{pp}}{4} \, .
\end{equation}
Together with (\ref{h02form}), this implies
\begin{equation} \label{h02minusformula}
 h^{0,2}_-(\Sigma) = \frac{1}{24}
\int_\Sigma c_1^2 + c_2 \, - \frac{n_{pp}}{8}.
\end{equation}
This formula holds in general (if $h^{0,1}=0$). For our example we have
$n_{pp}=8m(2m-8)$ and (\ref{chernSigma}).
Thus we find
\begin{equation}
 h^{0,2}_-(\Sigma)= \frac{4}{3} \, m^3 - 8 \, m^2 + \frac{59}{3} \, m \, ,
\end{equation}
in precise agreement with the direct counting (\ref{D7moddirect}).

We will now show more directly that $h^{0,2}_-$ indeed equals the
number of D7 moduli. The D7 is given by the equation
$P:=\eta^2-h\chi = 0$. Infinitesimal deformations are of the form
$\delta P = 2 \eta \delta \eta - h \delta \chi$, parametrized by
$(\delta \eta,\delta \chi)$ modulo the equivalence $(\delta
\eta,\delta \chi) \simeq (\delta \eta + h \delta \psi,\delta \chi +
2 \eta \delta \psi)$.

Now consider the Poincar\'e residue\footnote{see section 5.6 of \cite{Denef:2008wq} for notation and the general framework leading to this expression}
\begin{equation}
 \rho = \oint \oint \oint \frac{\omega \cdot V}{(\xi^2-h)(\xi
 t-\eta)(t^2-\chi)} \, \frac{\delta P}{\xi},
\end{equation}
where $\omega=du_1 \wedge du_2 \wedge du_3 \wedge du_4 \wedge d\xi
\wedge t$, $V=u_i \partial_{u_i} + 4 \xi \partial_\xi + (m-4) t
\partial_t$ and the integration contours are infinitesimal loops
around the zeros of the denominator dividing $\omega \cdot V$. Note
that this results in a well defined meromorphic $(2,0)$ form on
$\Sigma$, since the integrand is gauge invariant (invariant under the projective rescalings). In fact, despite
the $1/\xi$ factor, $\rho$ actually has no poles and hence is
holomorphic, not just meromorphic. To see this, note that inside the
residue integral we can replace
\begin{equation} \label{deltaPred}
 \frac{\delta P}{\xi} = \frac{h\delta \chi - 2 \eta \delta
 \eta}{\xi} \to \frac{\xi^2 \delta \chi - 2 \xi t \delta
 \eta}{\xi} = \xi \delta \chi - 2 t \delta
 \eta,
\end{equation}
where the substitutations $h \to \xi^2$, $\eta \to \xi t$ are
allowed because inside the residue integral anything with a factor
appearing in the denominator dividing $\omega \cdot V$ integrates to
zero as the integrand then becomes analytic inside one of the
contours. Finally, it is clear that $\rho$ is antisymmetric under
$\sigma:(\xi,t) \to (-\xi,-t)$, because the integrand is
antisymmetric.

Thus any infinitesimal deformation of the D7 maps to an element
$\rho$ of $H^{2,0}_-(\Sigma)$ by this map. Conversely, every element
$\rho$ of $H^{2,0}_-(\Sigma)$ can be written as a residue
\begin{equation}
 \rho = \oint \oint \oint \frac{\omega \cdot V}{(\xi^2-h)(\xi
 t-\eta)(t^2-\chi)} \, Q(u,\xi,t)
\end{equation}
where $Q(u,\xi,t)$ is a degree $2m-4$ polynomial antisymmetric under
$\sigma$. Using the fact that we can substitute $\xi^2 \to h(u)$,
$t^2 \to \chi(u)$ and $\xi t \to \eta(u)$, the most general such $Q$
is of the form $Q(u,\xi,t) = \xi \, Q_1(u) + t \, Q_2(u)$. Comparing
to (\ref{deltaPred}), we see that $\delta P/\xi$ parametrizes the
most general $Q$ of this kind, and therefore every $\rho \in
H^{2,0}_-(\Sigma)$ maps to a D7 deformation.

This completes the proof that there is a one to one map between
$H^{2,0}_-(\Sigma)$ and the infinitesimal deformations of the D7.
Although we set up the proof in the context of our example, it is
clear that the argument generalizes.

\subsection{Open string flux vacua}
\label{sec:fluxvacconstr}

Now we know how to correctly compute tadpoles, flux lattice
dimensions and moduli space dimensions, we can move on to
constructing open string flux vacua. These are configurations of D-branes carrying worldvolume magnetic gauge fluxes,
with all D-brane moduli frozen, in a given closed string background. We will restrict here
to the case without closed string background fluxes.

For our purposes, a supersymmetric open string flux vacuum, in the large volume limit, is hence given by a (possibly reducible, i.e.\ multi-component) holomorphic 4-cycle $\Sigma$ as described above, together with a $U(1)$ flux\footnote{Nonabelian fluxes are possible too, of course. For simplicity
we restrict to abelian fluxes in most of this section.} $F \in H^2_-(\Sigma)$, such that $F$ is anti-self dual: $F=-*F$, or equivalently:
 \begin{equation} \label{susyflux}
 F^{0,2}=F^{2,0}=0, \qquad F \wedge J = 0,
\end{equation}
where $J$ is the K\"ahler form on $X$ pulled back to $\Sigma$.
The flux is integrally quantized up to a shift\footnote{There might
be more subtle constraints on $F$ as well, along the lines of \cite{GarciaEtxebarria:2005qc}. \label{quantsubtlety}} $\Delta F = \frac{c_1(\Sigma)}{2}$ \cite{Minasian:1997mm,Freed:1999vc}.

Since cohomology classes have unique harmonic representatives, and $F \wedge J$ is harmonic if $F$ is
harmonic, it is sufficient to consider the equations (\ref{susyflux}) in cohomology. Note also that if $\Sigma$ is
irreducible, the second equation $F \wedge J = 0$ follows automatically from the fact that $F$ is odd
under $\sigma$ while $J$ is even. For a D7-image-D7 pair on the other hand, we can a priori
turn on an arbitrary flux $F$ on one D7 and the image flux $-\sigma^* F$ on the image D7. In this case the condition
$F \wedge J = 0$ must be enforced on the two components separately and becomes nontrivial (see section \ref{sec:stab}).

While the second equation can be thought of as a D-term constraint, which involves the background K\"ahler moduli, the first equation can be though of as an F-term constraint. Indeed it is the critical point equation $\partial_z W(\psi,z)=0$ of the following superpotential
\cite{Witten:1997ep,Gomis:2005wc,Martucci:2006ij,Denef:2007vg}, depending on 3-fold complex structure moduli $\psi$ and D7 deformation moduli $z$:
\begin{equation} \label{superpotential}
 W_F(\psi,z) = \int_{\Gamma_F(z)} \Omega(\psi)
\end{equation}
where $\Omega$ is the holomorphic 3-form on $X$ and $\Gamma_F$ is the 3-chain swept out by a 2-cycle Poincar\'e dual to $F$ on the D7-worldvolume $\Sigma$ by varying $\Sigma$ from some reference $\Sigma_0$ to $\Sigma_z$. The physical interpretation of this superpotential is
that its norm represents the tension of a domain wall interpolating between different open string flux vacua --- in the weak string coupling limit such a domain wall is a D5-brane wrapping $\Gamma_F$. See \cite{Denef:2008wq} for a detailed account of the relation of this superpotential to the Gukov-Vafa-Witten superpotential of F-theory.

Although elegant, (\ref{superpotential}) is of little practical use for explicit construction of flux vacua. For our example, we would need to identify a basis of the $23320$-dimensional flux lattice of $\Sigma$, compute its intersection form, compute the 23320 corresponding 3-chain periods of $\Omega$ as a function of the 3728 D7 moduli, and find critical points of the resulting superpotential for given flux quanta. Needless to say, this is not quite feasible.

There is however a much simpler, geometrical way to construct open string flux vacua, to which we now turn.

\subsection{Flux vacua and holomorphic curves} \label{sec:holocurves}

The F-term constraint $F^{0,2} = F^{2,0} = 0$ is equivalent to $F$ being Poincar\'e dual to a rational linear combination of holomorphic curves on $S$. Thus, flux vacua can be constructed by picking a (possibly multi-component) curve $\gamma$ in $X$, taking $S$ to contain $\gamma$, and putting
\begin{equation} \label{fluxvalue}
 F := {\rm PD}_\Sigma(\bar\gamma) - {\rm PD}_\Sigma(\bar\gamma') \, ,
\end{equation}
where $\Sigma$ is the parameter surface (\ref{Sigmadef}), $\bar\gamma$ is the lift of $\gamma$ to $\Sigma$, and $\bar \gamma'=\sigma \bar\gamma$ its orientifold image. This form guarantees the orientifold projection $F=-\sigma^* F$ is satisfied.

This construction (for D4-branes in non-orientifolded CY manifolds) was used extensively in \cite{Denef:2007vg,Gaiotto:2006wm} for the purpose of enumeration of supersymmetric D-brane configurations. For the purpose of explicitly constructing D7 flux vacua in IIB orientifolds, one could proceed as follows.

For concreteness we consider our basic example with a D7 equation of full degree 32 (so $m=16$), and restrict to the case where $\gamma$ is a rational curve of degree $d$, that is, in the coordinates used in (\ref{Sigmadef}):
\begin{equation}
 (\xi,\vec u) = \big( \Xi(x,y),\vec U(x,y) \big) \, , \qquad [x:y] \in \ICP^1 \, ,
\end{equation}
where $(\Xi,U_i)$ are homogeneous polynomials of degree $(4d,d)$ in $(x,y)$. Requiring the curve $\gamma$ to lie in $S$ is equivalent to imposing that for \emph{all} $(x,y)$:
\begin{eqnarray}
 \Xi^2(x,y) &=& h\big(\vec U(x,y)\big) \label{one} \\
 \eta^2\big( \vec U(x,y) \big) &=& \Xi^2(x,y) \, \chi\big(\vec U(x,y)\big) \, . \label{two}
\end{eqnarray}
The first equation is the condition for the curve to lie in the Calabi-Yau $X$. It is of degree $8d$ in $(x,y)$, so it amounts to $8d+1$ independent equations on the coefficients of $\Xi$ and the $U_i$. There are $(4d+1)+4 \times (d+1) = 8 d + 5$ such coefficients, of which $4$ can be set to zero by a $GL_2$ reparametrization of the $\ICP^1$, resulting in $8d+1$ independent curve deformations in the space parametrized by $(\xi,\vec u)$. Hence for a given $h$ we generically expect a discrete set of curves solving the first equation. The number of solutions is roughly speaking given by the genus zero Gopakumar-Vafa invariants. For our example this is $29,504$ for $d=1$ and 128,834,912 for $d=2$ \cite{Huang:2006hq}.

The second equation, (\ref{two}), is of degree $32 d$, and since we know $\Xi$ and $\vec U$ already from the previous step, this can be thought of as a set of
\begin{equation}
  N_{\rm constr}(\gamma) = 32 d + 1
\end{equation}
equations for the coefficients of $\chi$ and $\eta$ determining the D7 embedding. Thus we see explicitly that turning on flux freezes some of the moduli. Note that since the number of independent deformations of the D7 is 3728, we need at least $d \geq 117$ if we want to freeze all D7 deformation moduli in this way.

However, we must take into account D3 tadpole cancelation. Supersymmetric fluxes will induce positive D3-charge. The D3 charge induced by the flux (\ref{fluxvalue}) equals
$
 Q_{\rm D3}(\gamma) = -\int_\Sigma \frac{F^2}{2} = -\bar\gamma^2 \big|_{\Sigma} \, .
$
Here we dropped a term $+\bar\gamma \cdot \bar \gamma'|_\Sigma$ which we can typically expect to be zero, because for rational curves we generically only expect intersection points of $\gamma$ and $\gamma'$ at $\xi=0$, but these are split by the lifting to $\bar \gamma$ and $\bar \gamma'$, except in the nongeneric case in which they happen to coincide with pinch points. (If the intersection happens to be nonzero after all, it will only make $Q_{\rm D3}$ larger.) The self-interesection product can be computed using the adjunction formula: $\bar \gamma^2|_\Sigma = - \chi(\bar \gamma) + \bar \gamma \cdot c_1(\Sigma) = -2 -28 d$, where we used (\ref{chernSigma}) and $\chi(\ICP^1)=2$. We conclude
\begin{equation}
 Q_{\rm D3}(\gamma) = 28 d + 2 \, .
\end{equation}
So, if we take $d \geq 117$, as we saw necessary to freeze all moduli in this way, we get $Q_{\rm D3} > 3276$, which, interestingly, is quite a bit higher than the maximal value allowed by tadpole cancelation, $1944$ (minus the curvature induced D3 charge). We conclude that we cannot supersymmetrically freeze all D7 deformation moduli by turning on fluxes of the form (\ref{fluxvalue}) with $\gamma$ a rational curve. This behavior generally persists when $\gamma$ is taken to be a collection of curves of total euler characteristic positive or not too negative. For large negative Euler characteristics, the sitation is not clear, also because there may now be curve moduli. These observations suggest the possible existence of a no-go theorem forbidding moduli stabilization at weak coupling by worldvolume fluxes alone. It would be interesting to investigate this further.

\section{Weak coupling D9-anti-D9 picture} \label{sec:D9aD9}

Through tachyon condensation, D7-branes can be obtained as bound
states of (multiple) D9 branes carrying a certain vector bundle $E$
with their orientifold image anti-D9 branes, offering an
alternative description of D7-branes in orientifolds. We will see in this section
that this allows us to derive all of the formulae inferred in the previous section from
comparison with F-theory, as well as obtain further results on the systematics of D7-branes
in orientifolds, including determining enhanced gauge symmetries and charged particle spectra, all using nothing more than simple polynomial manipulations.

\subsection{D9-anti-D9 bound states: non-orientifold case}

\subsubsection{Tachyon quotient construction}

Bound states of D-branes on Calabi-Yau manifolds have been discussed
extensively in the context of non-orientifolded $\CN=2$ theories, beginning with the work \cite{Brunner:1999jq} and fitting at the most general level in the categorical framework of \cite{Douglas:2000gi}, as reviewed in \cite{Aspinwall:2004jr}. For the state of the art of this program, see \cite{Herbst:2008jq}.

In orientifolds, such bound states are less well understood, and the proper mathematical framework has not been developed in full generality. Let us therefore first briefly review some of what is known in
the non-orientifold case, while trying to make the discussion as
concrete and practical as possible. In the next subsection we will make the
necessary generalizations to the orientifold case.

We will first give the general abstract construction and then turn to concrete examples.

When D9-branes and anti-D9-branes, each carrying gauge fields
strengths (bundles) giving rise to lower dimensional charges, are
placed on top of each other, the combined system will not be
supersymmetric at large volume. Under favorable conditions however
(where ``favorable'' is determined by D-term/stability constraints), tachyonic open string modes $T$ exist
between the brane and the anti-brane systems, which can condense to
form a new supersymmetric bound state \cite{Sen:1998sm} (reviewed in \cite{Sen:2004nf}). This leads to the string
theoretic incarnation of K-theory (see e.g.\ \cite{atiyah}) as the proper classifications scheme for D-brane charges
\cite{Witten:1998cd}. A sampling of papers on this topic is \cite{Minasian:1997mm,Horava:1998jy,Gukov:1999yn,Hori:1999me,Bergman:1999ta,Olsen:1999xx,Moore:1999gb,Bouwknegt:2000qt,Diaconescu:2000wy,Witten:2000cn,Harvey:2000te,Uranga:2000xp,Maldacena:2001xj,Moore:2003vf}.

Classical supersymmetric bound states of a stack of D9-branes
carrying a bundle $F$ of rank $r$ and a stack of anti-D9 branes
carrying a bundle $E$ of rank $r' \leq r$ can be represented as follows (concrete examples will be
given below).\footnote{We will use $F$ both to denote bundles and gauge field curvatures. We regret the confusion this may cause.} The tachyon can be thought of as a section of $F
\otimes E^*$, or equivalently as a linear map
\begin{equation}
 T:E \to F,
\end{equation}
which locally can be represented by an $r \times r'$ matrix
function on the CY manifold $X$. To get a supersymmetric configuration, $T$ must be
holomorphic, which we will assume from now on.

If the tachyon is everywhere a one to one map between the vector
space fibers of $E$ and $F$, then $E$ and $F$ are isomorphic
--- in other words the second brane is the exact anti-brane of the
first --- and tachyon condensation annihilates them completely,
leaving the vacuum behind. If $T$ is not everywhere one to one, then
annihilation will only be partial, resulting in a D-brane described
by the quotient $G = F/TE$. What we mean by this
quotient\footnote{which we can also write as the short exact
sequence $0 \to E \xrightarrow{T} F \to G \to 0$.} is essentially
the fiberwise equivalence
\begin{equation} \label{FTEquotient}
 f \simeq f + T \cdot e
\end{equation}
for vectors $f$ and $e$ in the fibers of $F$ and $E$ respectively.
Indeed when $T$ is everywhere one to one, this quotient leaves the
zero element everywhere, i.e.\ the vacuum. When this is not the
case, the quotient amounts, loosely speaking, to annihilating the
``part'' of $E$ isomorphic to $F$. When $r' < r$, the quotient will
generically leave a fiber of dimension $r-r'$ behind, as expected
when $r'$ anti-D9-branes annihilate $r'$ out of $r$ D9-branes. When
$r=r'$, which is the case we will be interested in, $T$ will
generically be invertible everywhere except at points where $\det T
= 0$, so the quotient will be trivial except at this complex
codimension 1 locus.

A physical derivation of this prescription leading to K-theory, based on RG flow between boundary CFTs, can be found in \cite{Moore:2003vf}.

Since the dimension of the fiber is not constant, the quotient is
not a vector bundle. The proper mathematical description of this is
a sheaf. Physically thinking of sections as wave functions of a
hypothetical charged particle, a sheaf can essentially be thought of
as a description of a system by the set of all possible wave
functions with respect to all possible local observers. A bundle on
the other hand describes a system by specifying the gauge
transformations of wave functions between a number of local
observers which together cover the space.

Thus, in the quotient (\ref{FTEquotient}), we can think of $f$ as
the possible wave functions of (particles living on) the D9 and $e$
as the wave functions of the anti-D9, and the result are the wave
functions of the bound state. The fact that the quotient is zero
outside of
\begin{equation} \label{D7tacheq}
 S:\det T = 0
\end{equation}
just means that the wave functions are localized on this complex
codimension one locus, i.e.\ we get a D7-brane localized at $S$.

The D7 will in general again carry a holomorphic vector bundle
determined by (\ref{FTEquotient}). There is a slight subtlety, in
that the actual bundle carried by the worldvolume gauge fields on
the corresponding D7-brane is not exactly the bundle obtained by
taking the quotient (\ref{FTEquotient}) restricted to $S$, but
rather \cite{Katz:2002gh, Aspinwall:2004jr} this bundle tensored by the
``line bundle'' $K_S^{-1/2}$, where $K_S$ is the canonical line
bundle if $S$ (whose first Chern class is $-c_1(S)$). In practice
this can be thought of as the gauge flux $F$ being shifted by a
diagonal $U(1)$ flux $\Delta F = +\frac{c_1(S)}{2}$. The need for
this shift can easily be checked from charge conservation, as we
will illustrate in an example below.

For generic $T$, the fiber dimension of the resulting
bundle will be one, so we have a line bundle. Physically, this
corresponds to a single smooth D7 carrying a type (1,1) $U(1)$ flux,
equal to first chern class of the line bundle. A convenient and
useful way to describe this flux $F$ is by its Poincar\'e dual
2-cycle on the divisor $S$ wrapped by the D7. This was used
extensively in \cite{Gaiotto:2005rp,Gaiotto:2006aj}, whose construction we put in a more general framework here.
Because $F$ is a $(1,1)$-form, the 2-cycle will be a linear
combination of holomorphic curves, i.e.\ a divisor on $S$. Concretely the divisor is given by
the zeros and poles of a section of the line bundle corresponding to $F$.

To make this precise, assume for simplicity that $F$ restricted to $S$ has a holomorphic section $f(x)$ \footnote{Different
choices of $f(x)$ will lead to homologous divisors.}
and let $[f(x)]$ be its equivalence class under (\ref{FTEquotient}).
By construction, $[f(x)]$ is a section of a line bundle on $S$. As
such it is associated to a divisor $\gamma: [f(x)]=0$, i.e.
\begin{equation} \label{curvedef}
 \gamma = \{x \in S \,\, | \,\, \exists \, e(x): \, f(x)=T(x) \cdot e(x)
 \},
\end{equation}
where $e(x)$ denotes a local section of $E$ restricted to $S$. We can
rewrite the equation in (\ref{curvedef}) in the following useful
way. Let $\tilde{T}(x)$ be the matrix of cofactors ($(r-1) \times
(r-1)$ minors) obtained from $T(x)$. Then $\tilde{T}^t \cdot T =
(\det T) \, {\bf 1} = 0$ on $S$, so each $x \in \gamma$ satisfies
the system of holomorphic equations $\tilde{T}^t(x) \cdot f(x)=0$.
Conversely, if $x \in S$ is such that $\tilde{T}^t(x) \cdot f(x)=0$,
then $f(x)$ is in the image of $T(x)$. This can be checked most easily by going
to a basis where $T(x)$ is upper triangular with only the lowest diagonal element
equal to zero (which we can do since by assumption $\det T$ has a simple zero on $S$).
Thus
\begin{equation} \label{curvedef2}
 \gamma = \{x \in S \,\, | \,\, \tilde{T}^t(x) \cdot f(x)=0 \}.
\end{equation}
Taking the above mentioned flux shift into account, we thus find
that the flux carried by the D7-brane is
\begin{equation} \label{fluxBS}
 F = {\rm PD}_S(\gamma) + \frac{c_1(S)}{2} = {\rm PD}_S(\gamma) -
 \frac{\iota^*_S[S]}{2}.
\end{equation}
Here PD stands for Poincar\'e dual and $\iota_S^*[S]$ denotes the pullback of the 2-cohomology class
$[S]$ to $S$. In the last step we used that in a CY,
$c_1(TS)=-c_1(NS)=-\iota_S^*[S]$. This expression for $F$ can be
checked for example by matching the charge of the bound state to the
sum of the charges of the constituents, as we will illustrate below.

Finally, the moduli space of these supersymmetric bound states is
parametrized by different choices of $T$ (sections of $F \otimes
E^*$ invertible at generic points of $X$) modulo internal holomorphic complexified gauge transformations:
\begin{equation} \label{gaugesymm}
 T \to g_F T g_E^{-1}.
\end{equation}
Here $g_F:F \to F$ and $g_E:E \to E$ are (not necessarily constant) automorphisms.\footnote{An automorphism of the bundle $E$ is a change of basis respecting the bundle structure, i.e.\ an invertible linear map sending sections to sections. It is a holomorphic section of $E \otimes E^*$.}

The residual gauge symmetry is the set of gauge transformations (\ref{gaugesymm}), with $g_F$ and $g_E$ automorphisms \emph{independent} of the internal coordinates, that leave $T$ invariant. We usually express the residual gauge groups in their real rather than complexified forms (e.g.\ $U(N)$ instead of $GL(N)$). For generic $T$, the
gauge group is completely broken to a diagonal $U(1)$, as expected
for a single D-brane. For special choices of bundles and $T$, there
can be a larger residual gauge group. There can also be subspaces of
$X$ where locally the residual symmetry gets enhanced; if these loci
intersect $S$, this typically signals the presence of massless
matter arising e.g.\ at D-brane intersections.

\subsubsection{Bound states of fluxed D9-anti-D9 branes and brane recombination}

Let us now turn to some examples. The simplest possibility is to
take $E$ and $F$ to be two line bundles. For example on the quintic
Calabi-Yau, let us choose $E=\CO(-a)$, $F=\CO(b)$, where we take
$a,b>0$, so we have holomorphic sections for $E^* = \CO(a)$ and $F$. The holomorphic sections are
just the homogeneous polynomials of degree $a$ resp.\ $b$. Then $T$ is
a section of $F \otimes E^* = \CO(a+b)$, a polynomial of degree
$a+b$, hence the class of the D7 divisor $S$ is $[S]=(a+b)H$ with
$H$ the hyperplane class. Furthermore $\tilde{T}=1$, so we can take
$\gamma$ to be the zero locus of simply some degree $b$ polynomial
$f(x)$, and from (\ref{fluxBS})
\begin{equation}
 F = \iota_S^*(b H - \frac{[S]}{2}) = (b-\frac{a+b}{2}) \iota_S^* H = \frac{b-a}{2}
 \iota_S^* H.
\end{equation}
Charge conservation requires that the total charge of the
constituents as given by (\ref{D9charge}):
\begin{equation}
 ({\rm ch}\, F - {\rm ch}\, E) ( 1+\mbox{$\frac{c_2(X)}{24}$} ) =
 (e^{bH}-e^{-aH})(1+\mbox{$\frac{c_2(X)}{24}$})
\end{equation}
equals that of a D7 wrapping $S$ with flux $F$, as given by
(\ref{GamD7}) and (\ref{eulerformula}):
\begin{equation}
 (a+b)H + \frac{(b+a)(b-a)}{2} \, H^2 + \frac{(a+b)^3 H^3 + (a+b)
 c_2(X) \cdot H}{24} + \frac{(a+b)(b-a)^2 H^3}{8}.
\end{equation}
This is indeed the case, as can be checked by a short computation. (For the
quintic, $c_2(X) \cdot H=50$, but this is not needed to show the equality.)

The moduli space of such supersymmetric bound states is the
projectivization of the space of nonzero holomorphic sections of
$\CO(a+b)$ on the quintic, i.e.\ the space of nonzero degree $a+b$
homogeneous polynomials modulo $\IC^*$ rescalings and modulo the
defining equation of the quintic. This is indeed the moduli space of
the corresponding D7-brane wrapping $S$.

A less trivial and important class of examples is obtained by taking
$E$ and $F$ to be the direct sum of line bundles
\begin{equation}
 E = \bigoplus_{i=1}^r \CO(-U_i), \qquad F = \bigoplus_{i=1}^r \CO(V_i),
\end{equation}
where we take the $U_i$ and $V_i$ to be divisors such that there are
holomorphic sections of $E^*$ and $F$ (for example this is
guaranteed when they are very ample). In more basic terms, $F$
describes a superposition of $r$ D9 branes, each carrying a U(1)
flux $F_i = c_1(\CO(V_i))$, which is the Poincar\'e dual to the
divisor $V_i$. Similarly, $E$ describes a collection of anti-D9
branes.

Then $T$ is a $r \times r$ matrix with $T_{ij}$ a holomorphic
section of $\CO(V_i+U_j)$. The D7 divisor class is
\begin{equation}
 [S] = \sum_{i=1}^r V_i + U_i,
\end{equation}
and the curve $\gamma$ is given by the system of equations
\begin{equation}
 v_i(x) \, \tilde{T}_{ij}(x) = 0, \qquad x \in S,
\end{equation}
$v_i(x)$ being an arbitrary section of $\CO(V_i)$ and $\tilde{T}$ again the matrix of cofactors of $T$. This reproduces and clarifies the results of \cite{Gaiotto:2006aj}.

For a more explicit example (which will also be relevant further on), take $r=2$, $X$ the quintic, $E=\CO(-a) \oplus
\CO(-b)$, $F=\CO(a) \oplus \CO(b)$, with $a,b>0$. Then the most
general tachyon is of the form $T={\scriptsize \left(
\begin{array}{cc} P & Q \\ R & W
\end{array} \right)}$ with $(P,Q,R,W)$ polynomials of degree $(2a,a+b,a+b,2b)$.
The divisor $S$ wrapped by the D7 is given by the degree $2(a+b)$
equation $P(x) W(x) - Q(x) R(x)=0$ ($x \in X$), and
$\tilde{T}={\scriptsize \left(
\begin{array}{cc} W & -R \\ -Q & P
\end{array} \right)}$. Taking for instance $v(x)=(v_1(x),0)$, where $v_1$ is of
degree $a$, we get for the curve $\gamma: v_1(x) W(x) = 0, \, v_1(x)
R(x) = 0$ ($x \in S$). This splits in two components, because we
have chosen $v_2 \equiv 0$. The first is just the intersection of
the divisor $v_1(x)=0$ with $S$. The second is the curve
$R(x)=W(x)=0$, which indeed always lies on $S$, but not on an
arbitrary divisor in the class $[S]$. This is a manifestation of the
fact that the dual flux $F$ puts a restriction on the divisor
deformation moduli, as discussed at length in \cite{Gaiotto:2005rp,Gaiotto:2006aj}.

In the special case in which $T={\scriptsize \left(
\begin{array}{cc} P & 0 \\ 0 & W
\end{array} \right)}$, with $P \neq W$, the residual gauge symmetry
(\ref{gaugesymm}) is enhanced to $U(1) \times U(1)$. This case
corresponds to two intersecting D7 branes wrapping $S_1:P(x)=0$ and
$S_2:W(x)=0$. The curve construction prescription given above
becomes somewhat degenerate in this nongeneric case, so in order to
determine the flux carried by these D7-branes it is best to go back
to the original quotient prescription (\ref{FTEquotient}). On the
branch $P(x)=0$, $W(x) \neq 0$, we see that the equivalence relation
is fixed by putting $f_2 \equiv 0$. Hence on $S_1$ we get a line
bundle $\CO(a)|_{S_1} \otimes K_{S_1}^{-1/2} = \CO(a) \otimes
\CO(-a) = \CO(0)$, which corresponds to zero flux, and similarly on
$S_2$ we get zero flux. Thus, as in \cite{Gaiotto:2005rp,Gaiotto:2006aj}, we see that two
branes \emph{without} any flux can smoothly and physically
continuously recombine into a single brane \emph{with} flux. This
happens through condensation of bifundamental matter with charge
$(1,-1)$ and $(-1,1)$ under the $U(1) \times U(1)$ gauge group,
which corresponds to the off diagonal degree $a+b$ polynomials $Q$
and $R$.

When $a=b$ we can consider $T={\scriptsize \left(
\begin{array}{cc} P & 0 \\ 0 & P
\end{array} \right)}$. In this case we get two coincident D7-branes on
$P(x)=0$ and the residual gauge symmetry is enhanced to $U(2)$. The
bundle carried this system is easily seen from the quotient
construction to be $(\CO(a) \oplus \CO(a)) \otimes K_{S'}^{-1/2} =
(\CO(a) \oplus \CO(a)) \otimes \CO(-a) = \CO(0) \oplus \CO(0)$,
i.e.\ no flux.

\subsubsection{Other D9-anti-D9 bound states, ideal sheaves and Donaldson-Thomas invariants}

We could also start from more complicated D9-D7-D5-D3 bound states. In particular, even in the rank 1 case,
for suitable values of the moduli, supersymmetric stringy bound states may exist of a single D9 with a
``gas'' of D5 and D3 branes. Unlike the higher rank
case, these branes cannot disolve as flux into the D9, so these
bound states do not have a smooth bundle description. However they
can still be described by sheaves, more precisely they are rank 1
\emph{ideal} sheaves \cite{MNOP1,MNOP2}. These D9-branes and their charge conjugates in turn can be used to build large classes of D7-brane bound states through tachyon condensation. Since ideal sheaves are basically just collections of polynomials vanishing on the given loci, this can again be done very explicitly using only polynomial maniplualtions. Compared to the line bundle case, the tachyon matrix entries $T_{ij}$ must satisfy the additional constraint that they have to vanish on the D5 and D3 brane loci inside the $\overline{\rm D9}_i$ and the ${\rm D9}_j$ stacks. This implies in particular that on solutions the D7 locus $S:\det T=0$ must contain the D5 and D3 branes. This reproduces the picture of flux vacua of section \ref{sec:holocurves}. It is conceivable that in fact \emph{all} D7-brane vacua can be constructed in this way.

Ideal sheaves of a given charges are counted by (rank 1)
Donaldson-Thomas invariants, which are related to Gromov-Witten
invariants and hence to the topological string \cite{Iqbal:2003ds,MNOP1,MNOP2,Dijkgraaf:2006um}.
Bound states of ideal sheaves and their charge conjugates played a key role in \cite{Denef:2007vg} in enumerating D4 BPS states and proving a version of the OSV conjecture, suggesting a similar role for enumeration of D7 flux vacua. We leave this for future work.

\subsection{D7-branes as D9-D9$'$ bound states: the orientifold case} \label{sec:tachyonorientifold}

We will now generalize the results reviewed in the previous
subsection to orientifolds.  A  proper mathematical framework for the description of D-branes in terms of D9-D9$'$ in orientifolds has been studied in \cite{Hori:2006ic,Diaconescu:2006id}.  See also \cite{hori}.

\subsubsection{Tachyon quotient construction}

From section \ref{sec:oractions}, it follows that the orientifold image
of a D9 carrying a bundle $F$ is the charge conjugate of a brane with bundle
$F'=\sigma^* F^*$, where $F^*$ denotes the dual bundle, obtained by
from $F$ by inverting the gauge field. We can thus consider bound
states of such a D9 with its image anti-D9, resulting in a D7. This
gives us an alternative way to compute various topological
quantities such as RR charges.

Before tachyon condensation, the D9-D9$'$ system has a $G \times G$
gauge symmetry, where $G$ is the subgroup of $U(r)$ leaving the
bundle $F$ invariant. The orientifold projection reduces this to
$G$, since (\ref{sigmaonA2}) relates the gauge field $A'$ on the D9$'$ to the gauge
field $A$ on the D9 as $A'=-\sigma^*A^t$, and correspondingly the
gauge transformations as $g'=\sigma^* g^{t,-1}$ (the superscript $^t$
denotes the transpose). The tachyon $T$ therefore transforms as
\begin{equation} \label{orgaugetransf}
 T \to g \cdot T \cdot (g')^{-1} = g \cdot T \cdot \sigma^* g^t
\end{equation}
under $G$. $T$ is now a holomorphic linear map from $F'=\sigma^*
F^*$ to $F$, or in other words a section of $F \otimes {F'}^* = F \otimes \sigma^* F$.
We impose orientifold projection condition
\begin{equation} \label{tachproj}
 T = - \sigma^* T^t.
\end{equation}
As will see, this choice of sign corresponds to the O7$^-$ projection we want, while a plus sign corresponds to the O7$^+$ projection. Note that the projection is compatible with (\ref{orgaugetransf}).
It is  possible to deduce  such orientifold projection conditions from first principle using a more elaborate mathematical formalism \cite{hori}.

The resulting D7-brane will again be wrapped around the divisor $S:
\det T = 0$, and carries a
bundle given by the quotient $F/T F'$ shifted by $K_S^{-1/2}$.

For concreteness we specialize to the orientifold example described in
section \ref{sec:theorientifold} and to the case in which $F$ is the
direct sum of $r$ line bundles of degree $a_i \geq 0$,
\begin{equation}
 F = \CO(a_1) \oplus \cdots \oplus \CO(a_r),
\end{equation}
i.e. the superposition of $r$ D9 branes wrapping the Calabi-Yau
3-fold, each carrying a U(1) flux $F_i = a_i H$, $i=1,\ldots,r$,
where $H$ is as before the hyperplane class $[u_1=0]$. Its
orientifold image is the anti-brane of the D9 system with inverted
fluxes $F_i' = - a_i H$ (the action of $\sigma^*$ is trivial in the
example, as $H_-^2(X)=0$). By charge conservation, the total charge
of the D9-D9$'$ bound state (assuming it exists) is the sum of the
charges of the constituents, which is, using (\ref{D9charge}):
\begin{equation} \label{D9aD9charge}
 \Gamma_{{\rm D9-D9}'} = 2 \sum_i a_i H \oplus \biggl( \frac{1}{3} \sum_i a_i^3 +
 \frac{11}{6} \sum_i a_i \biggr) H^3.
\end{equation}
The tachyon $T$ is an $r \times r$ matrix with $T_{ij}$ a section of
$\CO(a_i) \otimes \CO(a_j) = \CO(a_i + a_j)$, i.e.\ a homogeneous
polynomial in $(u_1,u_2,u_3,u_4,\xi)$ of degree $a_i + a_j$. The
orientifold projection (\ref{tachproj})
constrains $T$ to be of the form
\begin{equation} \label{gentachformminusproj}
 T(u,\xi) = A(u) + \xi \, S(u),
\end{equation}
where $A_{ij}$ and $S_{ij}$ are sections of $\CO(a_i+a_j)$ resp.\
$\CO(a_i+a_j-4)$ satisfying
\begin{equation}
 A^t = -A, \qquad S^t = S.
\end{equation}
Terms of higher order in $\xi$ can be eliminated using the CY
equation $\xi^2=h(u)$. A modified projection condition
(\ref{tachproj}) with the plus sign instead would give $T=S+\xi A$.
Both give a D7 equation $\det T = 0$ invariant under $\sigma:\xi \to
-\xi$.

A crucial subtlety is that to avoid an uncanceled $\IZ_2$ tadpole, we must restrict to
\emph{even} $r$. One (slick\footnote{As stated before, probe arguments are not entirely satisfactory since it might be the probe that is inconsistent. They can in certain cases be shown to be equivalent to more direct K-theory arguments \cite{Maiden:2006qe}. It should be possible to provide such more direct arguments for our setup as well, perhaps using the framework of \cite{mooretexas}.}) way to see this is through a probe argument
\cite{Uranga:2000xp}: The worldvolume theory of a D3-probe placed on the
$O7^-$ in the presence of $r$ D9-branes is an $SU(2)$ $\CN=1$ gauge
theory coupled to $r$ chiral multiplets in the fundamental of
$SU(2)$ coming from open string stretching from the D3 (and its
image) to the D9;\footnote{Strings stretching to the D9$'$ are identified with string stretching to the
D9 and should therefore not be counted separately.} thus, if $r$ is odd, we have an odd number of Weyl
fermions in the fundamental of $SU(2)$, and this results in a
$\IZ_2$ anomaly \cite{Witten:1982fp}. Thus we conclude $r$ must be even. As will become clear below,
odd $r$ would moreover give results in contradiction with what we obtained in the previous sections.


\subsubsection{Bound states of fluxed D9-D9$'$ branes and brane recombination} \label{subsubsec:recombination}

Since $r$ must be even, the simplest possibilities are bundles of the form
$F=\CO(a) \oplus \CO(b)$.  These give rise to a D7 of charge $2 (a+b)
H$. The tachyon is of the form
\begin{equation} \label{D7D7tach}
 T = \left( \begin{array}{cc} 0 & \eta(u) \\ -\eta(u) & 0 \end{array} \right)
 + \xi \, \left( \begin{array}{cc} \rho(u) & \psi(u) \\ \psi(u) & \tau(u) \end{array} \right)
\end{equation}
with $(\eta,\rho,\tau,\psi)$ are homogeneous polynomials of degree $(a+b,2(a-2),2(b-2),a+b-4)$. The D7 is localized at $S:\det T = 0$, i.e.
\begin{equation} \label{D7imD7}
 S: \, \eta^2 = \xi^2(\psi^2-\rho \tau).
\end{equation}
Satisfyingly, this is precisely of the general allowed form (\ref{D7eq}). Note that if we had chosen the other sign in (\ref{tachproj}) or if we had started from an $r=1$ pair, this would not have been the case.\footnote{In particular taking the opposite sign in (\ref{tachproj}) would result in a D7 equation of the form $S: \xi^2 \eta^2 = (\psi^2 - \rho \tau)$, which generically does not have double D7-O7 intersections. This is compatible with the claim that this choice of sign corresponds to the O7$^+$ projection, which as discussed in section \ref{sec:pertexpl} is not expected to have the double intersection property.}

If $a<2$, $\rho$ must be zero, and the D7 splits in two components $\eta = \pm \xi \psi$. We will see in section \ref{sec:stab} that such configurations cannot be supersymmetric at large CY volume because they violate the D-term constraints.\footnote{Except when $a=b=1$, in which case only $\eta$ is nonzero and of degree 2.} Therefore we take $a \geq 2$ and similarly $b \geq 2$.

The largest number of tachyon degrees of freedom is obtained for the minimal case $a=2$. In this case (\ref{D7imD7}) gives the most general equation of the form (\ref{D7eq}). Moreover, the D3-charge for this case is, using (\ref{D9aD9charge}) and putting $m:=a+b=2+b$: $\Gamma|_{\rm D3} = (\frac{2}{3} m^3 - 4 \, m^2 + \frac{35}{3} m) \, \omega$. Happily, this is in exact agreement with our earlier proposed modified charge
formula (\ref{chiodef}) using (\ref{chioexample}) for $\chi_o$, assuming there is no flux on the D7. In the following we will see that for $a=2$ the latter is indeed true.

Let us compute the flux carried by the D7 for general $a$, $b$. For generic polynomials, (\ref{D7imD7}) describes a single D7, smooth everywhere except on the curve $C:\eta=\xi=0$, where we have double point
singularities. To deduce the line bundle carried by this D7, it is
convenient to use the $\Sigma$ parametrization of section
\ref{sec:modcharge} again. In this parametrization we can write the
tachyon $T$ and its matrix $\tilde{T}$ of cofactors as:
\begin{equation}
 T = \xi \left( \begin{array}{cc} \rho & \psi + t \\ \psi-t & \tau \end{array}
 \right), \qquad
 \tilde{T} = \xi \left( \begin{array}{cc} \tau & -\psi + t \\ -\psi-t & \rho \end{array}
 \right),
\end{equation}
and $\Sigma$ is given by
\begin{equation}
 \Sigma: t^2 = \psi^2 - \rho \tau.
\end{equation}
Now we consider the quotient $(F/TF')|_{\Sigma}$ defined by
(\ref{FTEquotient}) on $\Sigma$. More precisely we first remove the
locus $\xi=\eta=0$, compute the quotient bundle and then extend it
to the closure $\Sigma$; this amounts to throwing out
the degenerate $\xi=0$ branch. According to the general prescription, the line bundle carried
by the D7 is
\begin{equation}
 \CL = (F/TF')|_{\Sigma} \otimes K_\Sigma^{-1/2},
\end{equation}
which corresponds to a $U(1)$ flux $F = c_1(\CL)$
\begin{equation}
 F = {\rm PD}_{\Sigma}(\gamma) + \frac{1}{2} c_1(\Sigma)
\end{equation}
where $\gamma$ is a curve $(P_a,P_b) \cdot \tilde{T} = 0$ on
$\Sigma$ with $P_a$, $P_b$ arbitrary polynomials of degree $a$, $b$.
Taking $P_a \equiv 0$, we get $\gamma = \gamma_1 \cup \gamma_2$ with
\begin{equation} \label{gammadefeq}
 \gamma_1: P_b(u) = 0, \qquad \gamma_2: \rho(u)=0, \, t+\psi(u)=0,
\end{equation}
which are two smooth curves on $\Sigma$. Furthermore from
(\ref{chernSigma}) we get $c_1(\Sigma)=(4-2(a+b)) H_\Sigma$, with
$H_\Sigma$ a shorthand for $\iota_S^* \hat{H}$, the pullback of the
hyperplane class $\hat{H} = [u_1=0]$ to $\Sigma$. Therefore, using
${\rm PD}_{\Sigma}(\gamma_1) = b \, H_{\Sigma}$:
\begin{equation} \label{Fexprrr1}
  F = {\rm PD}_{\Sigma}(\gamma_2) - (a-2) H_{\Sigma}.
\end{equation}
This does not look like it respects the orientifold projection
$\sigma^* F = -F$. But it actually does, as we will now show. Let
$\gamma_2' = \sigma(\gamma_2)$ be the orientifold image of
$\gamma_2$, obtained by inverting $t \to - t$ in (\ref{gammadefeq}).
Now note that $\gamma_2 \cup \gamma_2'$ equals the complete
intersection of $\Sigma$ with the degree $2(a-2)$ divisor $\rho=0$.
Hence
\begin{equation}
 {\rm PD}_{\Sigma}(\gamma_2) + {\rm PD}_{\Sigma}(\gamma_2') = 2 (a-2)
 H_{\Sigma},
\end{equation}
and, substituting this in (\ref{Fexprrr1}),
\begin{equation} \label{Fgamgamp}
 F = \frac{1}{2} ( {\rm PD}_{\Sigma}(\gamma_2) - {\rm
 PD}_{\Sigma}(\gamma_2') ).
\end{equation}
This is manifestly antisymmetric under exchange of $\gamma_2$ and
$\gamma_2'$, so we see that, after all, $\sigma^* F = -F$.

We also see that in the minimal case $a=2$, we have $\gamma_2 = 0$
since $\rho$ is a constant in that case, and therefore $F=0$,
confirming our claim made above that the corresponding D7 has no
flux.

We can also study brane recombination in this framework. If we take
a nongeneric $T$ with $\rho=\tau=0$, the D7 splits into a
D7-image-D7 pair $\Sigma_{\pm}:t \pm \psi = 0$. In fact since now the
individual branes are smooth, it is no longer necessasry to
introduce the $\Sigma$ parametrization, and we can just consider the
surfaces $S_{\pm}: \eta \pm \xi \psi = 0$ in $X$. To compute the line
bundles $(F/TF')|_{S_{\pm}}$, we go back to the basic definition of
the quotient itself. On say $S_+$ the tachyon becomes
\begin{equation}
 T|_{S_+} = \left( \begin{array}{cc} 0 & 0 \\ 2 \xi \psi & 0 \end{array}
 \right).
\end{equation}
Hence the equivalence relation is simply
\begin{equation}
 (f_1,f_2) \simeq (f_1,f_2 +2 \xi \psi f'_1),
\end{equation}
so for generic $\psi$ we get a line bundle whose sections have
unique representatives $(f_1,0)$, i.e.\ it is identified with
$\CO(a)|_{S_+}$ inside $(\CO(a) \oplus \CO(b))|_{S_+}$. Furthermore
$c_1(S_+) = -(a+b) H_{S_+}$, so the flux on $S_+$ is
\begin{equation}
 F_{S_+} = a H_{S_+} - \frac{a+b}{2} H_{S_+} = \frac{a-b}{2}
 H_{S_+}.
\end{equation}
Similarly
\begin{equation}
 F_{S_-} = \frac{b-a}{2} H_{S_-}.
\end{equation}
So we get a D7 brane-image-brane pair with opposite fluxes turned
on, as expected. When the two branes coincide (i.e.\ $\psi \equiv
0$), one similarly computes that the rank two bundle on brane at
$\eta=0$ is $\CO(\frac{a-b}{2}) \oplus \CO(\frac{b-a}{2})$.

The charges of the two smooth D7-branes on $S_{\pm}$ are readily
computed using (\ref{GamD7}) and (\ref{eulerformula}):
\begin{equation} \label{GD7pm}
 \Gamma_{D7 \pm} = m H \, \oplus \, \pm \frac{a-b}{2} m H^2 \, \oplus \, \biggl( \frac{(m^2+22)m}{24}
 + \frac{1}{2} \left( \frac{a-b}{2} \right)^2 \biggr) m H^3,
\end{equation}
with $m=a+b$. As a check, summing these up agrees with the charges
obtained from (\ref{D9aD9charge}).

Note that when $a=b$, there is no flux on the separate branes. Yet
on the recombined branes, the flux (\ref{Fgamgamp}) is nonzero. Thus
we see again that fluxes can (and must, to respect charge conservation) be turned on by physically
smooth recombination processes.

\subsubsection{Moduli} \label{sec:nmoduliD9aD9}

The moduli space of these supersymmetric bound states is
parametrized by different choices of $T$ modulo internal holomorphic complexified gauge transformations:
\begin{equation}
 T \to g \cdot T \cdot \sigma^* g^{t} \, ,
\end{equation}
where $g:F \to F$ is an automorphism of $F$.

Specializing again to the case $F=\CO(a) \oplus \CO(b)$, we get that $g$ is a $2 \times 2$ matrix of polynomials of degrees
\begin{equation}
 \left(\begin{array}{cc}
  0 & a-b \\ b-a & 0
 \end{array}\right) \, .
\end{equation}
Note that if $b>a$, then $g_{12}=0$. Let the number of degree $k$ homogeneous polynomials on $\ICP^3$ be $N_k$, i.e.\ $N_k={k+3 \choose 3}$ (and $N_k \equiv 0$ if $k<0$). The
number of degrees of freedom of $T$ is $N_{a+b}+N_{2(a-2)}+N_{2(b-2)}+N_{a+b-4}$, the number of degrees of freedom of $g$ is $2N_{0}+N_{a-b}+N_{a-b-4}+N_{b-a}+N_{b-a-4}$, and we expect the dimension of the moduli space to be the difference of these. (Here $N_{a-b-4}$ and $N_{b-a-4}$ are the numbers of polynomials of degree $a-b$ and $b-a$ with one factor of $\xi$.). To compare to the results of section \ref{sec:D7modulicount}, we consider the fluxless D7 case $a=2$, $b=m-2$:
\begin{eqnarray}
 \# \mbox{moduli} &=& N_{m}+N_{0}+N_{2(m-4)}+N_{m-4}-2N_0-N_{m-4}-N_{m-8} \\
 &=& N_m + N_{2m-8} - N_{m-8} - N_0 \, ,
\end{eqnarray}
which reproduces (\ref{D7moddirect}).

\subsubsection{Enhanced gauge symmetries} \label{sec:enhanced}

After tachyon condensation, the residual gauge group in the four
dimensional low energy effective field theory is given by the gauge transformations
(\ref{orgaugetransf}) for automorphisms $g:F \to F$ \emph{independent} of the internal coordinates (so $\sigma^* g = g$) which leave $T$ invariant, i.e.\
\begin{equation}
 G = \{ g \, | \, g T g^t = T \} \, .
\end{equation}
At generic values
of $T$, this leaves only a discrete $O(1)=\IZ_2$. In particular, unlike
the non-orientifolded case, there will generically be no residual
$U(1)$s. From the geometric D7 point of view this is also clear: the generic D7 will be a single component brane invariant under the orientifold involution, so the photon polarized in the noncompact spacetime gets projected out.

At particular values of $T$, the gauge group may be enhanced.
Supersymmetric deformations away from these special loci correspond
to massless matter in the low energy effective field theory. Some
examples in the $r=2$, $F=\CO(a) + \CO(b)$ case, with $T$ as in
(\ref{D7D7tach}) are:
\begin{center}
\begin{tabular}{l|l|l|l}
  tachyon & D7 worldvolume & flux & gauge group \\
  \hline & & \\

  ${\scriptsize \left( \begin{array}{cc} \xi \rho & \xi \psi + \eta \\ \xi\psi - \eta & \xi \tau \end{array} \right)}$
  & $\eta^2 = \xi^2(\psi^2-\rho \tau)$
  & $\frac{1}{2} ( {\rm PD}_{\Sigma}(\gamma_2) - {\rm PD}_{\Sigma}(\gamma_2') )$
  & $O(1): g = {\scriptsize \left( \begin{array}{cc} -1 & 0 \\ 0 & -1 \end{array} \right) }$ \\

  ${\scriptsize \left( \begin{array}{cc} 0 & \xi \psi + \eta \\ \xi \psi - \eta & 0 \end{array} \right) }$
  & $\eta = \pm \xi \psi$
  & $\frac{a-b}{2} H_{\eta=\xi\psi} + \frac{b-a}{2} H_{\eta=-\xi\psi}$
  & $U(1): g = {\scriptsize \left( \begin{array}{cc} e^{i\theta} & 0 \\ 0 & e^{-i\theta} \end{array} \right) }$ \\

  ${\scriptsize \left( \begin{array}{cc} 0 & \eta \\ - \eta & 0 \end{array} \right)}$
  & $\eta = 0$
  & $\frac{a-b}{2} H_{\eta=0} \oplus \frac{b-a}{2} H_{\eta=0}$
  & $U(1): g ={\scriptsize \left( \begin{array}{cc} e^{i\theta} & 0 \\ 0 & e^{-i\theta} \end{array}  \right)}$ \\

  ${\scriptsize \left( \begin{array}{cc} 0 & \eta \\ - \eta & 0 \end{array} \right), a=b}$
  & $\eta = 0$
  & $0 \oplus 0$
  & $SU(2): \det g = 1$ \\

  ${\scriptsize \left( \begin{array}{cc} \xi & 0 \\ 0 & \xi \end{array} \right), a=b=2}$
 & $\xi = 0$
 & $0 \oplus 0$
 & $O(2): g = {\scriptsize \left( \begin{array}{cc} \cos \theta & \sin \theta \\ \mp \sin \theta & \pm \cos \theta \end{array}
 \right)}$ \\

  ${\scriptsize \left( \begin{array}{cc} \xi \rho & \eta \\ -\eta & \xi \rho \end{array} \right), a=b}$
 & $\eta = \pm i \xi \rho$
 & 0
 & $SO(2): g = {\scriptsize \left( \begin{array}{cc} \cos \theta & \sin \theta \\ -\sin \theta & \cos \theta \end{array}
 \right)}$ \\

\end{tabular}
\end{center}

\vskip2mm These are consistent with expectations from the D7 worldvolume
theory in the O7$^-$ projection; in particular we
expect orthogonal groups for flux-free branes coincident with the $O7^-$,
symplectic groups for flux-free, transversal, non-Higgsed brane stacks,
and subgroups of those for branes obtained by turning on
additional fluxes or by various Higgsings (i.e.\ deformations); in particular
brane-image-brane stacks will carry unitary groups.

Let us go through the above (non-exhaustive) list and check compatibility with the rules summarized in
footnote \ref{GGrules}:
\begin{itemize}
\item The first case is the most generic D7. It has a single component and maps to itself under the orientifold involution, so the photon is projected out in the four dimensional effective field theory and only a $O(1)=\IZ_2$ gauge symmetry remains.

\item The second case describes a brane-image-brane pair, each carrying some flux. The orientifold action maps one component to the other, so we get a single $U(1)$.

\item The third and fourth case are similar. The \emph{fourth} configuration corresponds to a stack of two flux free, coincident, transversal, involution invariant branes ($\eta^2=0$), hence the $SU(2)=USp(2)$ group. The \emph{third} case corresponds to the same rank two stack, however, with a non-trivial flux that breaks the $SU(2)$ down to $U(1)$.

\item The fifth case describes a stack of two flux-free branes coincident with the O-plane. Hence the $O(2)$ gauge group.

\item The sixth case describes a brane-image-brane pair, so the $SO(2)$ can be thought of as a $U(1)$ group, just as in the second case.

\end{itemize}
Note that the first case can be thought of as a Higgsing of the second case, whereby a brane-image-brane pair have recombined (i.e.\ whereby $\rho$ and $\tau$ acquire vev's), and the $U(1)$ group is broken to $\mathbb{Z}_2$. It can also be thought of as a Higgsing of any of the other cases with matching values of $(a,b)$. Fluxes can be induced by the recombination process, as required by charge conservation, and ensuring geometric and gauge theory degrees of freedom match each other.

If we had chosen the other sign in the orientifold projection condition (\ref{sec:pertexpl}), we would have found flipped roles for orthogonal and symplectic gauge groups, consistent with the O7$^+$ projection; for example two D7-branes coincident with the O7 would have been described by a tachyon ${\scriptsize T = \left( \begin{array}{cc} 0 & \xi \\ - \xi & 0 \end{array} \right)}$, with residual gauge group $SU(2)$ = ${\rm USp}(2)$.

We can also consider higher (even) rank $r$. For example we could
consider the bundle
\begin{equation}
 F = \CO(2) \oplus \CO(2) \oplus \cdots \oplus \CO(2)
\end{equation}
with $r$ terms in the sum. This gives a tachyon of the form
\begin{equation}
 T = A(u) + \xi S,
\end{equation}
with $A$ a $r \times r$ antisymmetric matrix with polynomial entries
of degree 4, and $S$ a constant symmetric matrix, which generically
after a suitable complexified gauge transformation we can take to be the unit
matrix. When $A \equiv 0$, this describes $r$ coincident D7 branes
on the O7 locus $\xi=0$, with zero flux, and $O(r)$ enhanced gauge
symmetry ($U(r)$ transformations $g$ satisfying $gg^t=1$). When $A
\neq 0$, the D7 is described by
\begin{equation}
 S:\det(A(u)+\xi {\bf 1}) = 0
\end{equation}
so for each $u$, the solution set for $\xi(u)$ is the set of
eigenvalues of $A(u)$, which is invariant under $\xi \to - \xi$.
Again there will be a particular flux on the worldvolume of this D7,
computed according to the general quotient prescription, which
restricts its moduli and balances the charges.

Note in particular that a configuration with $r=8$ will saturate the D7 tadpole,
leading to an $O(8)$ enhanced gauge symmetry at $A=0$, and a D3
charge $\frac{304}{3} \omega$, so the total D7+O7 charge is, using
(\ref{O7charge}), $\Gamma_{D7} + \Gamma_{O7} = 152 \, \omega$. This
can also be seen to agree with F-theory: in the $O(8)$ limit under
consideration the fibration of the fourfold degenerates to a $D_4$
singularity over $\xi=0$, so the Euler characteristic of the
fourfold is $\chi([\xi=0])\cdot \chi(D_4) = 304 \cdot 6 = 1824$ and
the D3-charge (in CY$_3$ units) equals $152 \, \omega$.

The number of moduli in this $O(8)$ sector equals the number of tachyon degrees of freedom minus the number of holomorphic complexified gauge transformations, i.e.
\begin{equation}
 \# \mbox{ moduli} = 28 \times 35 + 36 \times 1 - 64 = 952 \, .
\end{equation}
Notice that this is quite a bit less than the number of moduli in the generic D7 sector, 3728.

Similarly, we can describe $n$ coincident copies of a single generic D7-brane described by a $2 \times 2$ tachyon matrix  $T_0$ by the $2n \times 2n$ tachyon matrix
\begin{equation}
 T=T_0 \oplus \cdots \oplus T_0 \, .
\end{equation}
This will have gauge group $O(n)$.

\subsubsection{Charged matter fields} \label{sec:chargedmatter}

The massless matter representations at enhanced symmetry points can
be read off easily as well. For example at the $SU(2)$ locus, $\eta$
corresponds to matter in the antisymmetric (${\bf 1}$)
representation, and $\rho,\tau,\psi$ to matter in the symmetric
(${\bf 3}$) representation. At the $U(1)$ locus $T={\scriptsize
\left(
\begin{array}{cc} 0 & \xi \psi + \eta \\ \xi \psi - \eta & 0 \end{array} \right)}$,
$\eta$ and $\psi$ are neutral, while $\rho$ has charge $+2$ and
$\tau$ charge $-2$. At the $O(8)$ locus, $A$ is matter in the antisymmetric (${\bf 28}$) and $S$ in the symmetric (${\bf 36}$) representation.

The physical number of degrees of freedom of a given charge equals the number of linearized degrees of freedom of $T$ of that charge minus the number of those that can be absorbed in a linearized holomorphic complexified gauge transformation. Taking the example of the $U(1)$ locus $T_0={\scriptsize
\left(
\begin{array}{cc} 0 & \xi \psi + \eta \\ \xi \psi - \eta & 0 \end{array} \right)}$, the linearized degrees of freedom of $T$ are $\delta \rho$, $\delta \tau$, $\delta \psi$ and $\delta \eta$. The linearized gauge holomorphic complexified internal gauge transformations of $T_0$ for $b>a$ correspond to fluctuations
\begin{equation}
 \delta \rho = 0\, , \quad \delta \tau = \alpha \psi + \beta \eta \, , \quad \delta \eta = \gamma \eta \, ,
 \quad \delta \psi = \gamma \psi \, ,
\end{equation}
where $(\alpha,\beta,\gamma)$ are infinitesimal polynomials of degree $(b-a,b-a-4,0)$. Thus the number of physical degrees of freedom in each charge sector is
\vskip2mm
\begin{tabular}{ccc}
  $+2$ ($\delta \rho$): & $-2$ ($\delta \tau$): & $0$ ($\delta \psi$, $\delta \eta$): \\
  $N_{2(a-2)}$ & $N_{2(b-2)} - N_{b-a} - N_{b-a-4}$ & $N_{a+b-4} + N_{a+b} - N_0$
\end{tabular}
\vskip2mm
\noindent where as before $N_k:={k+3 \choose 3}$. Notice that this is $N_0=1$ more than the number of moduli computed in \ref{sec:nmoduliD9aD9}. The extra massless field gets eaten up by the Brout-Englert-Higgs mechanism when moving off the $U(1)$ locus.

From the D7 point of view, the charged massless fields arise from open strings stretching between the flux-carrying D7 and its orientifold image, given in the table in section \ref{sec:enhanced}.

The \emph{net} chirality of the spectrum is
\begin{equation} \label{chiralspectrum}
 \# (+2) - \# (-2) = N_{2(a-2)} - N_{2(b-2)} + N_{b-a} + N_{b-a-4} \, .
\end{equation}
This vanishes when $a=b$, i.e.\ when there is no flux on the D7 and its image. The index can be computed more directly from an index theorem as well, to which we now turn.

\subsection{Open string indices}

Index theorems give a powerful way to compute the number of massless degrees of freedom in various settings in string theory, or at least to compute such numbers counted with alternating signs according to some grading (referred to as ``net'' number). A thorough discussion of open string indices in the absence and presence of orientifold projections can be found e.g.\ in \cite{Brunner:2003zm}, part of whose results (adapted to our setup) we review here.  As in the rest of the paper, we will assume here that the $B$-field vanishes.

\subsubsection{General formulae}

In the non-orientifolded case, the Witten index counting the net
number of open string modes stretching between two D-branes, say between an anti-D9 carrying a
bundle $E_1$ and a D9 carrying a bundle $E_2$, is
\begin{equation} \label{chi12}
 I(1,2) = \sum_n (-)^n \, h^{0,n}(E_1 \otimes E^*_2) \, .
\end{equation}
By the Hirzebruch-Riemann-Roch formula
(\cite{Denef:2008wq} section 5.8), this equals
\begin{eqnarray}
I(1,2) &=& \int_X {\rm ch}(E_1 \otimes E^*_2) \, {\rm Td}(X) = \int_X   {\rm ch}(E_1) \,
{\rm ch}(E_2)^* \,  {\rm Td} (X)
\\
 &=& \int_X
 \left( {\rm ch}(E_1) \, \sqrt{{\rm Td} (X)}\,   \right) \,
 \left({\rm ch}(E_2) \, \sqrt{{\rm Td} (X)}\,  \right)^* \\
 &=& \int_X \Gamma_1 \wedge \Gamma_2^*.
\end{eqnarray}
Here $\Gamma^*$ is defined as the form obtained by flipping the sign
of the 2- and 6-form components of $\Gamma$, i.e.\ $\Gamma^*:=\sum_k
(-1)^k \Gamma^{(2k)}$ where $\Gamma^{(2k)}$ is the $(2k)$-form
component of $\Gamma$. This generalizes to arbitrary pairs of branes
of charge $\Gamma_1$ and $\Gamma_2$:
\begin{equation} \label{bifundindex}
 I(1,2) = \int_X \Gamma_1 \wedge \Gamma_2^* =: \langle
 \Gamma_1,\Gamma_2 \rangle.
\end{equation}
The symplectic product thus defined is called \emph{intersection
product} of the charges $\Gamma_1$ and $\Gamma_2$ of the two branes.
In the mirror IIA picture it is the geometric intersection product,
counting the number of 3-cycle intersection points with signs.

Let us now compute the analogous index of open strings between a D9 stack with bundle $F$ and its orientifold image (more accurately from the latter to the former), i.e.\ the strings giving rise to our tachyon field $T$. This is
\begin{equation} \label{IoE}
 I_o(F) := \sum_n {\rm Tr}_{H^{0,n}(X,F \otimes \sigma^* F)} \, (-)^n \, \mbox{$\frac{1}{2}$}(1-\sigma) \, .
\end{equation}
Here $\sigma$ denotes the full orientifold $\IZ_2$ action including worldsheet orientation reversal.
The insertion of $\mbox{$\frac{1}{2}$}(1-\sigma)$ is equivalent to imposing the O7$^-$ tachyon projection (\ref{tachproj}), since
\begin{equation}
 \sigma(T) = \sigma^* T^t \, .
\end{equation}

The charge of the D9, $\Gamma_{\rm D9}$, and that of its image, $\Gamma_{\rm D9'}$, are
\begin{equation} \label{D9charges}
 \Gamma_{\rm D9} = {\rm ch}(F) \sqrt{{\rm Td} (X) } \, , \qquad \Gamma_{\rm D9'} = -\Gamma_{D9}^* =- {\rm ch}(F^*) \sqrt{{\rm Td} (X) }.
\end{equation}
The charge of a O7 plane wrapping $X^\sigma$  is
\begin{equation}
\Gamma_{\rm  O7}=-{8  \sqrt{\frac{{L}(\frac{1}{4} T_{ X^\sigma})}{{L}(\frac{1}{4}  N_{X^\sigma})      }
 }} \    \,      [X^\sigma].
\end{equation}
Using (\ref{bifundindex}), we rewrite the index $I_o(F)$ as
\begin{equation}
 I_o(F) =  \frac{1}{2} \big( \langle \Gamma_{\rm D9'}, \Gamma_{D9} \rangle -\sum_n {\rm Tr}_{H^{0,*}(X,F \otimes \sigma^* F)} \, (-)^n \, \sigma \big) \, .
\end{equation}
Following  section    3.3 of \cite{Brunner:2003zm}, the second term can be written as
\begin{equation}
{\rm Tr}_{H^{0,*}(X,F \otimes \sigma^* F)} \, (-)^p \, \sigma =
2^{\rm{dim}_\mathbb{R} X^\sigma-\frac{1}{2} \mathrm{dim}_{\mathbb{R} }  X}
\int_{ X^\sigma}   {\rm
 ch}(F^*) \sqrt{\rm{Td(X)} }\sqrt{
\frac{{L}(\frac{1}{4} T_{ X^\sigma})}{{L}(\frac{1}{4}  N_{X^\sigma})}
 }.
 \end{equation}
For an ${\rm O7}$-plane wrapping a complex surface of a Calabi-Yau three-fold, we have ${\rm dim}_{\mathbb{R}} X^\sigma=4$   and ${\rm dim}_{\mathbb{R}} X=6$. It follows that
\begin{eqnarray}
{\rm Tr}_{H^{0,*}(X,F \otimes \sigma^* F)} \, (-)^p \, \sigma
&=& 2   \int_{ X^\sigma}    {\rm
 ch}(F^*) \sqrt{\rm{Td (X)} }\sqrt{
\frac{{L}(\frac{1}{4} T_{ X^\sigma})}{{L}(\frac{1}{4}  N_{X^\sigma})}
 }\\
 &=& - \frac{1}{4} \int_{X}     \Big({\rm
 ch}(F) \sqrt{{\rm Td}(X)}\     \Big)^*\        \Big(-{8  \sqrt{\frac{{L}(\frac{1}{4} T_{ X^\sigma})}{{L}(\frac{1}{4}  N_{X^\sigma})      }
 }} \    \,      [X^\sigma]  \Big)       \nonumber\\
 &=&  -\frac{1}{4} \int_X \Gamma_{\rm D9}^* \wedge  \Gamma_{\rm O7}   \nonumber\\
& =&   -\frac{1}{4} \langle \Gamma_{\rm  O7}, \Gamma_{\rm D9}\rangle .
 \end{eqnarray}

We finally get the simple result
\begin{equation}
 I_o(F) = \frac{1}{2} \biggl( \langle \Gamma_{\rm D9'}, \Gamma_{D9} \rangle + \frac{1}{4} \langle \Gamma_{\rm O7}, \Gamma_{\rm D9} \rangle \biggr) \, .
\end{equation}
This immediately generalizes to arbitrary $(\Gamma,\Gamma')$ brane-image-brane systems and to the case where $X^\sigma$ has several different components corresponding to O$p$-planes:
\begin{equation} \label{Ioformula}
 I_o(\Gamma) = \frac{1}{2} \biggl( \langle \Gamma', \Gamma\rangle +\frac{1}{4}  \langle \Gamma_{\rm O}, \Gamma \rangle \biggr) \, ,\quad  \Gamma_{\rm O}=\sum  \Gamma_{\rm Op},
\end{equation}
where $\Gamma_{\rm O}$ denotes the sum of all Op-planes\footnote{This formula also holds for any choice of sign in the projection (\ref{tachproj}). The other sign would change $\sigma$ to $-\sigma$ in (\ref{IoE}), but this gets compensated by the fact that we also have $\Gamma_{{\rm Op}^+} = - \Gamma_{{\rm Op}^-}$. The universality of this formula can be understood physically from the ``orientihole'' picture, which will be discussed elsewhere.}.
Along the same lines, more refined indices can be derived counting the net number of bifundamental, symmetric and antisymmetric matter representations with respect to the various brane gauge groups \cite{Blumenhagen:2002wn} (see \cite{Marchesano:2007de} or \cite{Douglas:2006xy} section 5.1.3. for a summary). The indices derived here trace over these representations.

\subsubsection{Application to the example}

It is instructive to make contact with the direct polynomial counting we did in the previous subsections.

For our example, taking $\Gamma$ to be a D9 stack with $F=\CO(a) \oplus \CO(b)$, we have $\Gamma_{\rm D9}=\big( e^{aH} + e^{bH} \big) \big( 1 + \frac{c_2}{24} \big)$, and $c_2 = 22 \, H^2$, $H^3=2$ and the O7-charge is given by (\ref{O7charge}), so (\ref{Ioformula}) gives $I_o(D9) = \frac{5 a^3}{3}+\frac{5 b^3}{3}+ba^2+b^2 a-4 a^2 -4
b^2 +\frac{22 a}{3}+\frac{22 b}{3}-2$, which can be checked to be the total number of coefficients of the polynomials appearing in the tachyon matrix $T$ given in (\ref{D7D7tach}). The latter equals the number of holomorphic sections of $F \otimes \sigma^* F$ satisfying (\ref{tachproj}). The reason why this total number matches the index is the Kodaira vanishing theorem, implying $h^{0,q}(F \otimes F)=0$ for $q>0$ in the case at hand.

The D7-D7$'$ degeneration $\eta = \pm \xi \psi$ discussed above has $D7$-$D7'$ charges given in (\ref{GD7pm}), leading to a D7-D7$'$ index $I_o(D7) = a^3-b^3+ a^2 b - b^2 a -4 a^2 +4 b^2$, in agreement with the net chirality index (\ref{chiralspectrum}). However, as is clear from the table in section \ref{sec:chargedmatter}, the \emph{total} number of charged matter fields generally exceeds the index. The advantage of the direct tachyon matrix analysis is that it allows straightforward enumeration of all charged matter fields using elementary methods.

\subsection{Stability / D-term constraints} \label{sec:stab}

Four dimensional $\CN=1$ supersymmetric vacua have vanishing $F$- and $D$-term potentials. So far we have only studied the $F$-flatness constraints for our D-brane configurations, which are encoded in holomorphic equations and depend on the complex structure moduli. The $D$-flatness constraints are nonholomorphic and depend on the K\"ahler moduli. Mathematically, these correspond to stability conditions, most generally to $\Pi$-stability \cite{Douglas:2000ah,Douglas:2000gi,bridgeland}. They are also related to stability conditions of BPS black hole bound states \cite{Denef:2000nb,Denef:2007vg}.

This subject is rather involved. We will restrict ourselves here to a few simple observations.

In an O3/O7 type orientifold, part of the D-flatness condition is that all D-brane components must have zero central charge phase. The central charge of a D-brane of charge $\Gamma$ is, in the large radius regime and for zero $B$-field
\begin{equation}
 Z(\Gamma) = \langle \Gamma , -e^{i J} \rangle = \int \Gamma \wedge \big(-1+iJ+\frac{J^2}{2}-i \frac{J^3}{6}\big) \, .
\end{equation}
Zero phase at large $J$ therefore means $\Gamma|_{\rm D7} \cdot J^2 > 0$ and
\begin{equation}
 \Gamma|_{\rm D5} \cdot J - \Gamma|_{\rm D9} \, J^3 = 0 \, ,
\end{equation}
where we can replace $\Gamma|_{\rm D5}$ by the $\sigma$-even part $\Gamma|_{\rm D5,+} \in H^2_+$ because $\Gamma|_{\rm D5,-} \cdot J = 0$. Thus, in the large radius limit $J \to \infty$, no brane configuration which includes a component of nonzero D9-charge can be supersymmetric. When there is more than one K\"ahler modulus, there can be special real codimension 1 loci in the K\"ahler moduli space where D-brane components with $\Gamma|_{\rm D5,+} \neq 0$ are supersymmetric. These are typically walls of marginal stability, with supersymmetric brane recombination occurring on one side of the wall.

In our example, there is only one modulus, so any brane configuration with a component of nonzero D5-charge will be nonsupersymmetric. In particular D7-D7 brane configurations such as (\ref{GD7pm}) will in fact be nonsupersymmetric when $a \neq b$: they are F-flat, but not D-flat. This implies also that when $a<2$ or $b<2$ and $a \neq b$, there are no supersymmetric D9-D9$'$ bound states of the type considered in section \ref{subsubsec:recombination}, because in this case $\rho \tau = 0$ so the D7 condensate necessarily splits with charges as in (\ref{GD7pm}). Furthermore, when $a=b \leq 0$, the condition $\Gamma|_{\rm D7} \cdot J^2 > 0$ is violated, leaving $a=b=1$ as a special acceptable case. This justifies the claim made in the beginning of section \ref{subsubsec:recombination}.

\section{Some generalizations} \label{sec:generalizations}

Throughout this paper, we have often focused on our basic example to illustrate the various computational geometrical methods that can be used to analyze D-branes in compact orientifolds. It is clear however that many of the results we derived for this example can be generalized and systematized.

Let us for example give general formulae for the curvature induced D3 charge, the number of moduli and the flux lattice dimension for arbitrary D7-branes in O7$^-$ orientifolds. As we have argued, the D7 will in general obey an equation of the form  $S: \eta^2 = \xi^2 \chi$. Let $D$ be the divisor $\xi=0$ wrapped by the O7$^-$. To construct the parameter surface $\Sigma$, we blow up the double intersection curve $\xi=\eta=0$, by introducing an additional coordinate $t$ and imposing the equation $\eta = \xi t$. Consistency requires $t$ to be a holomorphic section of $\CO(\frac{S}{2}-D)$. The existence of such a section in turn requires
\begin{equation} \label{poscond}
 S - 2 D \geq 0 \, ,
\end{equation}
in the sense of positivity of line bundles. If this is not the case, the D7 doubly wraps the smooth divisor $S':\eta=0$. We will discuss this case separately below, and assume for now that (\ref{poscond}) is satisfied.

The surface $\Sigma$ is the proper transform of $S$ in the blown up space, which is given by the equation $\Sigma:t^2=\chi$. Denoting the divisor class $[\eta = 0]$ by $[\eta]$ and similarly for the other polynomials and coordinates, we thus have
\begin{equation}
 [\eta]= \frac{S}{2}, \quad [\xi] = D, \quad [\chi] = S-2D, \quad [t] = \frac{S}{2}-D \, .
\end{equation}
The number of pinch points is
\begin{equation}
 n_{pp} = \int_X [\eta] [\xi] [\chi] = \frac{1}{2} \int_X \frac{1}{2} SD(S-2D) \, .
\end{equation}
The Chern class of $\Sigma$ can be computed using the adjunction formula:
\begin{equation}
 c(\Sigma) = \frac{c(X) \, (1+[t])}{(1+[\eta])(1+[\chi])} = 1 + (D-S) +
 \big( S^2 + c_2 + 2 D^2 - \mbox{$\frac{5}{2}$} D S \big) \, .
\end{equation}
This allows us to compute the orientifold Euler characteristic of $S$:
\begin{eqnarray}
 \chi_o(S) &=& \int_\Sigma c_2(\Sigma) - n_{pp} \\
 &=& \int_X S^3 + c_2 \cdot S + 3 D S(D-S) \, .
\end{eqnarray}
The curvature induced D3 charge $-Q_c(S)$ and dimension $b(S)$ of the flux lattice are derived from this using (\ref{Q3cformula}) and (\ref{b2minformula}):
\begin{equation} \label{bSformula}
 Q_c(S) = \frac{\chi_o(S)}{24} \, , \qquad b(S) = b^2_-(\Sigma) = \frac{\chi_o(S)}{2} \, .
\end{equation}
Similarly, assuming $h^{0,1}(\Sigma)=0$ (as will be automatic if $\Sigma$ is ample, by the Lefshetz hyperplane theorem), the number of D7 deformation moduli $N_{\rm def}(S)$ is given by (\ref{h02minusformula}):
\begin{eqnarray}
 N_{\rm def}(S) &=& h^{2,0}_-(\Sigma) = \frac{1}{24} \int_\Sigma c_1(\Sigma)^2 + c_2(\Sigma) - \frac{n_{pp}}{8}\\
 &=& \int_X \frac{S^3}{12} + \frac{c_2 S}{24} + \frac{1}{4} DS(D-S) \, . \label{Ndeffff}
\end{eqnarray}
If we take the D7 to cancel the O7 tadpole, we have
\begin{equation} \label{Sis8D}
 S = 8 D \, .
\end{equation}
Plugging this in the above formulae and recalling (\ref{GamO7}) gives for the total curvature induced D3 charge $-Q$ on D7, O7 and O3 planes\footnote{Here and in what follows we assume the ``standard'' O-plane projection, i.e.\ giving rise to O7$^-$ and O3$^+$, which have signs of D7 resp.\ D3 charge opposite to that of the D7 and D3 in our charge sign conventions.} (for generic D7):
\begin{equation} \label{Qcall}
 Q = \frac{N_{\rm O3}}{2} + \frac{29 \, D^3}{2} + \frac{c_2 D}{2} \,  \qquad \mbox{(generic)}.
\end{equation}
Similar formulae are obtained for the total number of D7-worldvolume fluxes and moduli for the most generic D7, by substituting (\ref{Sis8D}) in (\ref{bSformula}) and (\ref{Ndeffff}). These formulae allow for efficient computation of these physical topological quantities.

As promised we now briefly return to the case when (\ref{poscond}) is \emph{not} satisfied and the D7 hence doubly wraps the surface $S':\eta=0$ (so there is an enhanced gauge symmetry $SU(2)$). Since this is generically smooth, we can simply use the standard formulae. For example, $Q_c(S) = 2 \times \frac{(S')^3 + c_2 S'}{24} = \frac{S^3}{96} + \frac{c_2 S}{24}$, and instead of (\ref{Qcall}) we now get
\begin{equation}
 Q = \frac{N_{\rm O3}}{2} + \frac{11 \, D^3}{2} + \frac{c_2 D}{2} \,  \qquad \mbox{($SU(2)$ sector)}.
\end{equation}
Similarly, in the $O(8)$ sector, with all D7 branes coincident with the O7$^-$, we get
\begin{equation}
 Q = \frac{N_{\rm O3}}{2} + \frac{D^3}{2} + \frac{c_2 D}{2} \,  \qquad \mbox{($O(8)$ sector)}.
\end{equation}

Finally, by making use of index theory, some further interesting relations can be derived. From the Lefshetz fixed point index theorem (\ref{leffix}), we get
\begin{equation} \label{Xsigmaeq}
 2 h + 4 = \chi(X^\sigma) \, ,
\end{equation}
where $h$ can be expressed in terms of the $\sigma^*$ even and odd Hodge numbers of $X$, or in terms of the corresponding numbers of massless \emph{closed} string modes of the orientifold compactification:
\begin{eqnarray}
 h&:=&h^{1,1}_+ - h^{1,1}_- + h^{2,1}_- - h^{2,1}_+ \label{Qform0} \\
 &=& \mbox{K\"ahler moduli $-$ (B,C)-axions $+$ compl.\ struct.\ moduli $-$ U(1) vectors.}
\end{eqnarray}
This is also equal to the number of geometric moduli of the original $\CN=2$ theory minus twice the number of these moduli which get projected out by the orientifold projection. In (\ref{Xsigmaeq}), $X^\sigma$ is the fixed point set of $\sigma$, i.e.\ the collection of all O3-planes and O7-planes. Thence:
\begin{equation}
 D^3 + c_2 D + N_{\rm O3} = 2h+4 \, .
\end{equation}
From the holomorphic Lefshetz fixed point theorem (\ref{holoLef}) with trivial bundle, we get similarly
\begin{equation}
 -D^3 + c_2 D + 3 N_{\rm O3} = 48 \, .
\end{equation}
These two equations can be solved to eliminate $D^3$ and $c_2 D$ in the above equations. For example with this and (\ref{Qcall}) we find for the total D3 tadpole from O3, O7 and D7:
\begin{equation} \label{Qform1}
 Q = -306 + 15 \, h + 14 \, N_{\rm O3} \, \qquad \mbox{(generic)}.
\end{equation}
For our basic example, this curious formula indeed gives $Q = -306 + 15 \times(1-0+149-0) + 0 = 1944$. Similarly
\begin{equation} \label{Qform2}
 Q = -108 + 6 \, h + 5 \, N_{\rm O3} \, \quad \mbox{($SU(2)$ sector)}, \qquad Q = h+2 \, \quad \mbox{($O(8)$ sector)} \, .
\end{equation}
The last case can be compared to the resolved $T^6/\IZ_2 \times \IZ_2$ model of \cite{Denef:2005mm}, which indeed has $h=51-0+3-0=54$ and $Q=56$.

For the generic D7 case, we furthermore obtain by combining the above formulae the following relations between total D3-tadpole, D7 flux lattice dimension, number of D7 moduli and $h$:
\begin{equation} \label{Qform3}
  N_{\rm def} + h + 10 = 2 \, Q \, , \qquad 7 \, b = 2 \, h + 82 \, Q - 668   \, , \qquad 2 \, N_{\rm def} = 86 \, Q - 7 \, b - 688 \, . \end{equation}

These generalizations can also be studied in the tachyon condensation picture, where the generic, D7-tadpole canceling, flux free D7-configuration is given by the D9-D9$'$ condensate for a D9 carrying the rank two bundle
\begin{equation}
 F = \CO\left(\mbox{$\frac{D}{2}$}\right) \oplus \CO\left( \mbox{$\frac{7 D}{2}$} \right) \, .
\end{equation}
In particular this reproduces the curvature induced D3 charges given above. Note however that $D$ might not be even, in which case the above ``minimal'' bundle $F$ does not exist. In fact, this is as it should: On the D7-side, this corresponds to the Minasian-Moore-Freed-Witten effect, i.e.\ the fact that if the first Chern class of the wrapped 4-cycle is odd, the brane must carry some compensating half-integral flux \cite{Minasian:1997mm,Freed:1999vc}.

Further generalizations will be explored in \cite{ACDE}.

\section{Summary and conclusions} \label{sec:summary}

With model building applications in mind, we have laid out and clarified several different approaches to the analysis of type IIB O3/O7 orientifold compactifications, with emphasis on global aspects of the D7-brane sector. In particular:
\begin{enumerate}
 \item We pointed out that generic D7-branes have singular double intersections with O7$^-$ planes. This means that on an CY orientifold $X$ given by $X:\xi^2 = h(\vec u)$ and involution $\sigma:\xi \to -\xi$, D7-branes
     are given by equations of the form
     \begin{equation} \label{D7eqformc}
      S:\eta^2(\vec u) = \xi^2 \chi(\vec u),
     \end{equation}
     where $h$, $\eta$ and $\chi$ homogeneous polynomials. One obtains this form from Sen's weak coupling limit of F-theory \cite{Sen:1997gv} (this was also observed in \cite{Braun:2008ua}), and we argued that it is implied for O7$^-$ but not for O7$^+$ by the Dirac quantization condition. We also showed that it is reproduced by D9-anti-D9 tachyon condensation in the O7$^-$ projection, taking into account a subtlety requiring the number of D9-D9$'$ pairs to be even. The surface $S$ has double points along $\xi=\eta=0$, and pinch point singularities at $\xi=\chi=\eta=0$. A local model for it is the Whitney umbrella in $\IC^3$.
 \item The singular nature of $S$ invalidates the standard formulae for physical topological properties such as curvature induced D3 charge $-Q_{c \, \rm D7}$, number of moduli $N_{\rm D7 \, def}$ and flux lattice dimension $b$. Naive application of the standard formulae or naive modifications thereof results in large discrepancies with F-theory. We gave a prescription for computing these numbers directly in perturbative IIB theory by introducing an auxiliary parameter surface $\Sigma$ splitting up the double intersection locus, namely the proper transform of the D7 worldvolume in the CY blown up in the double intersection curve. In particular we found\footnote{Here and in what follows we use IIB conventions for the D3-charge; see footnote \ref{IIBFconv}.}
     \begin{equation}
      Q^{(\rm IIB)}_{c \, \rm D7} = \frac{\chi(\Sigma)-n_{pp}}{24} \, , \qquad N_{\rm D7 \, def} = h^{2,0}_-(\Sigma) \, ,  \qquad b = b^2_-(\Sigma) = 12 \, Q_{c, \rm D7} \, ,
     \end{equation}
     where $n_{pp}$ is the number of pinch points and the minus subscript on the Hodge numbers refers to the $\sigma^*$ parity. These results were derived independently from F-theory, from D7 geometry and from tachyon condensation, and we found exact agreement between these different approaches.

     Using index theorems, we gave general explicit formulae for the above numbers. In particular this led to curious formulae relating for example the total D3-tadpole induced by O3 and O7 and D7 curvature to the closed string massless spectrum, as well as to various nontrivial relations between D3-tadpole, D7 flux lattice dimension and the open and closed string massless spectrum in four dimensions; see eqs.\ (\ref{Qform0}), (\ref{Qform1}), (\ref{Qform2}) and (\ref{Qform3}).


 \item We explained how to explicitly construct D7 flux vacua from holomorphic curve embeddings, by picking a curve $\gamma \in X$ together with its orientifold image $\gamma'$, requiring $S$ to contain these, and setting
     \begin{equation} \label{fluxcurverep}
      F = {\rm PD}_\Sigma(\bar \gamma) - {\rm PD}_\Sigma(\bar \gamma')
     \end{equation}
     where $\bar \gamma$ is the lift of $\gamma$ to $\Sigma$. We pointed out that at least for rational curves $\gamma$, this procedure consumes the available D3 charge tadpole (i.e.\ $- Q_{c \, \rm D7} - Q_{c \, \rm O7} -\frac{1}{2} \int_\Sigma F^2 > 0$) before it freezes all D7 moduli, suggesting a potential problem for supersymmetric stabilization of D7 moduli in the weak coupling limit using only worldvolume fluxes and taking tadpole cancelation into account.

 \item We emphasized in particular the practical power of the tachyon condensation picture, which allows one in principle to compute charges, moduli, enhanced gauge groups and matter spectra using elementary methods --- essentially just polynomial manipulations.

     The tachyon arises from open strings to a stack of D9 branes with bundle $E$ from their image anti-D9 branes with bundle $E'=\sigma^* E^*$ and is formalized as a linear map
     \begin{equation}
       T:E' \to E \, ,
     \end{equation}
     i.e.\ a section of $E \otimes \sigma^* E$. The O7$^-$ projection imposes $\sigma^* T = - T^t$ and an anomaly argument implies that the rank $r$ of $E$ must be even in this case. The D7 locus is given by $S:\det T=0$, reproducing embedding equations of the form (\ref{D7eqformc}) in this case.

     When $E$ is the direct sum of line bundles $\CO(D_i)$, the tachyon entries $T_{ij}$ are holomorphic sections of $\CO(D_i+D_j)$, which when $X$ is given as a submanifold of a toric variety are simply polynomials of degrees determined by $D_i+D_j$. More generally we could have a gas of D5- and D3-branes wrapped on curves and points in each of the D9-branes (i.e.\ consider ideal sheaves), and then $T_{ij}$ must vanish on the curves and points wrapped by the brane gas in $D9_i$ and $D9_j$, implying in particular that $S$ must contain all of these curves and points. This makes contact with (\ref{fluxcurverep}), and, since ideal sheaves are counted by Donaldson-Thomas invariants, suggests a role for the latter in counting D7 flux vacua similar to their role in counting D4-D0 black hole microstates in \cite{Denef:2007vg,Gaiotto:2006wm}.

 \item Even when $E$ is just the sum of line bundles, the resulting D7 bound state will in general carry flux. We illustrated how to explicitly compute this flux (on $\Sigma$ as defined above). In general the D7 will carry a bundle
     \begin{equation}
      L = E/TE'|_\Sigma \otimes K_\Sigma^{-1/2} \, ,
     \end{equation}
     where the quotient is simply the fiberwise equivalence $e \simeq e + T e'$. For generic $S$ this will be a line bundle, corresponding to $U(1)$ flux $F=c_1(L)$. More concretely, we found
     \begin{equation}
      F = \frac{1}{2} \big( {\rm PD}_\Sigma(\bar\gamma) - {\rm PD}_\Sigma(\bar\gamma') \big) \, ,
     \end{equation}
     with $\bar \gamma$ the lift (proper transform) of the curve obtained as the zero locus of for example the first row of cofactors (maximal minors) of the matrix $T_{ij}$:
     \begin{equation}
      \gamma = \{ x \in S \, \, |\, \, \tilde{T}_{1j}(x) = 0 \,\, \forall j \} \, .
     \end{equation}
     We observed  in this framework that D7 brane recombination can induce fluxes on branes even if these were originally absent on the constituent branes. This is needed for charge conservation, and explains from the geometric point of view the reduction in number of D7 moduli in sectors with enhanced gauge symmetry points. This generalizes the observations of \cite{Gaiotto:2005rp,Gaiotto:2006aj}, for space-localized D4 branes on non-orientifolded Calabi-Yau manifolds, to D7-branes in orientifolds. Note that this implies that similarly in F-theory, 4-form fluxes will in general be induced when deforming away from a singular fourfold with enhanced gauge symmetry.

\item We illustrated how to obtain residual gauge symmetries, moduli and charged matter multiplets from the tachyon picture.

     The (F-term) D7 moduli space is parametrized by tachyon configurations $T$ modulo
     \begin{equation} \label{equivlences}
      T \to g \cdot T \cdot \sigma^* g^t \, ,
     \end{equation}
     for automorphisms $g:E \to E$. The physical moduli space also must take into account D-term constraints, which we briefly discussed.

     The residual gauge symmetries $G$ for a given $T$ are the automorphisms $g:F \to F$ independent of the internal coordinates, for which $T = g \cdot T \cdot g^t$. Generically this is $G=\IZ_2$, but at special values of $T$ this can be enhanced. Charged matter then corresponds to linearized fluctuations of particular entries in the tachyon matrix, modulo linearized equivalences (\ref{equivlences}).

\item Finally, we compared these results to general open string index computations, in particular to the index counting the net number of open string modes between a brane of charge $\Gamma$ and its image brane of charge $\Gamma'$:
    \begin{equation}
      I_o(\Gamma) = \frac{1}{2} \left( \langle \Gamma', \Gamma \rangle + \frac{1}{4} \langle \Gamma_{\rm O}, \Gamma \rangle \right) \, ,
    \end{equation}
    where $\Gamma_O$ is the total charge of the orientifold planes. This agrees with the net number found in the tachyon picture, but the latter also gives the absolute number, which typically exceeds the net number.

\end{enumerate}

These results remove some of the obstacles to developing a systematic understanding of the landscape of IIB orientifold vacua. In particular our analysis clarifies the relation between gauge theory and geometry. The framework we presented is in principle sufficient for detailed enumeration of supersymmetric D7 flux vacua in IIB orientifolds in a way similar to how D4 flux vacua are enumerated in the context of type IIA D4-D2-D0 black hole microstate counting. We plan to return to this in a future publication.

One important element which we did not address however is the effect of bulk flux on the D-brane sector, and the interplay between bulk and brane sectors as far as moduli stabilization is concerned. It seems to us that this is best addressed in the full F-theory framework. On the other hand, the D9-D9$'$ tachyon condensation picture is the most efficient and concrete framework to derive gauge theory content and connect gauge theory to geometry, but it is a perturbative string construction. It would therefore be desirable to find the analog of this tachyon condensation mechanism in F-theory.

\section*{Acknowledgements}

\noindent We would like to thank D.~Anninos, K.~Hori, M.~Kreuzer, F.~Marchesano, B.~Stefanski, D.~Van den Bleeken, and B.~Vercnocke for useful discussions, and P.~Aluffi for initial collaboration and crucial help in understanding topological invariants of singular varieties. A.C. would like to thank especially T.~Wyder for numerous long and useful discussions. F.D. would like to thank G.~Moore in particular for many illuminating comments and conversations about orientifolds. M.E. is grateful to the Max-Planck-Institute f\"ur Mathematik  (Bonn),  the 2007
Simons Workshop in Mathematics and Physics, and the Department
of Mathematics of  Florida State University for their
hospitality.

This work was supported in part by the European Community's Human Potential Programme under contract
MRTN-CT-2004-005104 `Constituents, fundamental forces and symmetries of the
universe', by the FWO - Vlaanderen, project G.0235.05 and by the Federal Office
for Scientific, Technical and Cultural Affairs through the `Interuniversity
Attraction Poles Programme - Belgian Science Policy' P6/11-P, as well as by
the Milton Fund, the Austrian Research Funds FWF under grant number P19051-N16, and a bake sale.


\appendix


\begin{thebibliography}{99}



\bibitem{Kachru:2003aw}
  S.~Kachru, R.~Kallosh, A.~Linde and S.~P.~Trivedi,
  ``De Sitter vacua in string theory,''
  Phys.\ Rev.\  D {\bf 68}, 046005 (2003)
  [arXiv:\hepth{0301240}].

\bibitem{Balasubramanian:2005zx}
  V.~Balasubramanian, P.~Berglund, J.~P.~Conlon and F.~Quevedo,
  ``Systematics of moduli stabilisation in Calabi-Yau flux
  compactifications,''
  JHEP {\bf 0503}, 007 (2005)
  [arXiv:\hepth{0502058}].

\bibitem{Douglas:2006es}
  M.~R.~Douglas and S.~Kachru,
  ``Flux compactification,''
  Rev.\ Mod.\ Phys.\  {\bf 79}, 733 (2007)
  [arXiv:\hepth{0610102}].

\bibitem{Grana:2005jc}
  M.~Grana,
  ``Flux compactifications in string theory: A comprehensive review,''
  Phys.\ Rept.\  {\bf 423}, 91 (2006)
  [arXiv:\hepth{0509003}].

\bibitem{Denef:2008wq}
  F.~Denef,
  ``Les Houches Lectures on Constructing String Vacua,''
  \arXivid{0803.1194} [hep-th].

\bibitem{Bousso:2000xa}
  R.~Bousso and J.~Polchinski,
  ``Quantization of four-form fluxes and dynamical neutralization of the
  cosmological constant,''
  JHEP {\bf 0006}, 006 (2000)
  [arXiv:\hepth{0004134}].

\bibitem{Denef:2004ze}
  F.~Denef and M.~R.~Douglas,
  ``Distributions of flux vacua,''
  JHEP {\bf 0405}, 072 (2004)
  [arXiv:\hepth{0404116}].

\bibitem{Giddings:2001yu}
  S.~B.~Giddings, S.~Kachru and J.~Polchinski,
  ``Hierarchies from fluxes in string compactifications,''
  Phys.\ Rev.\  D {\bf 66}, 106006 (2002)
  [arXiv:\hepth{0105097}].

\bibitem{Klebanov:2000hb}
  I.~R.~Klebanov and M.~J.~Strassler,
  ``Supergravity and a confining gauge theory: Duality cascades and
  chiSB-resolution of naked singularities,''
  JHEP {\bf 0008}, 052 (2000)
  [arXiv:\hepth{0007191}].

\bibitem{Kachru:2003sx}
  S.~Kachru, R.~Kallosh, A.~Linde, J.~M.~Maldacena, L.~P.~McAllister and S.~P.~Trivedi,
  ``Towards inflation in string theory,''
  JCAP {\bf 0310}, 013 (2003)
  [arXiv:\hepth{0308055}].

\bibitem{Baumann:2007np}
  D.~Baumann, A.~Dymarsky, I.~R.~Klebanov, L.~McAllister and P.~J.~Steinhardt,
  ``A Delicate Universe,''
  Phys.\ Rev.\ Lett.\  {\bf 99}, 141601 (2007)
  [\arXivid{0705.3837} [hep-th]].

\bibitem{Baumann:2007ah}
  D.~Baumann, A.~Dymarsky, I.~R.~Klebanov and L.~McAllister,
  ``Towards an Explicit Model of D-brane Inflation,''
  \arXivid{0706.0360} [hep-th].

\bibitem{Krause:2007jk}
  A.~Krause and E.~Pajer,
  ``Chasing Brane Inflation in String-Theory,''
  \arXivid{0705.4682} [hep-th].

\bibitem{Sen:1996vd}
  A.~Sen,
  ``F-theory and Orientifolds,''
  Nucl.\ Phys.\  B {\bf 475}, 562 (1996)
  [arXiv:\hepth{9605150}].

\bibitem{Blumenhagen:2006ci}
  R.~Blumenhagen, B.~Kors, D.~Lust and S.~Stieberger,
  ``Four-dimensional String Compactifications with D-Branes, Orientifolds   and
  Fluxes,''
  Phys.\ Rept.\  {\bf 445}, 1 (2007)
  [arXiv:\hepth{0610327}].

\bibitem{Marchesano:2007de}
  F.~Marchesano,
  ``Progress in D-brane model building,''
  Fortsch.\ Phys.\  {\bf 55}, 491 (2007)
  [arXiv:\hepth{0702094}].

\bibitem{Donagi:2008ca}
  R.~Donagi and M.~Wijnholt,
  ``Model Building with F-Theory,''
  \arXivid{0802.2969} [hep-th].

\bibitem{Beasley:2008dc}
  C.~Beasley, J.~J.~Heckman and C.~Vafa,
  ``GUTs and Exceptional Branes in F-theory - I,''
  \arXivid{0802.3391} [hep-th].



\bibitem{Blumenhagen:2003su}
  R.~Blumenhagen,
  ``Supersymmetric orientifolds of Gepner models,''
  JHEP {\bf 0311}, 055 (2003)
  [arXiv:\hepth{0310244}].

\bibitem{Blumenhagen:2004cg}
  R.~Blumenhagen and T.~Weigand,
  ``Chiral supersymmetric Gepner model orientifolds,''
  JHEP {\bf 0402}, 041 (2004)
  [arXiv:\hepth{0401148}].

\bibitem{Brunner:2003zm}
  I.~Brunner and K.~Hori,
  ``Orientifolds and mirror symmetry,''
  JHEP {\bf 0411}, 005 (2004)
  [arXiv:\hepth{0303135}].

\bibitem{Brunner:2004zd}
  I.~Brunner, K.~Hori, K.~Hosomichi and J.~Walcher,
  ``Orientifolds of Gepner models,''
  JHEP {\bf 0702}, 001 (2007)
  [arXiv:\hepth{0401137}].

\bibitem{Dijkstra:2004cc}
  T.~P.~T.~Dijkstra, L.~R.~Huiszoon and A.~N.~Schellekens,
  ``Supersymmetric standard model spectra from RCFT orientifolds,''
  Nucl.\ Phys.\  B {\bf 710}, 3 (2005)
  [arXiv:\hepth{0411129}].

\bibitem{Anastasopoulos:2006da}
  P.~Anastasopoulos, T.~P.~T.~Dijkstra, E.~Kiritsis and A.~N.~Schellekens,
  ``Orientifolds, hypercharge embeddings and the standard model,''
  Nucl.\ Phys.\  B {\bf 759}, 83 (2006)
  [arXiv:\hepth{0605226}].

\bibitem{GatoRivera:2007yi}
  B.~Gato-Rivera and A.~N.~Schellekens,
  ``Non-supersymmetric Tachyon-free Type-II and Type-I Closed Strings from
  RCFT,''
  Phys.\ Lett.\  B {\bf 656}, 127 (2007)
  [\arXivid{0709.1426} [hep-th]].

\bibitem{Becker:2006ks}
  K.~Becker, M.~Becker, C.~Vafa and J.~Walcher,
  ``Moduli stabilization in non-geometric backgrounds,''
  Nucl.\ Phys.\  B {\bf 770}, 1 (2007)
  [arXiv:hep-th/0611001].



\bibitem{Aldazabal:2000sa}
  G.~Aldazabal, L.~E.~Ibanez, F.~Quevedo and A.~M.~Uranga,
  ``D-branes at singularities: A bottom-up approach to the string  embedding of
  the standard model,''
  JHEP {\bf 0008}, 002 (2000)
  [arXiv:\hepth{0005067}].

\bibitem{Berenstein:2001nk}
  D.~Berenstein, V.~Jejjala and R.~G.~Leigh,
  ``The standard model on a D-brane,''
  Phys.\ Rev.\ Lett.\  {\bf 88}, 071602 (2002)
  [arXiv:\hepph{0105042}].

\bibitem{Verlinde:2005jr}
  H.~Verlinde and M.~Wijnholt,
  ``Building the standard model on a D3-brane,''
  JHEP {\bf 0701}, 106 (2007)
  [arXiv:\hepth{0508089}].

\bibitem{Buican:2006sn}
  M.~Buican, D.~Malyshev, D.~R.~Morrison, H.~Verlinde and M.~Wijnholt,
  ``D-branes at singularities, compactification, and hypercharge,''
  JHEP {\bf 0701}, 107 (2007)
  [arXiv:\hepth{0610007}].

\bibitem{Malyshev:2007zz}
  D.~Malyshev and H.~Verlinde,
  ``D-branes at Singularities and String Phenomenology,''
  Nucl.\ Phys.\ Proc.\ Suppl.\  {\bf 171}, 139 (2007)
  [\arXivid{0711.2451} [hep-th]].

\bibitem{Heckman:2007zp}
  J.~J.~Heckman, C.~Vafa, H.~Verlinde and M.~Wijnholt,
  ``Cascading to the MSSM,''
  \arXivid{0711.0387} [hep-ph].

\bibitem{Wijnholt:2007vn}
  M.~Wijnholt,
  ``Geometry of Particle Physics,''
  arXiv:\hepth{0703047}.



\bibitem{Franco:2006es}
  S.~Franco and A.~M.~..~Uranga,
  ``Dynamical SUSY breaking at meta-stable minima from D-branes at obstructed
  geometries,''
  JHEP {\bf 0606} (2006) 031
  [arXiv:\hepth{0604136}].

\bibitem{GarciaEtxebarria:2006rw}
  I.~Garcia-Etxebarria, F.~Saad and A.~M.~Uranga,
  ``Local models of gauge mediated supersymmetry breaking in string theory,''
  JHEP {\bf 0608}, 069 (2006)
  [arXiv:\hepth{0605166}].


\bibitem{Ooguri:2006pj}
  H.~Ooguri and Y.~Ookouchi,
  ``Landscape of supersymmetry breaking vacua in geometrically realized gauge
  theories,''
  Nucl.\ Phys.\  B {\bf 755} (2006) 239
  [arXiv:\hepth{0606061}].


\bibitem{Ooguri:2006bg}
  H.~Ooguri and Y.~Ookouchi,
  ``Meta-stable supersymmetry breaking vacua on intersecting branes,''
  Phys.\ Lett.\  B {\bf 641} (2006) 323
  [arXiv:\hepth{0607183}].


\bibitem{Argurio:2006ny}
  R.~Argurio, M.~Bertolini, S.~Franco and S.~Kachru,
  ``Gauge/gravity duality and meta-stable dynamical supersymmetry breaking,''
  JHEP {\bf 0701} (2007) 083
  [arXiv:\hepth{0610212}].

\bibitem{Aganagic:2006ex}
  M.~Aganagic, C.~Beem, J.~Seo and C.~Vafa,
  ``Geometrically induced metastability and holography,''
  Nucl.\ Phys.\  B {\bf 789}, 382 (2008)
  [arXiv:\hepth{0610249}].

\bibitem{Giveon:2007fk}
  A.~Giveon and D.~Kutasov,
  ``Gauge symmetry and supersymmetry breaking from intersecting branes,''
  Nucl.\ Phys.\  B {\bf 778}, 129 (2007)
  [arXiv:\hepth{0703135}].

\bibitem{Argurio:2007qk}
  R.~Argurio, M.~Bertolini, S.~Franco and S.~Kachru,
  ``Metastable vacua and D-branes at the conifold,''
  JHEP {\bf 0706}, 017 (2007)
  [arXiv:\hepth{0703236}].


\bibitem{Aharony:2007db}
  O.~Aharony, S.~Kachru and E.~Silverstein,
  ``Simple Stringy Dynamical SUSY Breaking,''
  Phys.\ Rev.\  D {\bf 76}, 126009 (2007)
  [\arXivid{0708.0493} [hep-th]].


\bibitem{Aganagic:2007kd}
  M.~Aganagic, C.~Beem and B.~Freivogel,
  ``Geometric Metastability, Quivers and Holography,''
  Nucl.\ Phys.\  B {\bf 795}, 291 (2008)
  [\arXivid{0708.0596} [hep-th]].

\bibitem{Aganagic:2007py}
  M.~Aganagic, C.~Beem and S.~Kachru,
  ``Geometric Transitions and Dynamical SUSY Breaking,''
  Nucl.\ Phys.\  B {\bf 796}, 1 (2008)
  [\arXivid{0709.4277} [hep-th]].

\bibitem{Aganagic:2007zm}
  M.~Aganagic and C.~Beem,
  ``Geometric Transitions and D-Term SUSY Breaking,''
  Nucl.\ Phys.\  B {\bf 796}, 44 (2008)
  [\arXivid{0711.0385} [hep-th]].




\bibitem{Douglas:2003um}
  M.~R.~Douglas,
  ``The statistics of string / M theory vacua,''
  JHEP {\bf 0305}, 046 (2003)
  [arXiv:\hepth{0303194}].

\bibitem{Ashok:2003gk}
  S.~Ashok and M.~R.~Douglas,
  ``Counting flux vacua,''
  JHEP {\bf 0401}, 060 (2004)
  [arXiv:\hepth{0307049}].

\bibitem{Denef:2004cf}
  F.~Denef and M.~R.~Douglas,
  ``Distributions of nonsupersymmetric flux vacua,''
  JHEP {\bf 0503}, 061 (2005)
  [arXiv:\hepth{0411183}].

\bibitem{Acharya:2005ez}
  B.~S.~Acharya, F.~Denef and R.~Valandro,
  ``Statistics of M theory vacua,''
  JHEP {\bf 0506}, 056 (2005)
  [arXiv:\hepth{0502060}].

\bibitem{Douglas:2005df}
  M.~R.~Douglas, B.~Shiffman and S.~Zelditch,
  ``Critical points and supersymmetric vacua. III: String/M models,''
  Commun.\ Math.\ Phys.\  {\bf 265}, 617 (2006)
  [arXiv:math-ph/0506015].


\bibitem{Dienes:2006ut}
  K.~R.~Dienes,
  ``Statistics on the heterotic landscape: Gauge groups and cosmological
  constants of four-dimensional heterotic strings,''
  Phys.\ Rev.\  D {\bf 73}, 106010 (2006)
  [arXiv:\hepth{0602286}].


\bibitem{Blumenhagen:2004xx}
  R.~Blumenhagen, F.~Gmeiner, G.~Honecker, D.~Lust and T.~Weigand,
  ``The statistics of supersymmetric D-brane models,''
  Nucl.\ Phys.\  B {\bf 713}, 83 (2005)
  [arXiv:\hepth{0411173}].

\bibitem{Gmeiner:2005vz}
  F.~Gmeiner, R.~Blumenhagen, G.~Honecker, D.~Lust and T.~Weigand,
  ``One in a billion: MSSM-like D-brane statistics,''
  JHEP {\bf 0601}, 004 (2006)
  [arXiv:\hepth{0510170}].

\bibitem{Gmeiner:2005nh}
  F.~Gmeiner,
  ``Standard model statistics of a type II orientifold,''
  Fortsch.\ Phys.\  {\bf 54}, 391 (2006)
  [arXiv:\hepth{0512190}].

\bibitem{Gmeiner:2006vb}
  F.~Gmeiner and M.~Stein,
  ``Statistics of SU(5) D-brane models on a type II orientifold,''
  Phys.\ Rev.\  D {\bf 73}, 126008 (2006)
  [arXiv:\hepth{0603019}].

\bibitem{Douglas:2006xy}
  M.~R.~Douglas and W.~Taylor,
  ``The landscape of intersecting brane models,''
  JHEP {\bf 0701}, 031 (2007)
  [arXiv:\hepth{0606109}].

\bibitem{Gomis:2005wc}
  J.~Gomis, F.~Marchesano and D.~Mateos,
  ``An open string landscape,''
  JHEP {\bf 0511} (2005) 021
  [arXiv:\hepth{0506179}].




\bibitem{Brunner:1999jq}
  I.~Brunner, M.~R.~Douglas, A.~E.~Lawrence and C.~Romelsberger,
  ``D-branes on the quintic,''
  JHEP {\bf 0008}, 015 (2000)
  [arXiv:\hepth{9906200}].

\bibitem{Douglas:2000gi}
  M.~R.~Douglas,
  ``D-branes, categories and N = 1 supersymmetry,''
  J.\ Math.\ Phys.\  {\bf 42} (2001) 2818
  [arXiv:\hepth{0011017}].

\bibitem{Aspinwall:2004jr}
  P.~S.~Aspinwall,
  ``D-branes on Calabi-Yau manifolds,''
  [arXiv:\hepth{0403166}].

\bibitem{Herbst:2008jq}
  M.~Herbst, K.~Hori and D.~Page,
  ``Phases Of N=2 Theories In 1+1 Dimensions With Boundary,''
  \arXivid{0803.2045} [hep-th].





\bibitem{Blumenhagen:2002wn}
  R.~Blumenhagen, V.~Braun, B.~Kors and D.~Lust,
  ``Orientifolds of K3 and Calabi-Yau manifolds with intersecting D-branes,''
  JHEP {\bf 0207}, 026 (2002)
  [arXiv:\hepth{0206038}].

\bibitem{Blumenhagen:2003vr}
  R.~Blumenhagen, D.~Lust and T.~R.~Taylor,
  ``Moduli stabilization in chiral type IIB orientifold models with fluxes,''
  Nucl.\ Phys.\  B {\bf 663}, 319 (2003)
  [arXiv:\hepth{0303016}].

\bibitem{Lust:2004fi}
  D.~Lust, S.~Reffert and S.~Stieberger,
  ``Flux-induced soft supersymmetry breaking in chiral type IIb  orientifolds
  with D3/D7-branes,''
  Nucl.\ Phys.\  B {\bf 706}, 3 (2005)
  [arXiv:\hepth{0406092}].

\bibitem{Marchesano:2004xz}
  F.~Marchesano and G.~Shiu,
  ``Building MSSM flux vacua,''
  JHEP {\bf 0411}, 041 (2004)
  [arXiv:\hepth{0409132}].

\bibitem{Denef:2004dm}
  F.~Denef, M.~R.~Douglas and B.~Florea,
  ``Building a better racetrack,''
  JHEP {\bf 0406}, 034 (2004)
  [arXiv:\hepth{0404257}].

\bibitem{Denef:2005mm}
  F.~Denef, M.~R.~Douglas, B.~Florea, A.~Grassi and S.~Kachru,
  ``Fixing all moduli in a simple F-theory compactification,''
  Adv.\ Theor.\ Math.\ Phys.\  {\bf 9}, 861 (2005)
  [arXiv:\hepth{0503124}].

\bibitem{Lust:2005bd}
  D.~Lust, P.~Mayr, S.~Reffert and S.~Stieberger,
  ``F-theory flux, destabilization of orientifolds and soft terms on
  D7-branes,''
  Nucl.\ Phys.\  B {\bf 732}, 243 (2006)
  [arXiv:\hepth{0501139}].

\bibitem{Lust:2005dy}
  D.~Lust, S.~Reffert, W.~Schulgin and S.~Stieberger,
  ``Moduli stabilization in type IIB orientifolds. I: Orbifold limits,''
  Nucl.\ Phys.\  B {\bf 766}, 68 (2007)
  [arXiv:\hepth{0506090}].

\bibitem{Lust:2006zg}
  D.~Lust, S.~Reffert, E.~Scheidegger, W.~Schulgin and S.~Stieberger,
  ``Moduli stabilization in type IIB orientifolds. II,''
  Nucl.\ Phys.\  B {\bf 766}, 178 (2007)
  [arXiv:\hepth{0609013}].

\bibitem{Diaconescu:2005pc}
  D.~E.~Diaconescu, B.~Florea, S.~Kachru and P.~Svrcek,
  ``Gauge - mediated supersymmetry breaking in string compactifications,''
  JHEP {\bf 0602}, 020 (2006)
  [arXiv:\hepth{0512170}].

\bibitem{Diaconescu:2006nk}
  D.~E.~Diaconescu, A.~Garcia-Raboso and K.~Sinha,
  ``A D-brane landscape on Calabi-Yau manifolds,''
  JHEP {\bf 0606}, 058 (2006)
  [arXiv:\hepth{0602138}].

\bibitem{Diaconescu:2007ah}
  D.~E.~Diaconescu, R.~Donagi and B.~Florea,
  ``Metastable quivers in string compactifications,''
  Nucl.\ Phys.\  B {\bf 774}, 102 (2007)
  [arXiv:\hepth{0701104}].

\bibitem{Conlon:2005ki}
  J.~P.~Conlon, F.~Quevedo and K.~Suruliz,
  ``Large-volume flux compactifications: Moduli spectrum and D3/D7 soft
  supersymmetry breaking,''
  JHEP {\bf 0508}, 007 (2005)
  [arXiv:\hepth{0505076}].


\bibitem{Conlon:2006gv}
  J.~P.~Conlon,
  ``Moduli stabilisation and applications in IIB string theory,''
  Fortsch.\ Phys.\  {\bf 55}, 287 (2007)
  [arXiv:\hepth{0611039}].

\bibitem{AbdusSalam:2007pm}
  S.~S.~AbdusSalam, J.~P.~Conlon, F.~Quevedo and K.~Suruliz,
  ``Scanning the Landscape of Flux Compactifications: Vacuum Structure and Soft
  Supersymmetry Breaking,''
  JHEP {\bf 0712}, 036 (2007)
  [\arXivid{0709.0221} [hep-th]].


\bibitem{Blumenhagen:2007sm}
  R.~Blumenhagen, S.~Moster and E.~Plauschinn,
  ``Moduli Stabilisation versus Chirality for MSSM like Type IIB
  Orientifolds,''
  JHEP {\bf 0801}, 058 (2008)
  [\arXivid{0711.3389} [hep-th]].


\bibitem{Gaiotto:2005rp}
  D.~Gaiotto, M.~Guica, L.~Huang, A.~Simons, A.~Strominger and X.~Yin,
  ``D4-D0 branes on the quintic,''
  JHEP {\bf 0603}, 019 (2006)
  [arXiv:\hepth{0509168}].

\bibitem{Gaiotto:2006aj}
  D.~Gaiotto and L.~Huang,
  ``D4-branes on Complete Intersection in Toric Variety,''
  [arXiv:\hepth{0612295}].



\bibitem{Klemm:1996ts}
  A.~Klemm, B.~Lian, S.~S.~Roan and S.~T.~Yau,
  ``Calabi-Yau fourfolds for M- and F-theory compactifications,''
  Nucl.\ Phys.\  B {\bf 518}, 515 (1998)
  [arXiv:\hepth{9701023}].

\bibitem{Andreas:1999ng}
  B.~Andreas and G.~Curio,
  ``On discrete twist and four-flux in N = 1 heterotic/F-theory
  compactifications,''
  Adv.\ Theor.\ Math.\ Phys.\  {\bf 3}, 1325 (1999)
  [arXiv:hep-th/9908193].



\bibitem{Gaiotto:2006wm}
  D.~Gaiotto, A.~Strominger and X.~Yin,
  ``The M5-brane elliptic genus: Modularity and BPS states,''
  JHEP {\bf 0708}, 070 (2007)
  [arXiv:\hepth{0607010}].

\bibitem{Denef:2007vg}
  F.~Denef and G.~W.~Moore,
  ``Split states, entropy enigmas, holes and halos,''
  [arXiv:\hepth{0702146}].




















\bibitem{Vafa:1996xn}
  C.~Vafa,
  ``Evidence for F-Theory,''
  Nucl.\ Phys.\  B {\bf 469} (1996) 403
  [arXiv: \hepth{9602022}].




\bibitem{Sethi:1996es}
  S.~Sethi, C.~Vafa and E.~Witten,
  ``Constraints on low-dimensional string compactifications,''
  Nucl.\ Phys.\  B {\bf 480} (1996) 213
  [arXiv: \hepth{9606122}].


\bibitem{Sen:1997gv}
  A.~Sen,
  ``Orientifold limit of F-theory vacua,''
  Phys.\ Rev.\  D {\bf 55}, 7345 (1997)
  [arXiv:\hepth{9702165}].

\bibitem{Gimon:1996rq}
  E.~G.~Gimon and J.~Polchinski,
  ``Consistency Conditions for Orientifolds and D-Manifolds,''
  Phys.\ Rev.\  D {\bf 54}, 1667 (1996)
  [arXiv:\hepth{9601038}].

\bibitem{Braun:2008ua}
  A.~P.~Braun, A.~Hebecker and H.~Triendl,
  ``D7-Brane Motion from M-Theory Cycles and Obstructions in the Weak Coupling
  Limit,''
  \arXivid{0801.2163} [hep-th].

\bibitem{Morales:1998ux}
  J.~F.~Morales, C.~A.~Scrucca and M.~Serone,
  ``Anomalous couplings for D-branes and O-planes,''
  Nucl.\ Phys.\  B {\bf 552} (1999) 291
  [arXiv:\hepth{9812071}].

\bibitem{Stefanski:1998yx}
  B.~J.~Stefanski,
  ``Gravitational couplings of D-branes and O-planes,''
  Nucl.\ Phys.\  B {\bf 548}, 275 (1999)
  [arXiv:hep-th/9812088].


\bibitem{Scrucca:1999uz}
  C.~A.~Scrucca and M.~Serone,
  ``Anomalies and inflow on D-branes and O-planes,''
  Nucl.\ Phys.\  B {\bf 556} (1999) 197
  [arXiv:\hepth{9903145}].

\bibitem{Minasian:1997mm}
  R.~Minasian and G.~W.~Moore,
  ``K-theory and Ramond-Ramond charge,''
  JHEP {\bf 9711} (1997) 002
  [arXiv:\hepth{9710230}].

\bibitem{Freed:1999vc}
  D.~S.~Freed and E.~Witten,
  ``Anomalies in string theory with D-branes,''
  [arXiv:\hepth{9907189}].














\bibitem{Becker:1996gj}
  K.~Becker and M.~Becker,
  ``M-Theory on Eight-Manifolds,''
  Nucl.\ Phys.\  B {\bf 477}, 155 (1996)
  [arXiv:\hepth{9605053}].

\bibitem{Gukov:1999ya}
  S.~Gukov, C.~Vafa and E.~Witten,
  ``CFT's from Calabi-Yau four-folds,''
  Nucl.\ Phys.\  B {\bf 584}, 69 (2000)
  [Erratum-ibid.\  B {\bf 608}, 477 (2001)]
  [arXiv:\hepth{9906070}].







\bibitem{Aluffi:2007sx}
  P.~Aluffi and M.~Esole,
  ``Chern class identities from tadpole matching in type IIB and F-theory,''
  [\arXivid{0710.2544}].


\bibitem{GH}
  P.~Griffiths and J.~Harris,
  \textit{Principles of Algebraic Geometry},
  Wiley, 1978.

\bibitem{wikiwhitney} \href{http://en.wikipedia.org/wiki/Whitney_umbrella}{Wikipedia: Whitney Umbrella}


\bibitem{GarciaEtxebarria:2005qc}
  I.~Garcia-Etxebarria and A.~M.~Uranga,
  ``From F/M-theory to K-theory and back,''
  JHEP {\bf 0602}, 008 (2006)
  [arXiv:\hepth{0510073}].

\bibitem{Witten:1997ep}
  E.~Witten,
  ``Branes and the dynamics of {QCD},''
  Nucl.\ Phys.\  B {\bf 507} (1997) 658
  [arXiv:\hepth{9706109}].


\bibitem{Martucci:2006ij}
  L.~Martucci,
  ``D-branes on general N = 1 backgrounds: Superpotentials and D-terms,''
  JHEP {\bf 0606} (2006) 033
  [arXiv:\hepth{0602129}].



\bibitem{Huang:2006hq}
  M.~x.~Huang, A.~Klemm and S.~Quackenbush,
  ``Topological String Theory on Compact Calabi-Yau: Modularity and Boundary
  arXiv:\hepth{0612125}.



\bibitem{Sen:1998sm}
  A.~Sen,
  ``Tachyon condensation on the brane antibrane system,''
  JHEP {\bf 9808} (1998) 012
  [arXiv:\hepth{9805170}].

\bibitem{Sen:2004nf}
  A.~Sen,
  ``Tachyon dynamics in open string theory,''
  Int.\ J.\ Mod.\ Phys.\  A {\bf 20}, 5513 (2005)
  [arXiv:\hepth{0410103}].

\bibitem{atiyah}
M.F.~Atiyah, ``K-Theory,'' (1967) W.A.~Benjamin, Inc. New York


\bibitem{Witten:1998cd}
  E.~Witten,
  ``D-branes and K-theory,''
  JHEP {\bf 9812}, 019 (1998)
  [arXiv:\hepth{9810188}].

\bibitem{Horava:1998jy}
  P.~Horava,
  ``Type IIA D-branes, K-theory, and matrix theory,''
  Adv.\ Theor.\ Math.\ Phys.\  {\bf 2}, 1373 (1999)
  [arXiv:\hepth{9812135}].

\bibitem{Gukov:1999yn}
  S.~Gukov,
  ``K-theory, reality, and orientifolds,''
  Commun.\ Math.\ Phys.\  {\bf 210}, 621 (2000)
  [arXiv:\hepth{9901042}].

\bibitem{Hori:1999me}
  K.~Hori,
  ``D-branes, T-duality, and index theory,''
  Adv.\ Theor.\ Math.\ Phys.\  {\bf 3}, 281 (1999)
  [arXiv:hep-th/9902102].


\bibitem{Bergman:1999ta}
  O.~Bergman, E.~G.~Gimon and P.~Horava,
  ``Brane transfer operations and T-duality of non-BPS states,''
  JHEP {\bf 9904}, 010 (1999)
  [arXiv:hep-th/9902160].

\bibitem{Olsen:1999xx}
  K.~Olsen and R.~J.~Szabo,
  ``Constructing D-branes from K-theory,''
  Adv.\ Theor.\ Math.\ Phys.\  {\bf 3}, 889 (1999)
  [arXiv:\hepth{9907140}].

\bibitem{Moore:1999gb}
  G.~W.~Moore and E.~Witten,
  ``Self-duality, Ramond-Ramond fields, and K-theory,''
  JHEP {\bf 0005}, 032 (2000)
  [arXiv:\hepth{9912279}].

\bibitem{Bouwknegt:2000qt}
  P.~Bouwknegt and V.~Mathai,
  ``D-branes, B-fields and twisted K-theory,''
  JHEP {\bf 0003}, 007 (2000)
  [arXiv:\hepth{0002023}].

\bibitem{Diaconescu:2000wy}
  D.~E.~Diaconescu, G.~W.~Moore and E.~Witten,
  ``E(8) gauge theory, and a derivation of K-theory from M-theory,''
  Adv.\ Theor.\ Math.\ Phys.\  {\bf 6}, 1031 (2003)
  [arXiv:\hepth{0005090}].

\bibitem{Witten:2000cn}
  E.~Witten,
  ``Overview of K-theory applied to strings,''
  Int.\ J.\ Mod.\ Phys.\  A {\bf 16}, 693 (2001)
  [arXiv:\hepth{0007175}].

\bibitem{Harvey:2000te}
  J.~A.~Harvey and G.~W.~Moore,
  ``Noncommutative tachyons and K-theory,''
  J.\ Math.\ Phys.\  {\bf 42}, 2765 (2001)
  [arXiv:\hepth{0009030}].

\bibitem{Oz:2000bs}
  Y.~Oz, T.~Pantev and D.~Waldram,
  ``Brane-antibrane systems on Calabi-Yau spaces,''
  JHEP {\bf 0102} (2001) 045
  [arXiv:\hepth{0009112}].

\bibitem{Maiden:2006qe}
  J.~Maiden, G.~Shiu and B.~J.~Stefanski,
  ``D-brane spectrum and K-theory constraints of D = 4, N = 1 orientifolds,''
  JHEP {\bf 0604} (2006) 052
  [arXiv:\hepth{0602038}].

\bibitem{Uranga:2000xp}
  A.~M.~Uranga,
  ``D-brane probes, RR tadpole cancellation and K-theory charge,''
  Nucl.\ Phys.\  B {\bf 598}, 225 (2001)
  [arXiv:\hepth{0011048}].

\bibitem{Maldacena:2001xj}
  J.~M.~Maldacena, G.~W.~Moore and N.~Seiberg,
  ``D-brane instantons and K-theory charges,''
  JHEP {\bf 0111}, 062 (2001)
  [arXiv:\hepth{0108100}].


\bibitem{Moore:2003vf}
 G.~W.~Moore,
 ``K-theory from a physical perspective,''
 arXiv:\hepth{0304018}.



\bibitem{Katz:2002gh}
  S.~H.~Katz and E.~Sharpe,
  ``D-branes, open string vertex operators, and Ext groups,''
  Adv.\ Theor.\ Math.\ Phys.\  {\bf 6} (2003) 979
  [arXiv:\hepth{0208104}].



\bibitem{Iqbal:2003ds}
  A.~Iqbal, N.~Nekrasov, A.~Okounkov and C.~Vafa,
  ``Quantum foam and topological strings,''
  [arXiv:\hepth{0312022}].

\bibitem{MNOP1}
 D.~Maulik, N.~Nekrasov, A.~Okounkov and R.~Pandharipande,
 ``Gromov-Witten theory and Donaldson-Thomas theory,
 I,'' [arXiv:\Math{AG}{0312059}].

\bibitem{MNOP2}
 D.~Maulik, N.~Nekrasov, A.~Okounkov and R.~Pandharipande, ``Gromov-Witten theory and Donaldson-Thomas theory,
 II,'' [arXiv:\Math{AG}{0406092}].

\bibitem{Dijkgraaf:2006um}
  R.~Dijkgraaf, C.~Vafa and E.~Verlinde,
  ``M-theory and a topological string duality,''
  [arXiv:\hepth{0602087}].

\bibitem{Hori:2006ic}
  K.~Hori and J.~Walcher,
  ``D-brane categories for orientifolds: The Landau-Ginzburg case,''
  JHEP {\bf 0804}, 030 (2008)
  [arXiv:hep-th/0606179].


\bibitem{Diaconescu:2006id}
  D.~E.~Diaconescu, A.~Garcia-Raboso, R.~L.~Karp and K.~Sinha,
  ``D-brane superpotentials in Calabi-Yau orientifolds (projection),''
  arXiv:hep-th/0606180.





\bibitem{Witten:1982fp}
  E.~Witten,
  ``An SU(2) anomaly,''
  \plb{117}{1982}{324}.


\bibitem{Douglas:2000ah}
  M.~R.~Douglas, B.~Fiol and C.~Romelsberger,
  ``Stability and BPS branes,''
  JHEP {\bf 0509}, 006 (2005)
  [arXiv:\hepth{0002037}].

\bibitem{bridgeland}
 T.~Bridgeland, ``Spaces of stability conditions,''
 arXiv:\Math{AG}{0611510}.

\bibitem{Denef:2000nb}
  F.~Denef,
  ``Supergravity flows and D-brane stability,''
  JHEP {\bf 0008}, 050 (2000)
  [arXiv:\hepth{0005049}].

\bibitem{ACDE}
  P.~Aluffi, A.~Collinucci, F.~Denef and M.~Esole,
  to appear

\bibitem{mooretexas}
 J.~Distler, D.~Freed and G.~W.~Moore, to appear. See also
 \href{http://golem.ph.utexas.edu:2500/jacques/s5/Orientifolds+and+Twisted+KR+Theory }{http://golem.ph.utexas.edu:2500/jacques/s5/Orientifolds+and+Twisted+KR+Theory}

\bibitem{hori}
 K.~Hori  and D.~Gao,
  to appear. See also
 \href{http://online.kitp.ucsb.edu/online/strings05/hori/oh/01.html}{http://online.kitp.ucsb.edu/online/strings05/hori/oh/01.html}

\end{thebibliography}
\end{document}